\documentclass[useAMS,usenatbib,fleqn]{mn2e}
\usepackage{graphicx}
\usepackage{amsmath}
\usepackage{subfigure}

\def\gsim{\;\lower4pt\hbox{${\buildrel\displaystyle >\over\sim}$}\;}
\def\lsim{\;\lower4pt\hbox{${\buildrel\displaystyle <\over\sim}$}\;}
\def\grls{\;\lower4pt\hbox{${\buildrel\displaystyle >\over <}$}\;}

\title[Coupled Magnetized Scale-Free Discs]
{Global structures in a composite system of two\\
scale-free discs with a coplanar magnetic field }
\author[Y.-Q. Lou, X. N. Bai]{
Yu-Qing Lou$^{1,2,3}$\thanks{E-mail: xiaobai\_626@163.com;
louyq@mail.tsinghua.edu.cn
and lou@oddjob.uchicago.edu}
and Xue-Ning Bai$^{1}$ \\
$^{1}$Physics Department and Tsinghua Centre for Astrophysics
(THCA), Tsinghua University, Beijing 100084, China;\\
$^{2}$Department of Astronomy and Astrophysics, The University
of Chicago, 5640 South Ellis Avenue, Chicago, IL 60637, USA;\\
$^{3}$National Astronomical Observatories, Chinese Academy
of Science, A20, Datun Road, Beijing 100012, China. }
\begin{document}

\date{Accepted  . Received ; in original form }

\pagerange{\pageref{firstpage}--\pageref{lastpage}} \pubyear{2006}

\maketitle

\label{firstpage}

\begin{abstract}

We investigate a theoretical magnetohydrodynamic (MHD) disc
problem involving a composite disc system of gravitationally
coupled stellar and gaseous discs with a coplanar magnetic
field in the presence of an axisymmetric dark matter halo.
The two discs are expediently approximated as razor-thin,
with a barotropic equation of state,
a power-law surface mass density,
a ring-like magnetic field,
and a power-law rotation curve in radius $r$.
By imposing the scale-free condition, we construct analytically stationary
global MHD perturbation configurations for both aligned and logarithmic
spiral patterns using our composite MHD disc model. MHD perturbation
configurations in a composite system of partial discs in the presence
of an axisymmetric dark matter halo are also considered. Our study
generalizes the previous analyses of Lou \& Shen and Shen \& Lou
on the unmagnetized composite system of two gravitationally
coupled isothermal and scale-free discs, of Lou and Shen et al.
on the cases of a single coplanarly magnetized isothermal and
scale-free disc, and of Lou \& Zou
on magnetized two coupled singular isothermal discs. We derive
analytically the stationary MHD dispersion relations for both
aligned and unaligned perturbation structures and analyze the
corresponding phase relationships between surface mass densities
and the magnetic field. Compared with earlier results, we obtain
three solution branches corresponding to super fast MHD density
waves (sFMDWs), fast MHD density waves (FMDWs) and slow MHD
density waves (SMDWs), respectively. We examine the $m=0$ cases
for both aligned and unaligned MHD perturbations. By evaluating
the unaligned $m=0$ case, we determine the marginal stability
curves where the two unstable regimes corresponding to Jeans
collapse instability and ring fragmentation instability are
identified. We find that the aligned $m=0$ case is simply the
limit of the unaligned $m=0$ case with the radial wavenumber
$\xi\rightarrow0$ (i.e., the breathing mode) which does not merely
represent a rescaling of the equilibrium state. We further show
that a composite system of partial discs behaves much differently
from a composite system of full discs in certain aspects. We
provide numerical examples by varying dimensionless parameters
$\beta$ (rotation velocity index), $\eta$ (ratio of effective
sound speed of the two discs), $\delta$ (ratio of surface mass
density of the two discs), $q$ (a measure of coplanar magnetic
field strength), $F$ (gravity potential ratio),
$\xi$ (radial wavenumber). Our formalism provides a useful
theoretical framework in the study of stationary global
perturbation configurations for MHD disc galaxies with bars,
spirals and barred spirals.


\end{abstract}


\section{Introduction}
The large-scale structure of disc galaxies has long been studied
observationally and theoretically by astrophysicists from various
complementary aspects. \citet{b4,b4a} and their co-workers
pioneered the classic density wave theory and achieved a great
success in understanding the dynamical nature of spiral galaxies
\citep{b3c,b1,b0c}.
The basic idea of analyzing such a large-scale density wave
problem is to treat coplanar perturbations in a background
axisymmetric rotating disc; this procedure has been proven to be
powerful in probing the galactic dynamics to various extents. In
such a model development, perturbations (either linear or
nonlinear) are introduced onto a background equilibrium to form
local or large-scale structures and to perform stability analysis
under various situations. Broadly speaking, axisymmetric and
non-axisymmetric perturbations may lead to aligned configurations
while non-axisymmetric perturbations can naturally produce spiral
wave patterns. Theoretically, a disc system may be treated as
razor-thin for simplicity \citep{b13,b10,b7,b11}, with a mass
density $\rho=\Sigma(r,\theta)\delta(z)$ under the situation where
the thickness of a galactic disc is sufficiently small compared
with its radial size scale; in this manner, the model problem
reduces to a two-dimensional one.

In theoretical model investigations, two classes of disc models
are frequently encountered. One class is the so-called singular
isothermal discs (SIDs), which bears a flat rotation curve and a
constant `temperature' with a diverging surface mass density
towards the centre [e.g., \cite{b12}]. Since the early study by
\citet{b7d} more than four decades ago, this idealized theoretical
SID model has attracted considerable interest among
astrophysicists in various contexts of disc dynamics [e.g.,
\citet{b14,b13b,b3b,b2b,b12,b5,b5c,b6,b7,b6a,b7a}]. The other
class is the so called scale-free discs [e.g.
\citet{b13,b10,b11,b13e}]
and is the main focus of this paper. Being of a more general form
for differentially rotating discs, a scale-free disc has a
rotation curve in the form of $v\propto r^{-\beta}$ (the case of
$\beta=0$ corresponds to a flat rotation curve) and a barotropic
equation of state in the form of $\Pi=K\Sigma^n$ where $\Pi$ is
the vertically integrated two-dimensional pressure. There is no
characteristic spatial scale and all quantities in the disc system
vary as powers of radius $r$. In our analysis, the rotation
velocity index $\beta$ satisfies $-1/4<\beta<1/2$ for warm discs.
Furthermore, it is possible to construct stationary (i.e.
$\omega=0$) density wave patterns in scale-free discs
\citep{b13,b10,b11,b13e},
clearly indicating that the pattern speed of a density wave is in
the opposite sense of the disc rotation speed.

Because of the self-gravity, one important technical aspect in
dealing with thin disc galaxies involves the Poisson integral,
relating the mass density distribution and gravitational
potential. In the early development of the density wave theory,
this Poisson integral is evaluated by an asymptotic analysis valid
in the large wavenumber regime of the
Wentzel-Kramers-Brillouin-Jeffreys (WKBJ) approximation. Based on
this, the pursuit of analytical solutions leads to a further
exploration of the problem. Several potential-density pairs are
generalized in Chapter 2 of \citet{b1}. One special utility is the
logarithmic spiral potential-density pair first derived by
\citet{b3}. This is a powerful tool in analyzing non-axisymmetric
perturbations in an axisymmetric background disc. Furthermore,
\citet{b8} made use of the techniques of \citet{b7c} and found a
larger family of potential-density pairs in terms of the
generalized hypergeometric functions. Using these results for
scale-free discs under our consideration, the Poisson integral is
solved analytically for both aligned and spiral perturbations
\citep{b10,b11,b13e}.

Magnetic field is an important ingredient in various astrophysical
disc systems. In the interstellar medium (ISM), the partially or
fully ionized gases make the cosmic space an ideal place for
applying the magnetohydrodynamics (MHD) equations. In many
astrophysical situations involving large-scale dynamics, the
magnetic field can be effectively considered as completely frozen
into the gas. While in some cases the magnetic force is relatively
weak and has little impact on the dynamical process. There do
exist certain situations when the magnetic field plays a crucial
role in both dynamics and diagnostics, such as in spiral galaxies
and accretion processes [e.g.
\citet{b0b,b0a,b2,b2a,b12,b5a,b5,b5d,b7,b7a,b11,b6a}
]. In terms of MHD model development, we need to prescribe the
magnetic field geometry.
\citet{b12a} introduced a model in which a disc is `isopedically'
magnetized such that the mass-to-flux ratio remains spatially
uniform and the effect of magnetic field is subsumed into two
parameters \citep{b12a,b12,b6a,b13e}.
In fact, \citet{b6a}
have shown explicitly that a constant mass-to-flux ratio is a
natural consequence of the frozen-in condition from the ideal MHD
equations. In parallel, the coplanar magnetic field also serves as
an interesting model \citep{b5,b7,b11} with the magnetic field
being azimuthally embedded into the disc system. In particular, with
the inclusion of an azimuthal magnetic field, one can construct the
so-called \emph{fast and slow MHD density waves} (FMDWs and SMDWs)
\citep{b2,b2a,b5a,b5} and the interlaced optical and magnetic spiral
arms in the nearby spiral galaxy NGC 6946 are sensibly explained
along this line.

The stability of such MHD discs is yet another lively debated
issue. In the WKBJ or tight-winding regime, the well-known $Q$
parameter criterion \citep{b8a,b13a} was suggested to determine
the galactic local axisymmetric stability in the absence of
magnetic field. Meanwhile, there have been numerous studies
concerning the global stability of the disc problem [e.g.,
\citet{b14,b13b,b3b,b1a,b1b,b2b,b12}].
For example, stationary perturbations
of the zero-frequency neutral modes are emphasized as the marginal
instability modes in scale-free discs. In this context, \citet{b13} 
made a breakthrough in obtaining semi-analytic solutions for 
stationary perturbation configurations in a class of
scale-free discs. \citet{b12} analyzed the stability 
of isopedically magnetized SIDs and derived stationary 
perturbation solutions. They interpreted these aligned 
and unaligned configurations as onsets of bar-like 
instabilities. \citet{b5} performed a coplanar MHD
perturbation analysis on azimuthally magnetized SIDs 
from a perspective of stationary FMDWs and SMDWs.

Our analysis on two-dimensional coplanar MHD perturbations has 
avoided at least two major issues. If perturbation velocity 
and magnetic field components perpendicular to the disc plane 
are allowed, it is then possible to describe Alfv\'enic 
fluctuations and model disc warping process. If one further
takes into account of vertical variations across the disc
(i.e., the disc thickness is not negligible), magneto-rotational
instabilities can develop (e.g., Balbus 2003). These two aspects 
are important and should bear physical consequences in modelling 
disc galaxies.

In a typical disc galaxy system, the basic component involves
stars, gases, dusts, cosmic rays, and a massive dark matter halo
\citep{b5a,b5d}. In terms of theoretical analysis, it would be a
great challenge to include all these factors into one single model
consideration. While limited in certain aspects, it remains
sufficiently challenging and interesting to consider a composite
system consisting of a coplanarly magnetized gas disc, a stellar
disc as well as an axisymmetric background of a massive dark
matter halo. A seminal analysis concerning a composite disc system
dates back to \citet{b4a,b4b} who combined a stellar distribution
function and a gas fluid description to derive a local dispersion
relation for galactic spiral density waves in the WKBJ
approximation. Since then, there have been extensive theoretical
studies on perturbation configurations and stability problems of 
the composite disc system.
\citet{b2d,b2e} investigated the growth of local
axisymmetric perturbations in a composite stellar and gaseous disc
system. \citet{b0d} considered the spiral modes containing gas in a 
two-fluid model. \citet{b13c,b13d} studied the effect of interstellar 
gas on oscillations and the stability of spheroidal galaxies. 
\citet{b8z} considered the stability of two-component fluid discs 
with finite thickness. Two-fluid approach was adopted into modal 
analysis morphologies of spiral galaxies by \citet{b7a1}, supporting 
the notion that spiral structures are long-lasting and slowly 
evolving. \citet{b1z} and \citet{b2c} suggested an effective
$Q_{eff}$ (or $Q_{s-g}$) parameter criterion \citep{b8a,b13a} for
local axisymmetric two-fluid instabilities of a disc galaxy.
\citet{b5b} explored basic properties of open and tight spiral
density-waves modes in a two-fluid model to describe a composite 
system of coupled stellar and gaseous discs. Recently,
\citet{b6} studied a composite SID system to derive
stationary global perturbation configurations and further explored
the axisymmetric instability properties \citep{b9} where they
proposed a fairly straightforward $D$ criterion for the axisymmetric
instability problem for a composite SID system. \citet{b10} extended
these analysis to a composite system of two scale-free discs and
carried out analytical analysis on both aligned and logarithmic
spiral perturbation configurations. By adding a coplanar magnetic
field to the background composite SIDs, \citet{b7} obtained MHD
perturbation configurations and further studied the axisymmetric
instability problem \citep{b7a}.

The main objective of this paper is to construct global scale-free
stationary configurations in a two-fluid gaseous and stellar disc
system with an embedded coplanar magnetic field in the gas disc.
Meanwhile, an axisymmetric instability analysis is also performed.
There are several new features compared with previous works
\citep{b7,b10,b11} that may provide certain new clues to
understand large-scale structures of disc galaxies. This paper is
structured as follows. In Section 2, we present the theoretical
formalism of the problem; both the stationary equilibrium state
and the linearized MHD perturbation equations are summarized. In
Section 3, we perform numerical calculations for the aligned
perturbation configurations. Both the dispersion relation and
phase relationship between density and the magnetic field
perturbations are deduced and evaluated. In Section 4, we apply
the same procedure to the analysis of global logarithmic spiral
configurations. The $m=0$ marginal stability is also discussed.
Finally, we summarize and discuss our results in Section 5.
Several technical details are included in Appendices A, B, and C.


\section[]{Fluid-Magnetofluid Discs}

We adopt the fluid-magnetofluid formalism in this paper to
construct large-scale stationary aligned and unaligned coplanar
MHD disturbances in a background MHD rotational equilibrium of
axisymmetry [\cite{b6,b7,b10,b6a,b7a,b13e}].
All the background physical quantities are assumed to be
axisymmetric and to scale as power laws in radius $r$.
Specifically, the rotation curves bear an index of $-\beta$ (viz.,
$\emph{v}\propto\emph{r}^{-\beta}$) and the vertically integrated
mass density has the form $\Sigma_{0}\propto\emph{r}^{-\alpha}$
with $-\alpha$ being another index. Physically, the magnetofluid
formalism is directly applicable to the magnetized gas disc, while
the fluid formalism is only an approximation when applied to the
stellar disc, where a distribution function approach would give a
more comprehensive description. For our purpose of modelling
large-scale stationary MHD perturbation structures and for
mathematical simplicity, as well as the similarity between the two
sorts of descriptions \citep{b10},
it suffices to work with the fluid-magnetofluid formalism.

In this section, we present basic MHD equations of the
fluid-magnetofluid description for a composite system consisting
of a scale-free stellar disc and a coplanarly magnetized gas disc.
In our approach, the two gravitationally coupled discs are treated
using the razor-thin approximation (i.e., we use vertically
integrated fluid-magnetofluid equations and neglect vertical
derivatives of physical variables along $z$) and the
two-dimensional barotropic equation of state to construct global
MHD perturbation structures. A coplanar magnetic field is involved
in the dynamics of the thin gas disc following the basic MHD
equations. The background state of axisymmetric rotational
equilibrium is first derived. We then superpose coplanar MHD
perturbations onto the equilibrium state and obtain linearized
equations for MHD perturbations in the composite disc system.

\subsection{Ideal Nonlinear MHD Equations}

By our conventions, we use either subscript or superscript `$s$'
and `$g$' to denote physical variables in association with the
stellar disc and the magnetized gas disc, respectively. For
large-scale MHD perturbations of our interest at this stage, all
diffusive effects such as viscosity, resistivity, thermal
conduction, and radiative losses etc. are ignored for simplicity.
In cylindrical coordinates (\emph{r},\ $\theta$,\ \emph{z}) and
coincident with the \emph{z}=0 plane, we readily write down the
basic nonlinear ideal MHD equations for the composite disc system,
namely

\begin{equation}
{\partial{\Sigma^{g}}\over{\partial{t}}}
+{1\over{r}}{\partial{}\over{\partial{r}}}
{(r\Sigma^{g}u^{g})}
+{1\over{r^{2}}}{\partial{}\over{\partial{\theta}}}
{(\Sigma^{g}j^{g})}=0\ , \label{eq:1}
\end{equation}

\begin{equation}
\begin{split}
{\partial{u^{g}}\over{\partial{t}}}
+{u^{g}}{\partial{u^{g}}\over{\partial{r}}}+
{{j^{g}}\over{r^{2}}}{\partial{u^{g}}\over{\partial{\theta}}}
-{(j^{g})^{2}\over{r^{3}}}
=-{1\over{\Sigma^{g}}}{\partial{\Pi^{g}}\over{\partial{r}}}
-{\partial{\phi}\over{\partial{r}}}&\\
-{1\over{\Sigma^{g}}}\int{{\emph{d}zB_\theta\over{4\pi r}}
\bigg[{\partial{(rB_\theta)}\over{\partial{r}}}
-{\partial{B_r}\over{\partial{\theta}}}\bigg]}\ ,
\end{split}\label{eq:2}
\end{equation}

\begin{equation}
\begin{split}
{\partial{j^{g}}\over{\partial{t}}}
+{u^{g}}{\partial{j^{g}}\over{\partial{r}}}+
{{j^{g}}\over{r^{2}}}{\partial{j^{g}}\over{\partial{\theta}}}=
-{1\over{\Sigma^{g}}}{\partial{\Pi^{g}}\over{\partial{\theta}}}
-{\partial{\phi}\over{\partial{\theta}}}\qquad &
\\ \qquad\qquad\qquad
+{1\over{\Sigma^{g}}}\int{{\emph{d}zB_r\over{4\pi}}
\bigg[{\partial{(rB_\theta)}\over{\partial{r}}}
-{\partial{B_r}\over{\partial{\theta}}}\bigg]}\
\end{split}\label{eq:3}
\end{equation}
for the magnetized gas disc, and

\begin{equation}
{\partial{\Sigma^{s}}\over{\partial{t}}}
+{1\over{r}}{\partial{}\over{\partial{r}}}{(r\Sigma^{s}u^{s})}
+{1\over{r^{2}}}{\partial{}\over{\partial{\theta}}}
{(\Sigma^{s}j^{s})}=0\ , \label{eq:4}
\end{equation}

\begin{equation}
{\partial{u^{s}}\over{\partial{t}}}
+{u^{s}}{\partial{u^{s}}\over{\partial{r}}}+
{{j^{s}}\over{r^{2}}}{\partial{u^{s}}\over{\partial{\theta}}}
-{(j^{s})^{2}\over{r^{3}}}=
-{1\over{\Sigma^{s}}}{\partial{\Pi^{s}}\over{\partial{r}}}
-{\partial{\phi}\over{\partial{r}}}\ ,\label{eq:5}
\end{equation}

\begin{equation}
{\partial{j^{s}}\over{\partial{t}}}
+{u^{s}}{\partial{j^{s}}\over{\partial{r}}}+
{{j^{s}}\over{r^{2}}}{\partial{j^{s}}\over{\partial{\theta}}}=
-{1\over{\Sigma^{s}}}{\partial{\Pi^{s}}\over{\partial{\theta}}}
-{\partial{\phi}\over{\partial{\theta}}}\label{eq:6}
\end{equation}
for the stellar disc in the `fluid' approximation, respectively.
Here, $\Sigma$ denotes the surface mass density, $u$ is the radial
component of the bulk flow velocity, $j\equiv rv$ is the specific
angular momentum in the vertical $z$-direction and $v$ is the
azimuthal bulk flow velocity, $\Pi$ is the vertically integrated
two-dimensional pressure, $B_r$ and $B_\theta$ are the radial and
azimuthal components of the coplanar magnetic field
$\textbf{\emph{B}}\equiv(B_r,\ B_\theta,\ 0)$, and $\phi$ is the
total gravitational potential. This $\phi$ can be expressed in
terms of the Poisson integral

\begin{equation}
\emph{F}\phi(r,\theta)=\oint{\emph{d}\varphi}
\int_{0}^{\infty}{-G{\Sigma(r^{\prime},\ \varphi,\ t)r^{\prime}
\emph{d}r^{\prime}}\over{[r^{\prime2}+r^{2}-2r^{\prime}
r\cos(\varphi-\theta)]^{1/2}}}\ ,\label{eq:7}
\end{equation}
where $G$ is the gravitational constant,
$\Sigma=\Sigma^{g}+\Sigma^{s}$ denotes the total surface mass
density, and $F$ is defined as the ratio of the gravitational
potential arising from the composite discs to that arising from
the entire system including an axisymmetric massive dark matter
halo that is presumed not to respond to the coplanar MHD
perturbations in the disc plane [\cite{b13,b12,b6,b7,b10}; Lou \&
Wu (2006); Lou \& Zou (2006)] with $F=1$ for a full composite disc
system and $0<F<1$ for a composite system of partial discs.

For a coplanar magnetic field $\textbf{\emph{B}}=(B_r,B_\theta,0)$ in
cylindrical coordinates $(r,\theta,z)$, the divergence-free condition
is
\begin{equation}
{\partial{(rB_r)}\over{\partial{r}}}
+{\partial{B_\theta}\over{\partial{\theta}}}=0\ .\label{eq:8}
\end{equation}
The radial and azimuthal components of
the magnetic induction equation are

\begin{equation}
{\partial{B_r}\over{\partial{t}}}
={1\over{r}}{\partial{}\over{\partial{\theta}}}
(u^gB_\theta-v^gB_r)\ ,\label{eq:9}
\end{equation}

\begin{equation}
{\partial{B_\theta}\over{\partial{t}}}
=-{\partial{}\over{\partial{r}}}(u^gB_\theta-v^gB_r)\ .\label{eq:10}
\end{equation}
Among $(\ref{eq:8})-(\ref{eq:10})$, only two of them are
independent. The barotropic equation of state for the scale-free
discs is
\begin{equation}
\Pi^i=K_i\Sigma^{n_i}\ ,\label{eq:11}
\end{equation}
where $K_i>0$ 
is a constant proportional coefficient and $n_i>0$ is the barotropic
index with subscript $i$ denoting either $g$ or $s$ for the two
discs (we use this convention for simplicity). An isothermal
equation of state has $n_i=1$.
For the stellar disc and the magnetized gas disc, $K_i$ are
allowed to be different, but we need to require $n_g=n_s=n$ in
order to meet the scale-free requirement (see the next
subsection). It follows that the sound speed $a_i$ (in the stellar
disc the velocity dispersion mimics the sound speed) for either
disc is readily defined by
\begin{equation}
a_i^2\equiv
{{d\Pi_0^i}\over{d\Sigma_0^i}}=nK_i(\Sigma_0^i)^{n-1}\ ,\label{eq:12}
\end{equation}
where the subscript $_0$ denotes the background equilibrium.

\subsection{Axisymmetric Background MHD Equilibria }

For the axisymmetric rotational MHD equilibrium, we set $u^g=u^s=0$ and
all terms involving $\partial/\partial{t}$ and $\partial/\partial{\theta}$
in equations $(\ref{eq:1})-(\ref{eq:10})$ to vanish. We now
determine the background axisymmetric equilibrium with physical variables
denoted by subscript `$_0$'. In our notations, a background equilibrium is
characterized by the following power-law radial scalings: both surface mass
densities $\Sigma_0^s,\ \Sigma_0^g\propto r^{-\alpha}$ carrying the same
exponent index $\alpha$ yet with different proportional coefficients.
The disc rotation curves $v_0^s$ and $v_0^g$ take the power-law form of
$\propto r^{-\beta}$ and therefore the $z-$component specific angular
momentum $j_0\equiv rv_0\propto r^{1-\beta}$ for both discs. In our
composite MHD disc system, a background magnetic field is purely
azimuthal in the gas disc, that is, $B_r^0=0$ and $B_\theta^0\equiv
B_0\propto r^{-\gamma}$. A substitution of these power-law radial
scalings into equations $(\ref{eq:1})-(\ref{eq:12})$ leads to the
following radial force balances
\begin{equation}
{v_0^g}^2=-\alpha a_g^2+rd\phi_0/{dr}+(1-\gamma)C_A^2\label{eq:13}
\end{equation}
in the magnetized gas disc, where $C_A$ is the Alfv\'en
wave speed in a thin magnetized gas disc defined by
\begin{equation}
C_A^2\equiv{1\over{4\pi \Sigma_0^g}}\int{B_0^2 dz}\label{eq:14}
\end{equation}
varying with $r$ in general, and
\begin{equation}
{v_0^s}^2=-\alpha a_s^2+rd\phi_0/{dr}\ \label{eq:15}
\end{equation}
in the stellar disc. Poisson integral (\ref{eq:7}) yields
\begin{equation}
F\phi_0=-2\pi Gr\Sigma_0\mathcal{P}_0(\alpha)\ ,\label{eq:16}
\end{equation}
where the numerical factor $\mathcal{P}_0(\alpha)$
is explicitly defined by
\begin{equation}
\mathcal{P}_0(\alpha)\equiv{{\Gamma(-\alpha/2+1)\Gamma(\alpha/2-1/2)}
\over{2\Gamma(-\alpha/2+3/2)\Gamma(\alpha/2)}}\label{eq:17}
\end{equation}
with $\Gamma(\cdots)$ being the standard gamma function.
Expression (\ref{eq:17}) can also be included in a more
general form of $\mathcal{P}_m(\beta)$ defined later in
equation (\ref{eq:54}).

In order to satisfy the scale-free condition, radial force
balances (\ref{eq:13}) and (\ref{eq:15}) should hold for all
radii, leading to the following simple relation among the
four indices
\begin{equation}
2\beta=\alpha(n-1)=2\gamma-\alpha=\alpha-1\ .\label{eq:18}
\end{equation}
This in turn immediately gives the explicit expressions of
indices $\alpha,\gamma$ and $n$ in terms of $\beta$, namely
\begin{equation}
\alpha=1+2\beta\ , \qquad \gamma={(1+4\beta)\over{2}}\ , \qquad
n={(1+4\beta)\over(1+2\beta)}\ .\label{eq:19}
\end{equation}
Since $n>0$ for warm discs, we have $\beta>-1/4$. Furthermore,
Poisson integral (\ref{eq:7}) converges when $1<\alpha<2$ which
corresponds to $0<\beta<1/2$. In addition, for a finite total
gravitational force a larger $\beta$-range of $-1/2<\beta<1/2$
is allowed. Therefore, the physical range for $\beta$ is
constrained by $-1/4<\beta<1/2$ \citep{b13,b10,b7,b11}.

For simplicity, we introduce the following
parameters $A_i$, $D_i$, and $q$ defined by
\begin{equation}
\begin{split}
&a_i^2=\frac{A_i^2}{(1+2\beta)r^{2\beta}}\ ,
\quad\qquad\qquad v_0^i =\frac{A_iD_i}{r^\beta}\ ,\\
&C_A^2\equiv q^2 a_g^2=\frac{q^2A_g^2}{(1+2\beta)r^{2\beta}}\ ,
\end{split}\label{eq:20}
\end{equation}
where the subscript or superscript $i$ denotes either $g$ or $s$ for
either the magnetized gas disc or the stellar disc, respectively.
We see that $A_i$ is related to a scaled sound speed (or velocity
dispersion of the stellar disc), the constant $D_i={v_0^i}/
{[a_i(1+2\beta)^{1/2}]}$ is a scaled rotational Mach number, and the
constant parameter $q$ represents the ratio of the Alfv\'en speed
$C_A$ to the sound speed in the magnetized gas disc \citep{b11}.

It is then straightforward to express other equilibrium physical
variables in terms of $A_i,\ D_i$ and $q$. For the disc angular
speed $\Omega_i$ and the epicyclic frequency $\kappa_i$, we have
respectively
\[
\Omega_i\equiv\frac{v_0^i}r=\frac{A_i D_i}{r^{1+\beta}}\ ,
\]
\begin{equation}
\kappa_i\equiv\bigg[\frac{2\Omega_i}{r}
\frac{d(r^2\Omega_i)}{dr}\bigg]^{1/2}
=[2(1-\beta)]^{1/2}\Omega_i\ .\label{eq:21}
\end{equation}
A combination of equations (\ref{eq:13}), (\ref{eq:15}),
(\ref{eq:16}) and (\ref{eq:20}) leads to the relation
among these parameters in the form of
\begin{equation}
\begin{split}
A_g^2\bigg(D_g^2+1+\frac{4\beta-1}{4\beta+2}q^2\bigg)=A_s^2(D_s^2+1)
\ \ \qquad\qquad\\ \qquad\qquad
=2\pi G{2\beta\mathcal{P}_0}\Sigma_0 r^{1+2\beta}/F\ .\label{eq:22}
\end{split}
\end{equation}
We now introduce two dimensionless parameters to compare the
properties of the two scale-free discs. The first one is simply
the surface mass density ratio $\delta\equiv\Sigma_0^g/\Sigma_0^s$,
and the second one is the square of the ratio of effective sound
speeds $\eta\equiv a_g^2/a_s^2=A_g^2/A_s^2$. For disc galaxies, the
ratio $\delta$ can be either greater (i.e., younger disc galaxies)
or less (i.e., older disc galaxies) than 1, depending on the type
and evolution stage of a disc galaxy. Meanwhile, the ratio $\eta$
can be generally taken as less than 1 because the sound speed
in the magnetized gas disc is typically less than the stellar
velocity dispersion (regarded as an effective sound speed).

With these notations, we now have from condition (\ref{eq:22})
\begin{equation}
\begin{split}
&\Sigma_0^g=\frac{FA_g^2[D_g^2+1+(4\beta-1)q^2/(4\beta+2)]\delta}
{2\pi G(2\beta\mathcal{P}_0)(1+\delta)r^{(1+2\beta )}}\ ,\\
&\Sigma_0^s=\frac{FA_g^2[D_g^2+1+(4\beta-1)q^2/(4\beta+2)]}
{2\pi G(2\beta\mathcal{P}_0)(1+\delta)r^{(1+2\beta )}}\ .
\end{split}\label{eq:23}
\end{equation}
Note that expressions (\ref{eq:22}) and (\ref{eq:23}) reduce to
expression (14) of \citet{b10} when the magnetic field vanishes
(i.e., $q=0$) and are also in accordance with \citet{b11} when
a single magnetized scale-free gas disc is considered.

Since the magnetic field effect is represented by the $q$
parameter multiplied by a factor $(1-\gamma)$ or $(4\beta-1)$ [see
eq. (\ref{eq:13}) or eq. (\ref{eq:22})], we know that in the
special case of $\beta=1/4$, the Lorentz force vanishes due to the
cancellation between the magnetic pressure and tension forces in
the background equilibrium \citep{b11}. Moreover, when
$-1/4<\beta<1/4$, the net Lorentz force arising from the azimuthal
magnetic field points radially inward, while for $1/4<\beta<1/2$,
the net Lorentz force points radially outward \citep{b11}.

Another point should also be noted here. From equation
(\ref{eq:22}), there exists another physical requirement for the
rotational Mach number $D_g$, i.e., it should be large enough to
warrant a positive $D_s^2$. This requirement for $D_g^2$ is simply
\begin{equation}
D_s^2=\eta\bigg(D_g^2+1+\frac{4\beta-1}
{4\beta+2}q^2\bigg)-1>0\ .\label{eq:22b}
\end{equation}

In the presence of magnetic field, the relation between $D_g$ and
$D_s$ becomes somewhat involved, as compared to the hydrodynamic
case studied by \citet{b10}. Since $\eta\leq1$, we see from
equation (\ref{eq:22b}) that when $\beta\leq1/4$, $D_s^2$ remains
always smaller than $D_g^2$, while for $\beta>1/4$, $D_g^2$ is
smaller than $D_s^2$ when $\eta$ is of order of 1.

\subsection{Coplanar MHD Perturbations}

In this subsection, we use subscript 1 along physical variables
to denote small coplanar MHD perturbations. For example,
$\Sigma^g=\Sigma_0^g+\Sigma_1^g$ for gas surface mass density with
$\Sigma_1^g$ a small perturbation quantity under consideration.
Similar notations hold for other physical variables. Now from
equations $(\ref{eq:1})-(\ref{eq:10})$ and the specification of
the background rotational MHD equilibrium, we readily obtain
linearized equations for coplanar MHD perturbations below, namely,

\begin{equation}
{\partial{\Sigma_1^{g}}\over{\partial{t}}}
+{1\over{r}}{\partial{}\over{\partial{r}}} {(r\Sigma_0^{g}u_1^{g})}+
\Omega_g\frac{\partial{\Sigma_1^g}}{\partial{\theta}}
+{\Sigma_0^{g}\over{r^{2}}}{\partial{j_1^{g}}
\over{\partial{\theta}}}=0\ ,\label{eq:24}
\end{equation}

\begin{equation}
\begin{split}
&{\partial{u_1^{g}}\over{\partial{t}}}+
{\Omega_g}{\partial{u_1^{g}}\over{\partial{\theta}}}
-2\Omega_g{j_1^{g}\over{r}} =-{\partial{}\over{\partial{r}}}
\bigg(a_g^2\frac{\Sigma_1^g}{\Sigma_0^g}+\phi_1\bigg)
\qquad \\ & \qquad\qquad
+\frac{(1-4\beta)C_A^2\Sigma_1^g}{2\Sigma_0^g r}
-{1\over{\Sigma_0^{g}}}\int{{\emph{d}zb_\theta\over{4\pi r}}
{\partial{(rB_0)}\over{\partial{r}}}}
\qquad \\ &\qquad\qquad
-{1\over{\Sigma_0^{g}}}\int{{\emph{d}zB_0\over{4\pi r}}
\bigg[{\partial{(rb_\theta)}\over{\partial{r}}}
-{\partial{b_r}\over{\partial{\theta}}}\bigg]}\ ,
\end{split}\label{eq:25}
\end{equation}

\begin{equation}
\begin{split}
{\partial{j_1^{g}}\over{\partial{t}}}+\frac{r\kappa_g^2}
{2\Omega_g}{u_1^{g}}+
{\Omega_g}{\partial{j_1^{g}}\over{\partial{\theta}}}=
-{\partial{}\over{\partial{\theta}}}
\bigg(a_g^2\frac{\Sigma_1^g}{\Sigma_0^g}+\phi_1\bigg)\qquad &\\
\qquad +{1\over{\Sigma_0^{g}}}\int{{\emph{d}zb_r\over{4\pi}}
{\partial{(rB_0)}\over{\partial{r}}}}
\end{split}\label{eq:26}
\end{equation}
in the magnetized gas disc and
\begin{equation}
\begin{split}
&{\partial{\Sigma_1^{s}}\over{\partial{t}}}
+{1\over{r}}{\partial{}\over{\partial{r}}}{(r\Sigma_0^{s}u_1^{s})}
+\Omega_s\frac{\partial{\Sigma_1^s}}{\partial{\theta}}
+{\Sigma_0^{s}\over{r^{2}}}{\partial{j_1^{s}}
\over{\partial{\theta}}}=0\ ,\\
&{\partial{u_1^{s}}\over{\partial{t}}}+
{\Omega_s}{\partial{u_1^{s}}\over{\partial{\theta}}}
-2\Omega_s{j_1^{s}\over{r}} =-{\partial{}\over{\partial{r}}}
\bigg(a_s^2\frac{\Sigma_1^s}{\Sigma_0^s}+\phi_1\bigg)\ ,\\
&{\partial{j_1^{s}}\over{\partial{t}}}+\frac{r\kappa_s^2}
{2\Omega_s}{u_1^{s}}+
{\Omega_s}{\partial{j_1^{s}}\over{\partial{\theta}}}=
-{\partial{}\over{\partial{\theta}}}
\bigg(a_s^2\frac{\Sigma_1^s}{\Sigma_0^s}+\phi_1\bigg)
\end{split}\label{eq:27}
\end{equation}
in the stellar disc, respectively. All dependent variables are
taken to the first-order smallness with nonlinear terms ignored.
The perturbed Poisson integral appears as
\begin{equation}
\phi_1=\oint{\emph{d}\varphi}\int_{0}^{\infty}
{{-G(\Sigma_1^g+\Sigma_1^s)r^{\prime}\emph{d}r^{\prime}}\over
{[r^{\prime2}+r^{2}-2r^{\prime}r\cos(\varphi-\theta)]^{1/2}}}\ .
\label{eq:28}
\end{equation}
Note that for a gravitational potential perturbation, we no longer
have the factor $F$ by the simplifying assumption that the dark
matter halo exists as a background and does not respond to
perturbations in the two discs. Together for a coplanar magnetic
field perturbation $\textbf{b}\equiv(b_r,b_\theta,0)$, we have
\begin{equation}
\begin{split}
&{\partial{(rb_r)}\over{\partial{r}}}
+{\partial{b_\theta}\over{\partial{\theta}}}=0\ ,\\
&{\partial{b_r}\over{\partial{t}}}
={1\over{r}}{\partial{}\over{\partial{\theta}}}
(u_1^gB_0-r\Omega_g b_r)\ ,\\
&{\partial{b_\theta}\over{\partial{t}}}
=-{\partial{}\over{\partial{r}}}(u_1^g B_0-r\Omega_g b_r)\
\end{split}\label{eq:29}
\end{equation}
from the divergence-free condition and the magnetic induction
equation.

\subsubsection{Non-Axisymmetric Cases of $m\geq 1$ }

As the background equilibrium state is stationary and
axisymmetric, these perturbed physical variables can be decomposed
in terms of Fourier harmonics with the periodic dependence
$\mathrm{exp}(\mathrm{i}\omega t-\mathrm{i}m\theta)$ where
$\omega$ is the angular frequency and $m$ is an integer to
characterize azimuthal variations. More specifically, we write
\begin{equation}
\begin{split}
\Sigma_1^i&=S^i(r)\mathrm{exp}(\mathrm{i}\omega
t-\mathrm{i}m\theta)\ ,
\quad
u_1^i=U^i(r)\mathrm{exp}(\mathrm{i}\omega t-\mathrm{i}m\theta)\ ,\\
j_1^i&=J^i(r)\mathrm{exp}(\mathrm{i}\omega t-\mathrm{i}m\theta)\ ,
\quad
\phi_1=V(r)\mathrm{exp}(\mathrm{i}\omega t-\mathrm{i}m\theta)\ ,\\
b_r&=R(r)\mathrm{exp}(\mathrm{i}\omega t-\mathrm{i}m\theta)\ ,
\quad
b_\theta=Z(r)\mathrm{exp}(\mathrm{i}\omega t-\mathrm{i}m\theta)\ ,
\end{split}\label{eq:30}
\end{equation}
where we use the italic superscript $i$ to indicate associations
with the two discs and the roman i for the imaginary unit.
We also define $V^i$ ($i=s,\ g$) for the corresponding 
gravitational potentials associated with the stellar 
and gaseous discs such that $V(r)\equiv V^s(r)+V^g(r)$.
By substituting expressions (\ref{eq:30}) into equations 
$(\ref{eq:24})-(\ref{eq:29})$, we readily attain
\begin{equation}
\mathrm{i}(\omega-m\Omega_g)S^g+\frac{1}{r}
\frac{d}{dr}(r\Sigma_0^gU^g)-\mathrm{i}m\Sigma_0^g
\frac{J^g}{r^2}=0\ ,\label{eq:31}
\end{equation}

\begin{equation}
\begin{split}
&\mathrm{i}(\omega-m\Omega_g)U^g-2\Omega_g\frac{J^g}{r}
=-\frac{d\Psi^g}{dr}+\frac{(1-4\beta)C_A^2 S^g}{2\Sigma_0^g r}
\\
&-{1\over{\Sigma_0^{g}}}\int{{\emph{d}zB_0\over{4\pi r}}
\bigg[{d(rZ)\over{dr}}+\mathrm{i}mR\bigg]}
-{1\over{\Sigma_0^{g}}}\int{{\emph{d}zZ\over{4\pi r}}
{d(rB_0)\over{dr}}}\ ,\label{eq:32}
\end{split}
\end{equation}

\begin{equation}
\mathrm{i}(\omega-m\Omega_g)J^g+\frac{r\kappa_g^2}
{2\Omega_g}U^g=\mathrm{i}m\Psi^g
+{1\over{\Sigma_0^{g}}}\int{{\emph{d}zR\over{4\pi
}}{d(rB_0)\over{dr}}}\label{eq:33}
\end{equation}
for the magnetized gas disc, and
\begin{equation}
\begin{split}
&\mathrm{i}(\omega-m\Omega_s)S^s+\frac{1}{r}\frac{d}{dr}
(r\Sigma_0^sU^s)-\mathrm{i}m\Sigma_0^s\frac{J^s}{r^2}=0\ ,\\
&\mathrm{i}(\omega-m\Omega_s)U^s-2\Omega_s\frac{J^s}{r}
=-\frac{d\Psi^s}{dr}\ ,\\
&\mathrm{i}(\omega-m\Omega_s)J^s
+\frac{r\kappa_s^2}{2\Omega_s}U^s=\mathrm{i}m\Psi^s
\end{split}\label{eq:34}
\end{equation}
for the stellar disc, respectively, where we define
\begin{equation}
\Psi^i\equiv a_i^2{S^i}/{\Sigma_0^i}+V\ .\label{eq:35}
\end{equation}
The perturbed Poisson integral now becomes
\begin{equation}
V(r)=\oint{d\varphi}\int_{0}^{\infty}{{-G(S^g+S^s) \cos
(m\varphi)\ r^{\prime}dr^{\prime}}\over{(r^{\prime2}
+r^{2}-2r^{\prime}r\cos\varphi)^{1/2}}}\label{eq:36}
\end{equation}
and perturbation magnetic field variations reduce to
\begin{equation}
\begin{split}
&{d{(rR)}\over{d{r}}}-\mathrm{i}mZ=0\ ,\\
&\mathrm{i}(\omega-m\Omega_g)R
+\frac{\mathrm{i}mB_0}{r}U^g=0\ ,\\
&\mathrm{i}\omega Z=\frac{d}{dr}(r\Omega_g R-B_0 U^g)\ .
\end{split} \label{eq:37}
\end{equation}
A combination of equations $(\ref{eq:31})-(\ref{eq:34})$ and
$(\ref{eq:36})-(\ref{eq:37})$ constitutes a complete description
of coplanar MHD perturbations in a composite disc system. From
equation (\ref{eq:37}), we obtain
\begin{equation}
Z=-\frac{\mathrm{i}}{m}\frac{d(rR)}{dr}\ ,
\qquad\qquad
R=-\frac{mB_0U^g}{r(\omega-m\Omega_g)}\ .\label{eq:38}
\end{equation}
A substitution of expressions (\ref{eq:38}) into the
radial and azimuthal components of the momentum
equation (\ref{eq:32}) and (\ref{eq:33}) yields
\begin{equation}
\begin{split}
\mathrm{i}(\omega-m\Omega_g)U^g-2\Omega_g\frac{J^g}{r}
=-\frac{d\Psi^g}{dr}+\frac{(1-4\beta)C_A^2 S^g}{2\Sigma_0^g r}
\qquad &\\
-C_A^2\bigg[\frac{d^2}{dr^2} +\frac{1-12\beta }{2r}
\frac{d}{dr} +\frac{2\beta(1+4\beta)-m^2}{r^2}\bigg]
\frac{\mathrm{i}U^g}{\omega-m\Omega_g}\ ,\label{eq:39}
\end{split}
\end{equation}
\begin{equation}
\mathrm{i}(\omega-m\Omega_g)J^g
+\frac{r\kappa_g^2}{2\Omega_g}U^g
=\mathrm{i}m\Psi^g-\frac{mC_A^2(1-4\beta)U^g}
{2r(\omega-m\Omega_g)}\ .\label{eq:40}
\end{equation}
We are primarily interested in stationary global configurations
of zero-frequency MHD perturbations yet without the usual WKBJ
approximation. By setting $\omega=0$, we can preserve the
scale-free condition. Stationary perturbation patterns in our
frame of reference have been studied previously [e.g.,
\citet{b13,b12,b5,b9,b10,b7,b11}; Lou \& Wu (2005)] and our
analysis here is more general, involving a composite system of two
scale-free discs with a coplanar magnetic field. For $m\geq1$, we
set $\omega=0$ in MHD perturbation equations to deduce
\begin{equation}
m\Omega_g S^g+\frac{1}{r}\frac{d}{dr}(r\Sigma_0^g
\mathrm{i}U^g)+m\Sigma_0^g\frac{J^g}{r^2}=0\ ,\label{eq:41}
\end{equation}
\begin{equation}
\begin{split}
&m\Omega_g\mathrm{i}U^g+2\Omega_g\frac{J^g}{r}=\frac{d\Psi^g}{dr}
-\frac{(1-4\beta)C_A^2 S^g}{2\Sigma_0^g r}\\&
-\frac{C_A^2}{m}\bigg[\frac{d^2}{dr^2}+\frac{1-12\beta}{2r}\frac{d}{dr}
+\frac{2\beta(1+4\beta)-m^2}{r^2}\bigg]
\frac{\mathrm{i}U^g}{\Omega_g}\ ,\label{eq:42}
\end{split}
\end{equation}
\begin{equation}
m\Omega_g J^g+\frac{r\kappa_g^2}{2\Omega_g}\mathrm{i}U^g=-m\Psi^g
+\frac{C_A^2(1-4\beta)}{2r}\frac{\mathrm{i}U^g}{\Omega_g}\label{eq:43}
\end{equation}
for the magnetized gas disc. Meanwhile, we derive
from two independent equations of (\ref{eq:38})
\begin{equation}
Z=-\frac{d}{dr}\bigg(\frac{B_0\mathrm{i}U^g}{m\Omega_g}\bigg)\ ,
\qquad\qquad R=\frac{B_0U^g}{r\Omega_g}\ ,\label{eq:44}
\end{equation}
for the magnetic field perturbation. For coplanar
perturbations in the stellar disc, we derive
\begin{equation}
\begin{split}
&m\Omega_s S^s+\frac{1}{r}\frac{d}{dr}(r\Sigma_0^s
\mathrm{i}U^s)+m\Sigma_0^s\frac{J^s}{r^2}=0\ ,\\
&m\Omega_s\mathrm{i}U^s+2\Omega_s\frac{J^s}{r}=\frac{d\Psi^s}{dr}\ ,\\
&m\Omega_s J^s+\frac{r\kappa_s^2}{2\Omega_s}\mathrm{i}U^s=-m\Psi^s\ .
\end{split}\label{eq:45}
\end{equation}
As part of the derivation, we deduce from the last
two equations of (\ref{eq:45}) for the expressions
of $U^s$ and $J^s$ as
\begin{equation}
\begin{split}
U^s&=\frac{-\mathrm{i}m\Omega_s}{m^2\Omega_s^2-\kappa_s^2}
\bigg(\frac{2}{r}+\frac{d}{dr}\bigg)\Psi^s\ ,\\
\frac{J^s}{r}&=\frac{-1}{m^2\Omega_s^2-\kappa_s^2}
\bigg(\frac{m^2\Omega_s}{r}+\frac{\kappa_s^2}{2\Omega_s}
\frac{d}{dr}\bigg)\Psi^s\ .
\end{split}\label{eq:46}
\end{equation}
A substitution of equation (\ref{eq:46}) into equation (\ref{eq:45})
gives a single equation relating $S^s$ and $\Psi^s$ for coplanar
perturbations in the stellar disc, namely
\begin{equation}
\begin{split}
&m\Omega_sS^s+\frac{1}{r}\frac{d}{dr}
\bigg[\frac{m\Omega_sr\Sigma_0^s}{m^2\Omega_s^2-\kappa_s^2}
\bigg(\frac{2}{r}+\frac{d}{dr}\bigg)\Psi^s\bigg]
\\ &\quad\qquad
-\frac{m\Sigma_0^s}
{m^2\Omega_s^2-\kappa_s^2}\bigg(\frac{m^2\Omega_s}{r^2}
+\frac{\kappa_s^2}{2r\Omega_s}\frac{d}{dr}\bigg)\Psi^s=0\ .
\end{split}\label{eq:47}
\end{equation}
Now equations $(\ref{eq:41})-(\ref{eq:43})$, (\ref{eq:47})
together with equation (\ref{eq:36}) form the coplanar MHD
perturbation equations for constructing global stationary
non-axisymmetric $m\geq1$ configurations for both aligned
and logarithmic spiral cases in a composite MHD scale-free
disc system.

\subsubsection{The Axisymmetric Case of  $m=0$}

We now examine the axisymmetric case of $m=0$ for which equations
$(\ref{eq:41})-(\ref{eq:43})$, (\ref{eq:47}) become degenerate.
Instead, we adopt a different limiting procedure by setting $m=0$
and a small $\omega\neq 0$ in equations (\ref{eq:31}),
(\ref{eq:39}), (\ref{eq:40}), (\ref{eq:37}) and (\ref{eq:34}) to
obtain
\begin{equation}
\omega S^g=\frac{1}{r}\frac{d}{dr}
(r\Sigma_0^g\mathrm{i}U^g)\ ,\label{eq:48}
\end{equation}
\begin{equation}
\begin{split}
\omega^2\mathrm{i}U^g-2\omega\Omega_g\frac{J^g}{r}
=-\omega\frac{d\Psi^g}{dr} +\omega\frac{(1-4\beta)C_A^2S^g}
{2\Sigma_0^g r}\ \quad &\\ \quad
-C_A^2\bigg[\frac{d^2}{dr^2}+\frac{1-12\beta}{2r}\frac{d}{dr}
+\frac{2\beta(1+4\beta)}{r^2}\bigg]\mathrm{i}U^g\ ,\label{eq:49}
\end{split}
\end{equation}
\begin{equation}
\omega J^g=\frac{r\kappa_g^2}{2\Omega_g}\mathrm{i}U^g\ ,\label{eq:50}
\end{equation}
\begin{equation}
R=0\ ,\qquad\qquad \omega
Z=\frac{d}{dr}(B_0\mathrm{i}U^g)\label{eq:51}
\end{equation}
for the magnetized gas disc. For coplanar perturbations
in the stellar disc, we have in parallel
\begin{equation}
\begin{split}
&\omega S^s=\frac{1}{r}\frac{d}{dr}(r\Sigma_0^s\mathrm{i}U^s)\ ,\\
&\omega\mathrm{i}U^s-2\Omega_s\frac{J^s}{r}=-\frac{d\Psi^s}{dr}\ ,\\
&\omega J^s=\frac{r\kappa_s^2}{2\Omega_s}\mathrm{i}U^s\ ,
\end{split}\label{eq:52}
\end{equation}
where the gravitational coupling between the two scale-free
discs is implicit in $\Psi^g$ and $\Psi^s$ expressions.

For global stationary axisymmetric MHD perturbations, we take
the limiting procedure $\omega\rightarrow0$ in equations
$(\ref{eq:48})-(\ref{eq:52})$ \citep{b7,b10,b11,b7a} rather than
setting $\omega=0$ in MHD perturbation equations and then let
$m\rightarrow0$. Note that these two procedures of taking limit
lead to different results except for certain special cases [e.g.,
\citet{b12}].

\section[]{Global Stationary Aligned
MHD Perturbation Configurations}

In the following two sections, we solve the Poisson integral by
introducing two kinds of scale-free global MHD perturbation patterns,
namely, the aligned and the logarithmic spiral configurations.
These two kinds
of MHD perturbation structures are the results of the special forms in
a list of the potential-density pairs \citep{b1,b8}. For the aligned
case, the maximum density perturbations at different radii line up
in the azimuth, while for the spiral case, the maximum density
perturbations involve a systematic phase shift at different radii
such that a logarithmic spiral pattern emerges.

In this section, we derive aligned stationary perturbation dispersion
relations on the basis of the results of the previous section and
discuss solution properties in detail including the dependence of disc
rotation speed on several parameters, the phase relationship among the
surface mass density perturbation in the two discs and the perturbed
magnetic field. As an example of illustration, we mainly focus on a
composite system of full scale-free discs (i.e., $F=1$).

\subsection{Dispersion Relation for Global\\
\ \qquad Aligned Coplanar MHD Perturbations}

For aligned MHD perturbations, we select those perturbations
that carry the same power-law variations in $r$ as those of the
background equilibrium, that is, in expressions (\ref{eq:30})
\begin{equation}
S^i(r)=\epsilon^i r^{-1-2\beta}\ ,\qquad\quad V^i(r)
=-2\pi GrS^i\mathcal{P}_m(\beta)\ ,\label{eq:53}
\end{equation}
where $\epsilon^{i}$ is a sufficiently small amplitude coefficient,
superscript $i$ denotes either $g$ or $s$ for either gas or stellar
discs, respectively. The numerical factor $\mathcal{P}_m(\beta)$ is
defined explicitly by
\begin{equation}
\mathcal{P}_m(\beta)\equiv \frac{\Gamma(m/2-\beta+1/2)
\Gamma(m/2+\beta)}{2\Gamma(m/2-\beta+1)
\Gamma(m/2+\beta+1/2)}\ ,\label{eq:54}
\end{equation}
where $-m/2<\beta<(m+1)/2$ \citep{b8,b13,b10,b11}. Note that the
requirement of $-1/4<\beta<1/2$ automatically satisfies this
inequality for $m\geq1$. In the limit $\beta\rightarrow0$, we
simply have $\mathcal{P}_m=1/m$. For $m=0$, equation (\ref{eq:54})
reduces exactly to equation (\ref{eq:17}) by setting
$\alpha=1+2\beta$; meanwhile, we also have the limiting result of
$2\beta\mathcal{P}_0(\beta) \rightarrow1$ as $\beta\rightarrow0$.

We start with equations $(\ref{eq:41})-(\ref{eq:43})$,
(\ref{eq:47}) and (\ref{eq:53}) by applying power-law
variations that satisfy scale-free conditions. Here, we
have $U\propto r^{-\beta}$ and $J\propto r^{1-\beta}$.
By proper combination and simplification, we then arrive at
\begin{equation}
m\Omega_g S^g-\frac{3\beta\Sigma_0^g}{r}\mathrm{i}U^g
+\frac{m\Sigma_0^g}{r^2}J^g=0\ ,\label{eq:55}
\end{equation}
\begin{equation}
\begin{split}
&\bigg\{m\Omega_g+\frac{C_A^2[(1-4\beta)^2-2m^2]} {2m\Omega_g
r^2}\bigg\}
\mathrm{i}U^g+\frac{2\Omega_g}{r}J^g=\\
&\bigg[4\pi G\beta\mathcal{P}_m-\frac{4\beta
a_g^2+(1-4\beta)C_A^2}{2r\Sigma_0^g}\bigg]S^g +4\pi
G\beta\mathcal{P}_mS^s\ ,\label{eq:56}
\end{split}
\end{equation}
%
\begin{equation}
\begin{split}
m\Omega_g &J^g+\bigg[(1-\beta)r\Omega_g-\frac{C_A^2(1-4\beta)}
{2r\Omega_g}\bigg]\mathrm{i}U^g=
\\ &\quad\qquad
mr\bigg[2\pi G\mathcal{P}_m-\frac{a_g^2}
{r\Sigma_0^g}\bigg]S^g+2\pi Gmr\mathcal{P}_mS^s\ ,
\end{split}\label{eq:57}
\end{equation}
\begin{equation}
Z=-\bigg(\frac{1}{2}-2\beta\bigg)
\frac{B_0\mathrm{i}U^g}{m\Omega_g r}\ ,
\qquad\qquad R=\frac{B_0U^g}{r\Omega_g}\label{eq:58}
\end{equation}
for the magnetized gas disc and
\begin{equation}
\begin{split}
&\bigg[\frac{r^2(m^2-2+2\beta)\Omega_s^2}{(m^2+4\beta-4\beta^2)}
-a_s^2+2\pi Gr\Sigma_0^s\mathcal{P}_m\bigg]S^s
\\ &\qquad\qquad\qquad\qquad\qquad
+2\pi Gr\Sigma_0^s\mathcal{P}_mS^g=0\label{eq:59}
\end{split}
\end{equation}
for the stellar disc. Note that equations
$(\ref{eq:55})-(\ref{eq:58})$ are the same as equation (43) of
\citet{b11} for a single magnetized scale-free disc by setting
$S^s=0$ and equation (\ref{eq:59}) is exactly the same as the
first one of equation (35) in \citet{b10}. A combination of
equations $(\ref{eq:55})-(\ref{eq:57})$ and (\ref{eq:59}) then
gives a complete solution for the stationary dispersion relation
in the gravitational coupled MHD discs that are scale free.
Because these equations are linear and homogeneous, to obtain
non-trivial solutions for $(S^g, S^s, U^g, J^g)$, the determinant
of the coefficients of equations $(\ref{eq:55})-(\ref{eq:57})$
and (\ref{eq:59}) should vanish. This actually gives rise to the
stationary MHD dispersion relation. Meanwhile, in order to get
a physical sense of the dispersion relation, we solve the above
MHD equations directly.

A combination of equations (\ref{eq:55}) and (\ref{eq:57})
produces relations of $\mathrm{i}U^g$ and $J^g$ in terms
of $S^g$ and $S^s$, namely
\begin{equation}
\begin{split}
&\mathrm{i}U^g=\frac{m\Omega_gr}{\Sigma_0^g}
[(\Theta_A+\Xi_A)S^g+\Theta S^s]\ ,\\
&J^g=\frac{\Omega_gr^2}{\Sigma_0^g}
\{[-1+3\beta(\Theta_A+\Xi_A)]S^g+3\beta\Theta_A S^s\}\ ,
\end{split}\label{eq:60}
\end{equation}
where for notational simplicity, we define
\begin{equation}
\begin{split}
&\Theta_A\equiv(2\pi Gr\mathcal{P}_m\Sigma_0^g)/\Delta_A\ ,\\
&\Xi_A\equiv (-a_g^2+\Omega_g^2r^2)/\Delta_A\ ,\\
&\Delta_A\equiv(1+2\beta)\Omega_g^2r^2+(2\beta-1/2)C_A^2\ .
\end{split}\label{eq:61}
\end{equation}
Here, we use the subscript $_A$ to indicate the `aligned case'.
In particular, $\Theta_A$ and $\Xi_A$ are two dimensionless
constant parameters. A substitution of equation (\ref{eq:60})
into equation (\ref{eq:56}) and a further combination with
equation (\ref{eq:59}) lead to
\begin{equation}
\begin{split}
[\mathcal{K}_A(\Theta_A+\Xi_A)+\mathcal{L}_A-2\pi
G\Sigma_0^gr(2\beta\mathcal{P}_m)]&S^g
\\ \quad
+[\mathcal{K}_A\Theta_A-2\pi G
\Sigma_0^gr(2\beta\mathcal{P}_m)]&S^s=0\ ,\\
2\pi G\Sigma_0^sr\mathcal{P}_mS^g+[\mathcal{M}_A +2\pi
G\Sigma_0^sr\mathcal{P}_m]&S^s=0\ ,
\end{split}\label{eq:62}
\end{equation}
where for notational simplicity we define
\begin{equation}
\begin{split}
&\mathcal{K}_A\equiv(m^2+6\beta)\Omega_g^2r^2-[m^2-(1-4\beta)^2/2]C_A^2\ ,\\
&\mathcal{L}_A\equiv(1/2-2\beta)C_A^2-2\Omega_g^2r^2+2\beta a_g^2\ ,\\
&\mathcal{M}_A\equiv{(m^2+2\beta-2)\Omega_s^2r^2}
/{(m^2+4\beta-4\beta^2)}-a_s^2\ .
\end{split}\label{eq:63}
\end{equation}
We now obtain the stationary MHD dispersion relation by
calculating the coefficient determinant of equation
(\ref{eq:62}). After a proper rearrangement, we obtain
\begin{equation}
\begin{split}
&\frac{\Delta_A}{\mathcal{K}_A}[\mathcal{K}_A(\Theta_A+\Xi_A)
+\mathcal{L}_A-2\pi G\Sigma_0^gr(2\beta\mathcal{P}_m)]\times\\
&(\mathcal{M}_A+2\pi G\Sigma_0^sr\mathcal{P}_m)
=4\pi^2G^2\Sigma_0^g\Sigma_0^sr^2\mathcal{P}_m^2
\bigg(1-\frac{2\beta\Delta_A}{\mathcal{K}_A}\bigg)\ ,
\end{split}\label{eq:64}
\end{equation}
where the left-hand side consists of two factors. One can show
that the left bracket is exactly the dispersion relation for a
single coplanar magnetized scale-free disc discussed by \citet{b11}
and the second parentheses denotes the dispersion relation for a
single hydrodynamic scale-free disc. The right-hand side denotes
the effect of gravitational coupling between perturbations in the
two scale-free discs. By setting $\beta=0$ for an isothermal
composite disc system, equation (\ref{eq:64}) then reduces to
dispersion relation (62) of \citet{b7} where MHD perturbations
in isothermal fluid-magnetofluid discs are investigated.

These calculations are straightforward but tedious and we
turn to the following subsections for further analyses.

\subsection{The  Aligned $m=0$ Case}

%
As already discussed earlier, the $m=0$ case is special and should
be treated in the procedure of setting $m=0$ and then taking the
limit of $\omega\rightarrow0$. By letting $\omega\rightarrow0$ in
equations $(\ref{eq:48})-(\ref{eq:52})$, we find $U=0$ and $R=0$
similar to the earlier work [e.g., \citet{b12,b6,b10,b7,b11}]. By
further requiring the scale-free condition, other physical
quantities should be in the forms of $Z\propto r^{-\gamma}$ and
$J\propto r^{1-\beta}$ which are exactly the same as those of the
background equilibrium. Since these perturbations are axisymmetric,
it turns out that such perturbations simply represent a sort of
rescaling of the axisymmetric background equilibrium. Nevertheless,
by carefully taking limits in calculations, we can also find a
stationary `dispersion relation'. We leave this analysis to
Appendix A for a further discussion and focus our attention
on $m\geq1$ cases in the next subsection.

\subsection{The Aligned $m\geq1$ Cases}

\subsubsection{$D_g^2$ Solutions for the Dispersion Relation}

We aim at evaluating the dispersion relation numerically to explore
the dependence of $D_g^2$ or $D_s^2$ on a set of dimensionless
parameters (i.e., $m,\ \beta,\ \eta,\ \delta,\ q,\ F$). One way
is to begin directly from equation (\ref{eq:64}), and the other
way is to calculate the coefficient determinant of equations
$(\ref{eq:55})-(\ref{eq:57})$ and (\ref{eq:59}). For both, we
use equations $(\ref{eq:20})-(\ref{eq:23})$ to make systematic
calculations. Making a full use of mathematical tools for
computations, we choose to follow the latter procedure. This
is a straightforward but onerous task; we show the results
below and leave the technical details to Appendix B. Here we
first compute $D_g^2$ because its expression appears simpler,
and then obtain the corresponding $D_s^2$ using equation
(\ref{eq:22}). When $D_s^2$ is smaller than $D_g^2$, we show
$D_s^2$ solutions in figures. Otherwise, we show both $D_s^2$
and $D_g^2$ in figures to identify physical solutions. We now
introduce a few handy notations below to simplify mathematical
manipulations, namely
\begin{equation}
\begin{split}
\mathcal{A}^A_m(\beta)&\equiv m^2+4\beta-4\beta^2\ ,\\
\mathcal{B}_m(\beta)&\equiv (1+2\beta)(m^2-2+2\beta)\ ,\\
\mathcal{C}(\beta)&\equiv {2\beta\mathcal{P}_0}/{[F(1+2\beta)]}\ ,\\
\mathcal{H}^A_m(\beta)&\equiv {A^A_m\mathcal{P}_m}/{\mathcal{C}}+B_m\ .\\
\end{split}\label{eq:65}
\end{equation}
In the following, the stationary MHD dispersion relation is
written as a cubic equation in terms of $y\equiv D_g^2$, namely
\begin{equation}
C^A_3y^3+C^A_2y^2+C^A_1y+C^A_0=0\ .\label{eq:66}
\end{equation}
The four coefficients $C^A_3$, $C^A_2$, $C^A_1$, and $C^A_0$ are
functions of dimensionless parameters $m,\ \beta,\ \delta,\ \eta$,
and $q^2$ and are defined explicitly by
\begin{equation}
\begin{split}
&C^A_3\equiv\mathcal{B}_m\mathcal{H}^A_m\eta\ ,\\
&C^A_2\equiv \bigg[(\mathcal{B}_m-\mathcal{A}^A_m)\mathcal{H}^A_m
+\frac{(\mathcal{A}^A_m+\mathcal{B}_m)
(\mathcal{H}^A_m-\mathcal{B}_m)\delta}{(1+\delta)}\bigg]\eta\\
&\qquad\qquad -\frac{(\mathcal{A}^A_m+\mathcal{B}_m)
(\mathcal{H}^A_m\delta+\mathcal{B}_m)}{(1+\delta)}+C^A_{2Q}q^2\ ,\\
&C^A_1\equiv\bigg[-\mathcal{A}^A_m\mathcal{H}^A_m
+\frac{(\mathcal{A}^A_m+\mathcal{B}_m)
(\mathcal{H}^A_m-\mathcal{B}_m)\delta}{(1+\delta)}\bigg]\eta\\
&+(\mathcal{A}^A_m+\mathcal{B}_m)^2
-\frac{(\mathcal{A}^A_m+\mathcal{B}_m)
(\mathcal{H}^A_m\delta+\mathcal{B}_m)}{(1+\delta)}+C^A_{1Q}q^2\ ,\\
&C^A_0\equiv C^A_{0Q3}q^6+C^A_{0Q2}q^4+C^A_{0Q1}q^2\ ,
\end{split}\label{eq:67}
\end{equation}
where the five new relevant coefficients
are further defined explicitly by
\begin{equation*}
\begin{split}
&C^A_{2Q}\equiv\frac{m^2(4\beta-1)-2\mathcal{A}^A_m
+8\beta^2-20\beta+6}{2(1+\delta)}(\mathcal{H}^A_m
+\mathcal{B}_m\delta)\eta
\\ &\qquad\qquad
-\frac{4(1-2\beta)\mathcal{A}^A_m-(1+2\beta)}
{(4\beta+2)(1+\delta)}\frac{\mathcal{B}_m
\mathcal{P}_m}{\mathcal{C}}\delta\eta\ ,\\
\end{split}
\end{equation*}
\begin{equation*}
\begin{split}
&C^A_{1Q}\equiv -\frac{\mathcal{A}^A_m+5\beta-2}{(1+\delta)}
(\mathcal{H}^A_m+\mathcal{B}_m\delta)\eta
\\ &\qquad
-\frac{\mathcal{A}^A_m(4\beta-3)+2\beta+1}
{(4\beta+2)(1+\delta)}\bigg(\mathcal{H}^A_m +\mathcal{B}_m\delta
-\frac{2\mathcal{B}_m\mathcal{P}_m\delta}{\mathcal{C}}\bigg)\eta
\\ &\qquad
-\frac{\mathcal{A}^A_m(4\beta-3)+2\beta+1}
{(4\beta+2)(1+\delta)}\frac{\mathcal{P}_m}
{\mathcal{C}}(\mathcal{A}^A_m+\mathcal{B}_m)\delta
\\ &\qquad
+(\mathcal{A}^A_m+\mathcal{B}_m)(\mathcal{A}^A_m+5\beta-2)\ ,\\
\end{split}
\end{equation*}
\begin{equation*}
\begin{split}
&C^A_{0Q3}\equiv -\frac{(4\beta-1)^3
(\mathcal{H}^A_m+\mathcal{B}_m\delta)\eta}{2(4\beta+2)^2(1+\delta)}
\\ &\qquad\ \ \
-\frac{(4\beta-1)^2(2\mathcal{A}^A_m-2\beta-1)}{(4\beta+2)^3(1+\delta)}
\frac{\mathcal{B}_m\mathcal{P}_m}{C}\delta\eta\ ,\\
\end{split}
\end{equation*}
\begin{equation*}
\begin{split}
&C^A_{0Q2}\equiv\frac{(4\beta-1)}{(4\beta+2)}\bigg[
\frac{(4\beta-1)}{2}
+\frac{(2\mathcal{A}^A_m-2\beta-1)}{(4\beta+2)(1+\delta)}
\frac{\mathcal{P}_m}{\mathcal{C}}\delta\bigg]
\\ &\qquad\quad
\times (\mathcal{A}^A_m+\mathcal{B}_m)-\frac{(4\beta-1)^2
(\mathcal{H}^A_m+\mathcal{B}_m\delta)\eta}
{2(4\beta+2)(1+\delta)}
\\ &\quad
+\frac{(4\beta-1)(2\mathcal{A}^A_m-2\beta-1)}
{(4\beta+2)^2(1+\delta)}\bigg(\mathcal{H}^A_m
+\mathcal{B}_m\delta-\frac{2\mathcal{B}_m\mathcal{P}_m\delta}
{\mathcal{C}}\bigg)\eta\ ,\\
\end{split}
\end{equation*}
\begin{equation}
\begin{split}
&C^A_{0Q1}\equiv\frac{(2\mathcal{A}^A_m-2\beta-1)}
{(4\beta+2)(1+\delta)}\bigg(\mathcal{H}^A_m+\mathcal{B}_m\delta
-\frac{\mathcal{B}_m\mathcal{P}_m\delta}{C}\bigg)\eta
\\ &\qquad
-\frac{(2\mathcal{A}^A_m-2\beta-1)}{(4\beta+2)(1+\delta)}
\bigg(1+\delta-\frac{\mathcal{P}_m}{\mathcal{C}}\delta\bigg)
(\mathcal{A}^A_m+\mathcal{B}_m)\ .
\end{split}\label{eq:68}
\end{equation}
Before solving cubic equation (\ref{eq:66}), we can examine
qualitatively several necessary requirements between $y\equiv D_g^2$
and other parameters involved. For $q=0$, we have $C^A_0=0$; meanwhile,
$C^A_3,\ C^A_2$, and $C^A_1$ become exactly the same as $C_2,\ C_1$,
and $C_0$, respectively in equation (39) of \citet{b10}. With the
introduction of a coplanar magnetic field, a new solution (SMDWs)
emerges and in the limit of $q\rightarrow0$, this solution shrinks
to zero. By letting $\eta=0$ and $\delta\rightarrow\infty$, we have
$C^A_3=0$ and other coefficients reduced back to the case of a single
MHD scale-free disc studied by \citet{b11} [see their coefficient
expressions (49)]. It is the presence of a stellar disc in the model
that results in an extra possible mode for stationary perturbations
(Lou \& Fan 1998b).


It is also useful to express this relation in terms of $D_s^2$
because a physical solution should have $D_g^2>0$ and $D_s^2>0$
simultaneously. We deduce from equation (\ref{eq:22}) that
\begin{equation}
\begin{split}
&D_g^2=D_s^2/\eta +\mathcal{X}\ ,\\
&\mathcal{X}\equiv (1-\eta)/\eta-(4\beta-1)q^2/(4\beta+2)\ .
\end{split}\label{eq:69}
\end{equation}
Substituting equation (\ref{eq:69}) into equation
(\ref{eq:68}), we immediately obtain a new
relation in terms of $y'\equiv D_s^2$, namely
\begin{equation}
C^{A'}_3y'^3+C^{A'}_2y'^2+C^{A'}_1y'+C^{A'}_0=0\ ,\label{eq:70}
\end{equation}
where the four primed coefficients are defined by
\begin{equation}
\begin{split}
&C^{A'}_3\equiv C^A_3/\eta^3\ ,\\
&C^{A'}_2\equiv (C^A_2+3C^A_3\mathcal{X})/\eta^2\ ,\\
&C^{A'}_1\equiv (C^A_1+2C^A_2\mathcal{X}+3C^A_3\mathcal{X}^2)/\eta\ ,\\
&C^{A'}_0\equiv C^A_0+C^A_1\mathcal{X}+C^A_2\mathcal{X}^2
+C^A_3\mathcal{X}^3\ .
\end{split}\label{eq:71}
\end{equation}

It would be straightforward but tedious to express the coefficients
of cubic equation (\ref{eq:70}) explicitly as we did before.
We would use equation (\ref{eq:66}) for $D_g^2$ computations
and equation (\ref{eq:70}) for theoretical inferences later.

\subsubsection{Dependence of $D_s^2$ and $D_g^2$ on
\\ \quad\qquad
Relevant Dimensionless Parameters }

We now perform numerical computations. Our main device is to
examine the solution of $D_g^2$ or $D_s^2$ versus $\sigma$
when other parameters are specified. Here $\sigma\equiv1/\eta$
is introduced merely for convenience (n.b., $\eta\leq1$ thus
$1\leq\sigma<+\infty$).
Recall that when $\beta\leq1/4$, $D_s^2$ is always smaller than
$D_g^2$ and when $\beta>1/4$, $D_g^2$ can be smaller than $D_s^2$
when $\sigma$ is sufficiently close to $1$; for this reason, we show the
dispersion relation in term of $D_s^2$ through equation (\ref{eq:22})
instead of $D_g^2$ in most cases for $\beta\leq1/4$. From now on, we
use $y$ instead of $y'$ to denote the $D_s^2$ solution for
convenience. When $\beta>1/4$, we will show both $D_s^2$ and
$D_g^2$ solutions with clear labels, respectively.

Typically, equations (\ref{eq:66}) and (\ref{eq:70}) are algebraic
cubic equations and have three roots and there exists at least one
real root because all coefficients are real. In most cases, the
three roots are all real and do not intersect with each other and
we use $y_1$, $y_2$ and $y_3$ to denote the upper, middle and
lower branches of the $D_s^2$ solution, respectively. Generally,
they correspond to different types of stationary MHD density waves
\citep{b5,b11}. As we have several parameters to choose freely in
our model, the solution behaviours are much more richer than
previous works as we will discuss presently.

\begin{figure}
    \centering
    \subfigure[$m=2,\ \beta=-0.24,\ q=0.3,\ F=1$]{
      \label{fig:1a}
      \includegraphics[width=75mm,height=51mm]{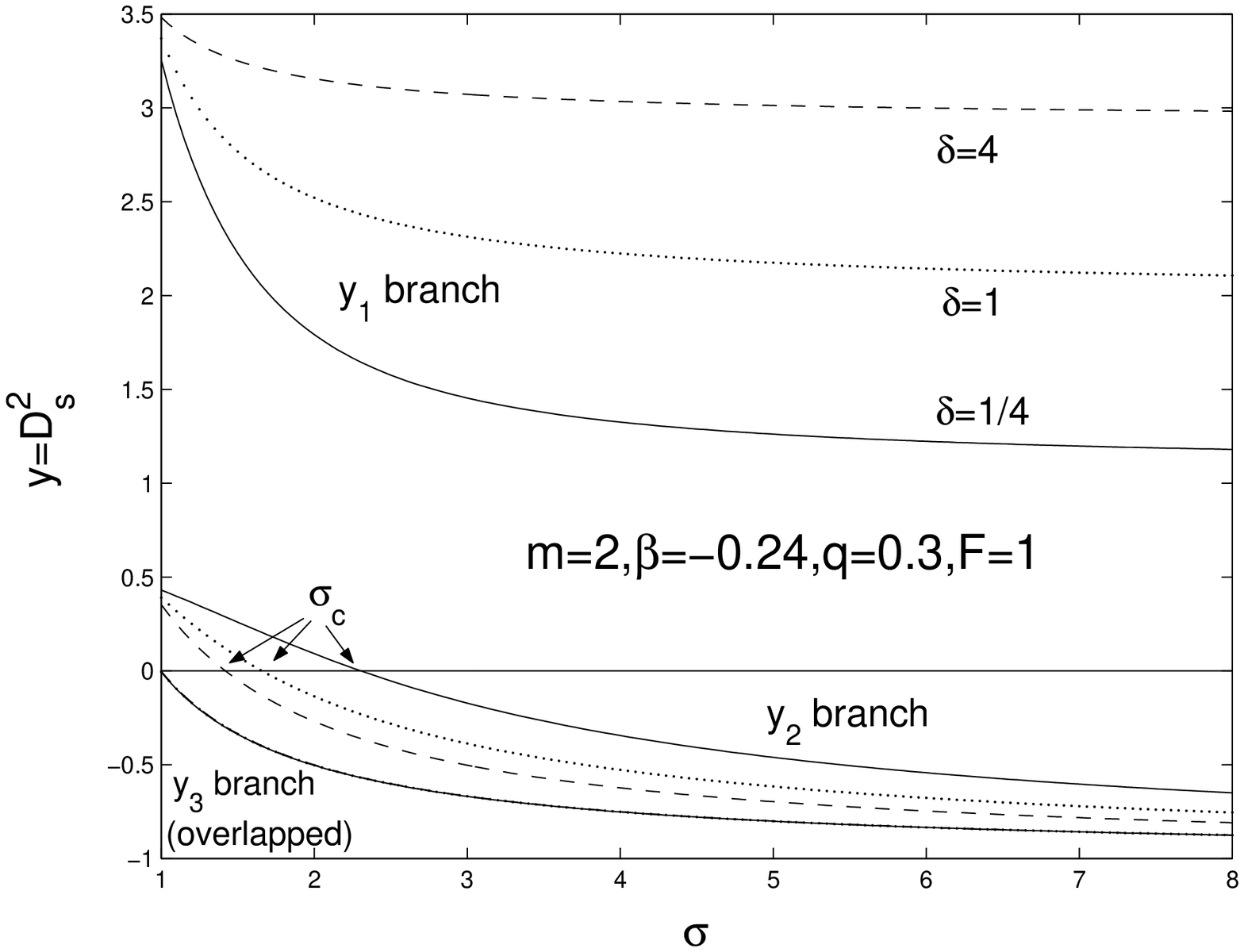}}
    \hspace{1in}
    \subfigure[$m=2,\ \beta=-0.24,\ q=1,\ F=1$]{
      \label{fig:1b}
      \includegraphics[width=75mm,height=51mm]{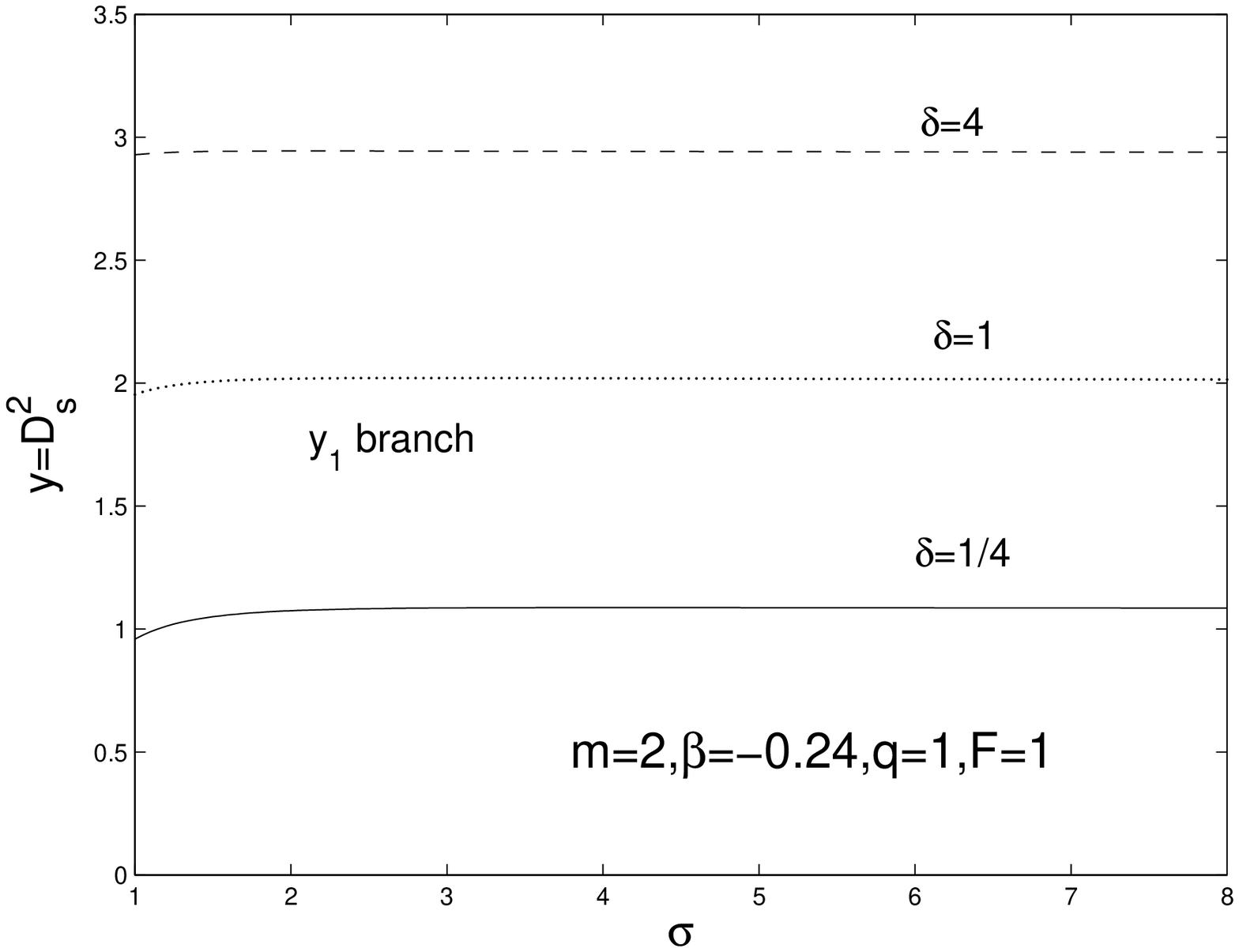}}
    \hspace{1in}
    \subfigure[$m=2,\ \beta=-0.24,\ q=3,\ F=1$]{
      \label{fig:1c}
      \includegraphics[width=75mm,height=51mm]{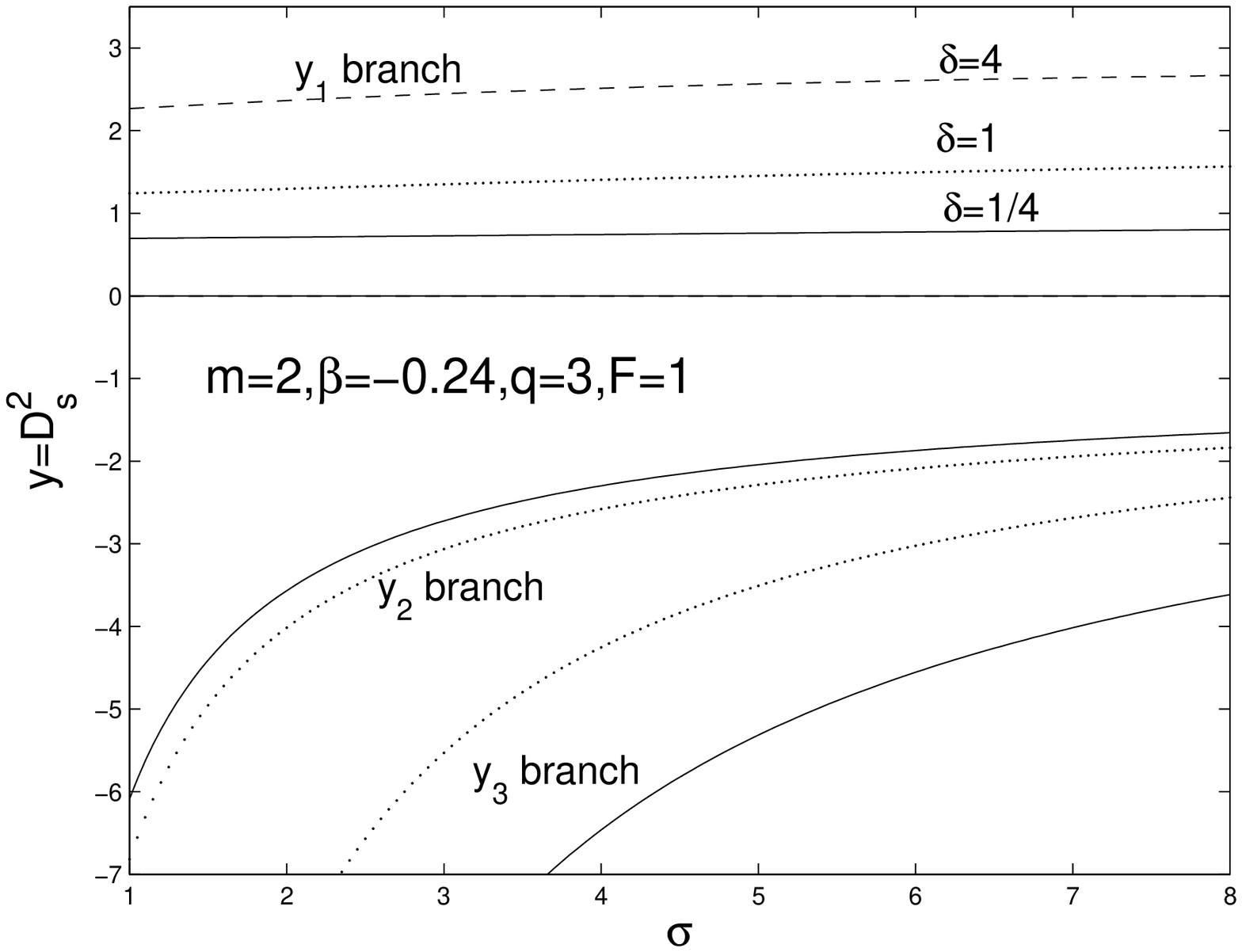}}
  \caption{Aligned $D_s^2$ solution branches derived from
  equations (\ref{eq:66}) and (\ref{eq:22}) as functions of
  $\sigma\equiv1/\eta$ by specifying $m=2$ (i.e., bar-like),
  $F=1$ (i.e., full discs) and $\beta=-0.24$. For each set
  of $\delta=1/4$ (solid curve), $1$ (dotted curve) and $4$
  (dashed curve), we choose $q=0.3,\ 1$ and $3$ and show
  corresponding results in panels (a), (b) and (c),
  respectively. In panel (a), the three curves of the $y_3$
  branch overlap with each other. Sometimes cubic equation
  (\ref{eq:66}) has only one real root. In panel (b), the
  $y_2$ and $y_3$ branches are complex for the three chosen
  $\delta$ values. In panel (c), the $y_2$ and $y_3$
  branches are complex for the $\delta=4$ case
  For $\beta<1/4$, $D_g^2$ remains always greater
  than $D_s^2$.}\label{fig:1}
\end{figure}

\begin{figure}
    \centering
    \subfigure[$m=2,\ \beta=0.4,\ \delta=1,\ q=0.3,\ F=1$]{
      \label{fig:4a}
      \includegraphics[width=75mm,height=55mm]{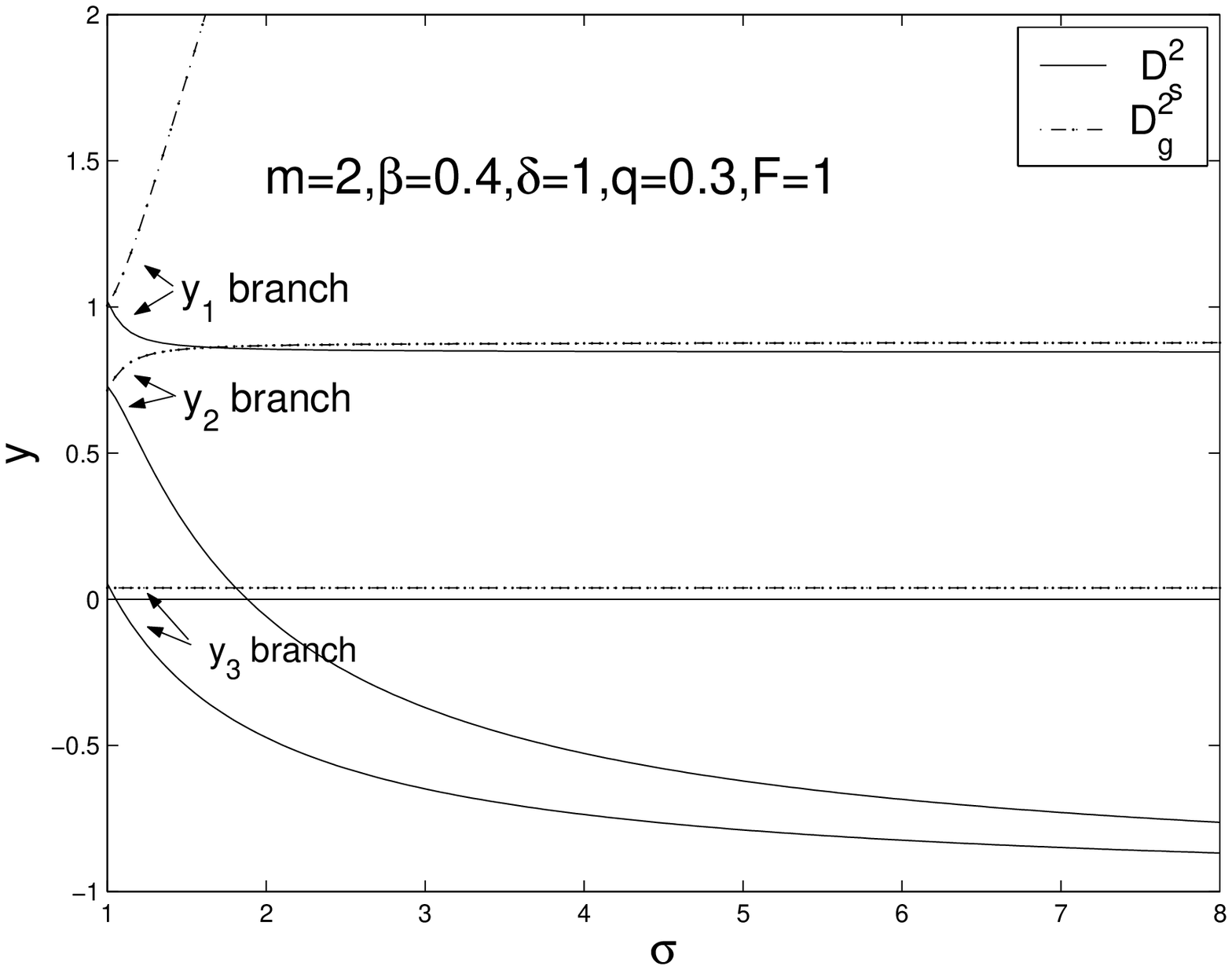}}
    \hspace{1in}
    \subfigure[$m=2,\ \beta=0.4,\ \delta=1,\ q=1,\ F=1$]{
      \label{fig:4b}
      \includegraphics[width=75mm,height=55mm]{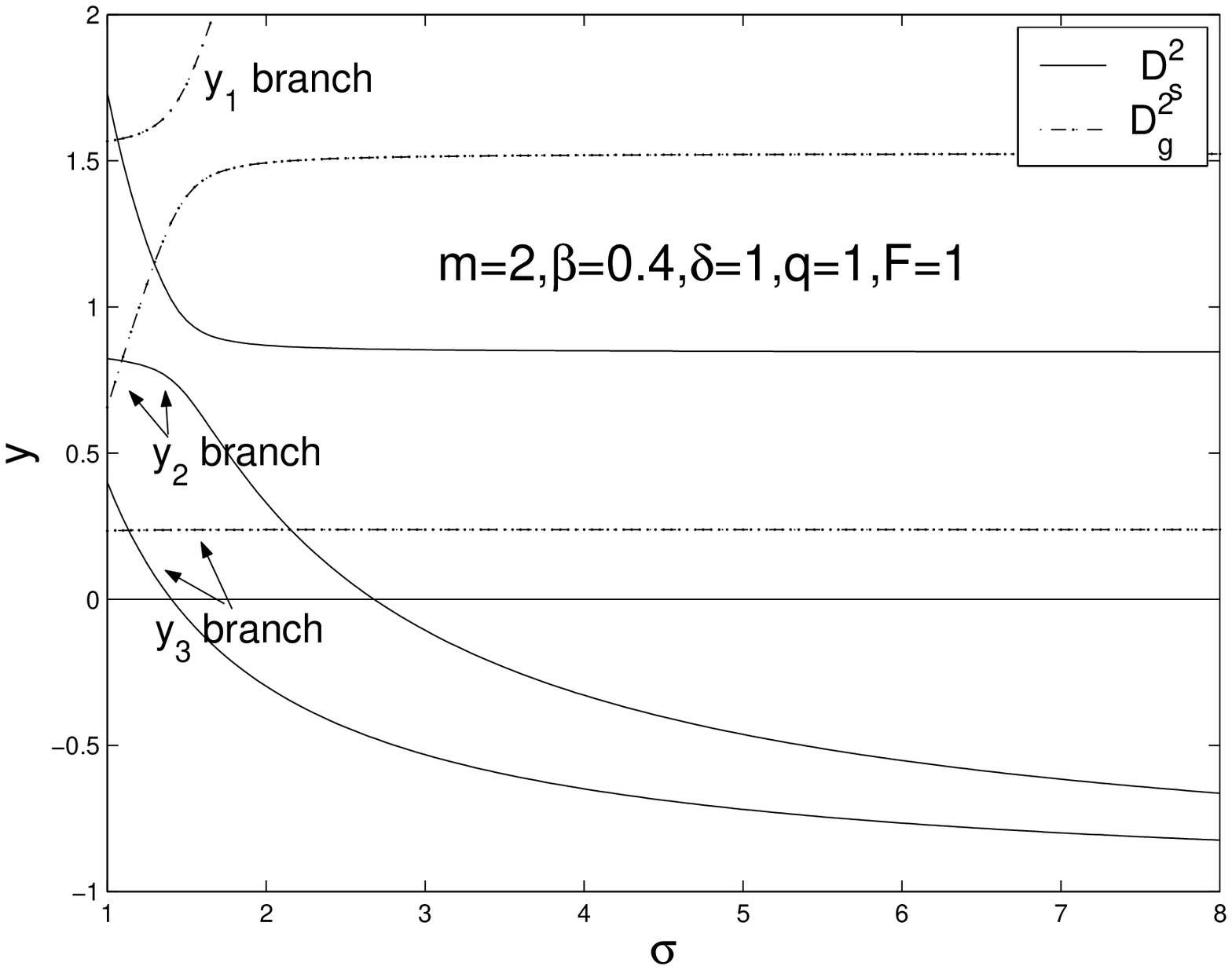}}
    \hspace{1in}
    \subfigure[$m=2,\ \beta=0.4,\ \delta=1,\ q=2.5,\ F=1$]{
      \label{fig:4c}
      \includegraphics[width=75mm,height=55mm]{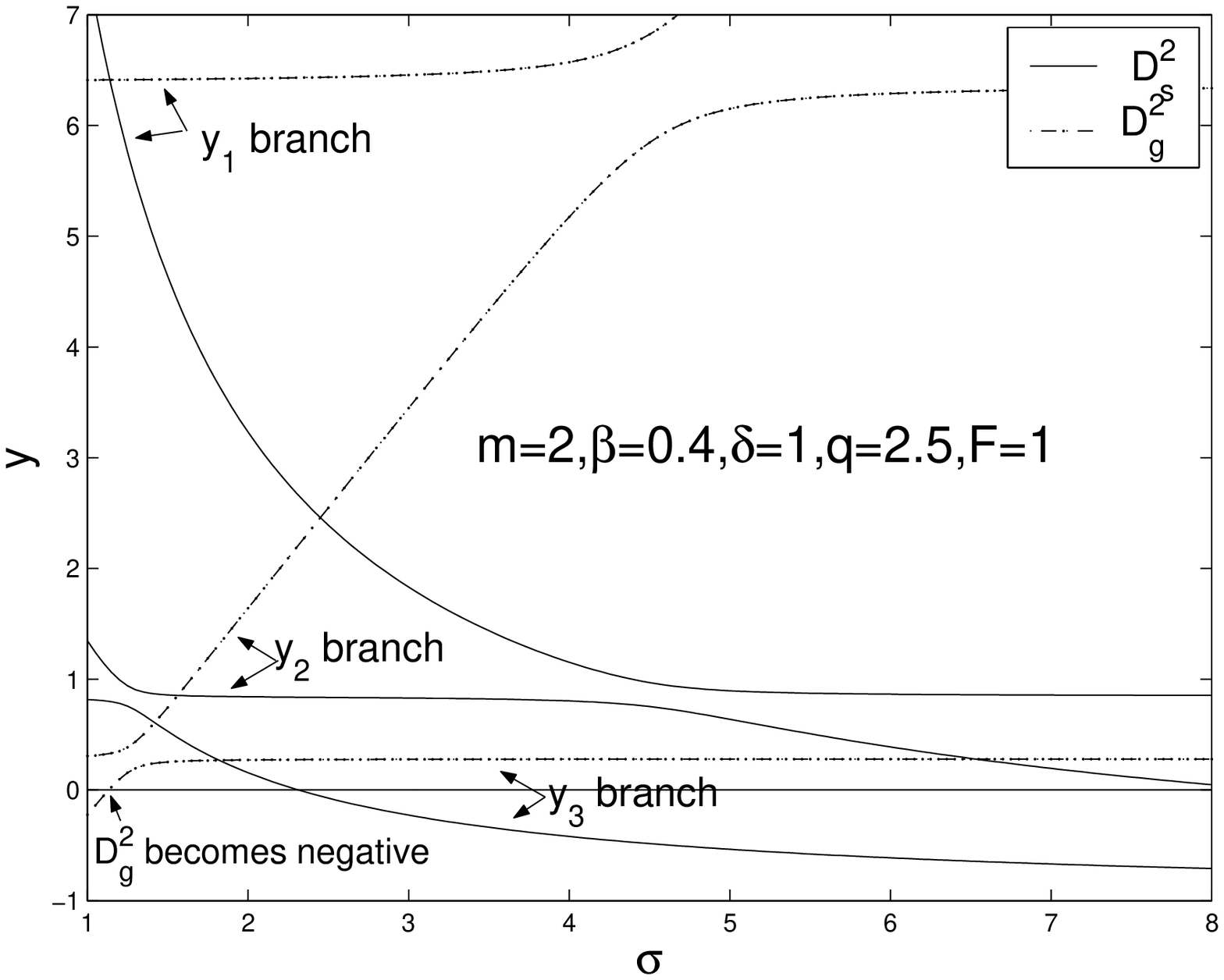}}
  \caption{The $D_s^2$ and $D_g^2$ solution curves versus $\sigma$
  for aligned $m=2$ (bar-like) full discs with $\beta=0.4$, $\delta=1$,
  $F=1$ and $q=0.3$, $1$ and $2.5$ shown in panels (a), (b) and (c),
  respectively. $D_s^2$ solutions are shown in solid curves, while
  $D_g^2$ solutions are shown in dashed curves. The corresponding
  $y_1$, $y_2$, and $y_3$ branches are ordered from up to down,
  respectively. In panel (c), the lowest $D_g^2$ branch becomes
  negative while the corresponding $D_s^2$ branch remains positive
  as $\sigma$ becomes sufficiently small.
  }\label{fig:4}
\end{figure}

As has mentioned before, the net magnetic force serves as either a
centripetal or a centrifugal force when $\beta<1/4$ or $\beta>1/4$,
respectively. We take $\beta=-0.24$, $\beta=-0.1$, $\beta=1/4$ and
$\beta=0.4$ in order and examine the dynamical influence of the
magnetic field on solution behaviours as shown in Fig.\ \ref{fig:1}$
-$Fig.\ \ref{fig:4}. For the moment, we focus mainly on the full
(i.e., $F=1$) disc system with $m=2$. Meanwhile, we specify different
values of mass density ratio $\delta=1/4$, $\delta=1$ and $\delta=4$
to assess the influence of the surface mass density ratio on the two
coupled discs.

\begin{figure}
    \centering
    \subfigure[$m=2,\ \beta=1/4,\ q=0.3,\ F=1$]{
      \label{fig:3a}
      \includegraphics[width=75mm,height=55mm]{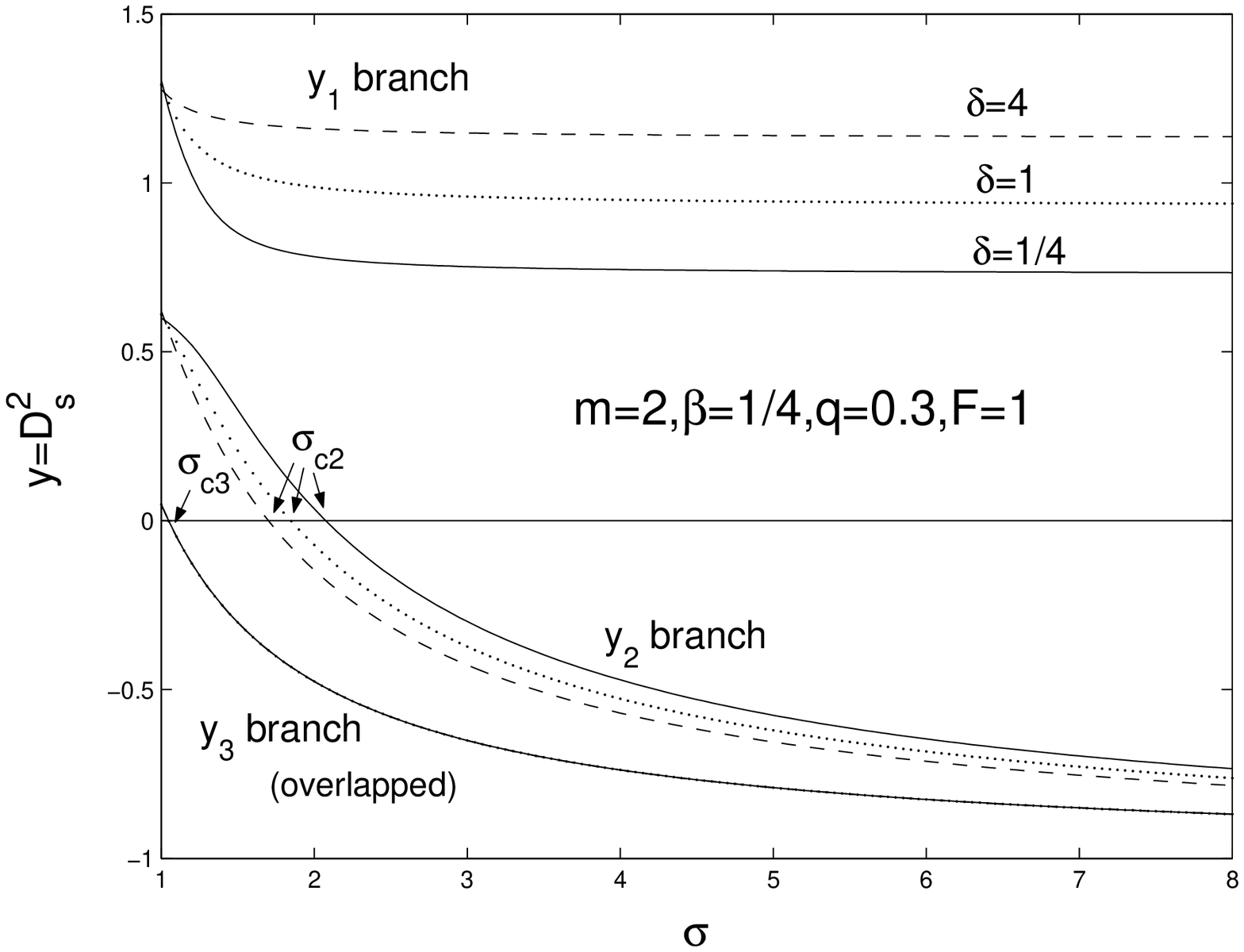}}
    \hspace{1in}
    \subfigure[$m=2,\ \beta=1/4,\ q=1,\ F=1$]{
      \label{fig:3b}
      \includegraphics[width=75mm,height=55mm]{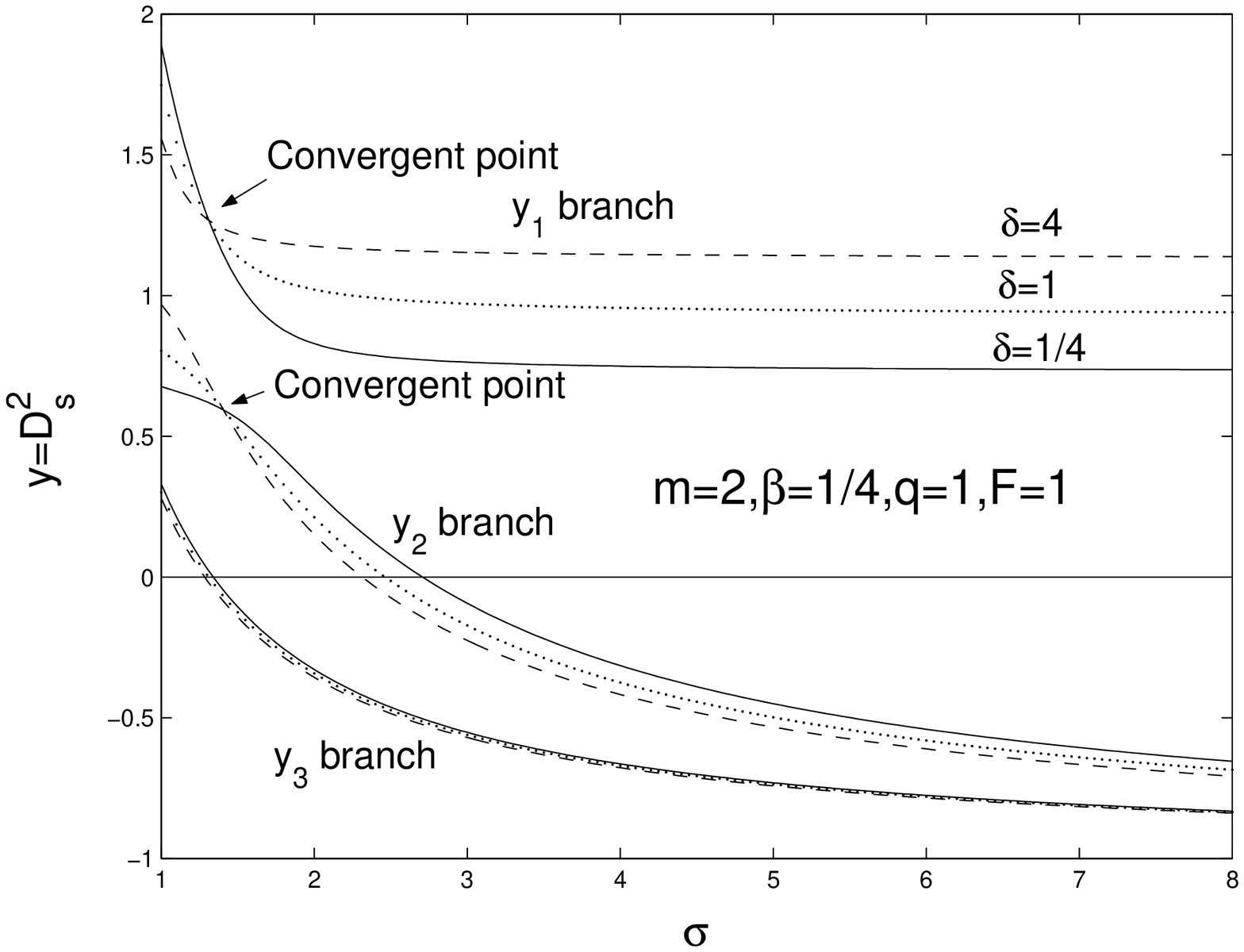}}
    \hspace{1in}
    \subfigure[$m=2,\ \beta=1/4,\ q=3,\ F=1$]{
      \label{fig:3c}
      \includegraphics[width=75mm,height=55mm]{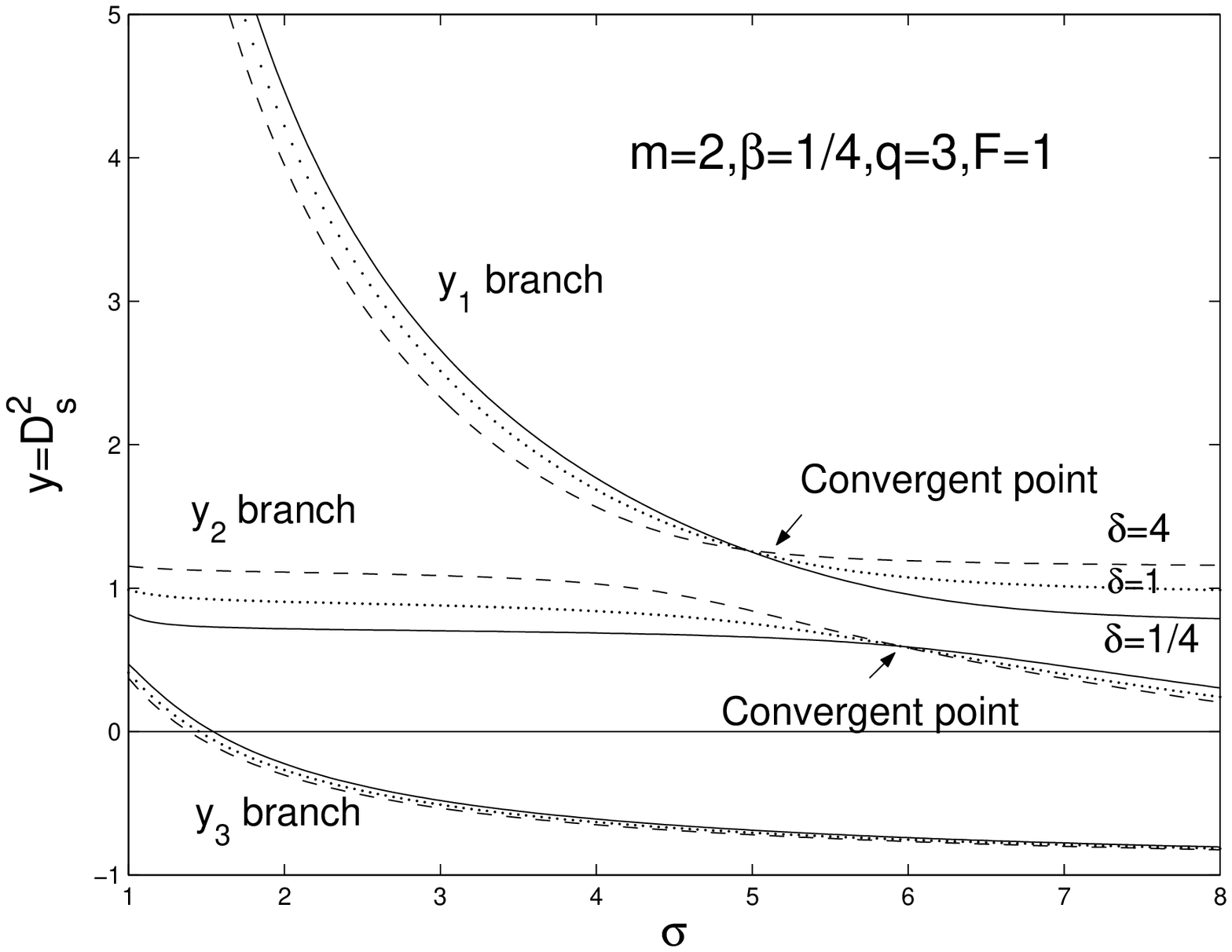}}
  \caption{The solution curves of $D_s^2$ versus $\sigma$ for aligned
  $m=2$ (bar-like) full discs with $F=1$, $\beta=1/4$, $\delta=1/4$
  (solid curve), $1$ (dotted curve) and $4$ (dashed curve) and $q=0.3$,
  $1$ and $3$ shown in panels (a), (b), and (c), respectively. Note
  that for small $q$, there are three physical $D_s^2$ solutions as
  $\sigma$ approaches 1 and when $q$ is large enough there exists a
  `convergent point' where branches with different $\delta$ values
  converge. Curves for different $\delta$ values in the $y_3$ branch
  almost merge with each other to various extents; in panel (a), the
  three curves of the $y_3$ branch overlap each other.
   }\label{fig:3}
\end{figure}

The $y_1$ and $y_2$ solution branches generally correspond to
the unmagnetized solutions [i.e., Lou \& Fan 1998b; \citet{b10}]
while the $y_3$ solution branch is additional (i.e., SMDWs by
nature) due to the very presence of a coplanar magnetic field;
these MHD density wave mode classifications are qualitatively
similar to the corresponding isothermal solution branches of
\citet{b7}. This identification of SMDWs for $y_3$ branch
can be seen clearly as the magnetic field becomes sufficiently
small (e.g., $q=0.3$). Here, the $y_3$ branch is almost independent
of $\delta$ variation (see Fig.\ \ref{fig:1}$-$Fig.\ \ref{fig:3}).
In fact, this root is exactly the newly emerged root as noted
after equation (\ref{eq:68}). The corresponding $D_g^2$ solution
of this branch remains almost constant and borders on zero because
coefficient $C^A_0$ is sufficiently small when magnetic field is
weak and this solution will become trivial
for $q=0$ because of zero $D_g^2$ solution.

Generally speaking, the $y_1$ branch tends to be positive (this may not
be always true as we shall show later), and the $y_2$ branch may be
either positive or negative. The $y_3$ branch is mostly negative except
for few cases. The critical value of $\sigma$ when we have zero root is
determined by $C^{A'}_0=0$ in equation (\ref{eq:71}). By a proper
substitution, we find that the result is a cubic equation of $\eta$
(or equivalently of $\sigma$) which indicates that theoretically there
might be three critical values of $\sigma$ labelled as $\sigma_{c1}$,
$\sigma_{c2}$ and $\sigma_{c3}$ for each branch of the zero root point.
Practically, we often see $\sigma_{c2}$ and sometimes $\sigma_{c3}$ and
seldom $\sigma_{c1}$ in the range of $1\leq\sigma<+\infty$. However, for
$D_g^2$ solutions, since $C^A_0$ is a linear function of $\eta$, there
can be at most only one critical value of $\eta$ or $\sigma$ within a
proper range.

We also note that there can be situations that only one real
root exists and the other two roots are complex conjugates
[e.g., Fig.\ \ref{fig:1b} and Fig.\ \ref{fig:1c}]. In previous
investigations [e.g., \citet{b6,b7,b10,b11}], we never encounter
complex roots. According to Fig.\ \ref{fig:1} (i.e., $\beta=-0.24$),
we see a stronger magnetic field tends to terminate the existence
of the physical part of the $y_2$ branch.
For a larger value of $\beta$, this tendency appears to diminish.

\begin{figure}
    \centering
    \subfigure[$m=2,\ \beta=-0.1,\ q=0.3,\ F=1$]{
      \label{fig:2a}
      \includegraphics[width=75mm,height=52mm]{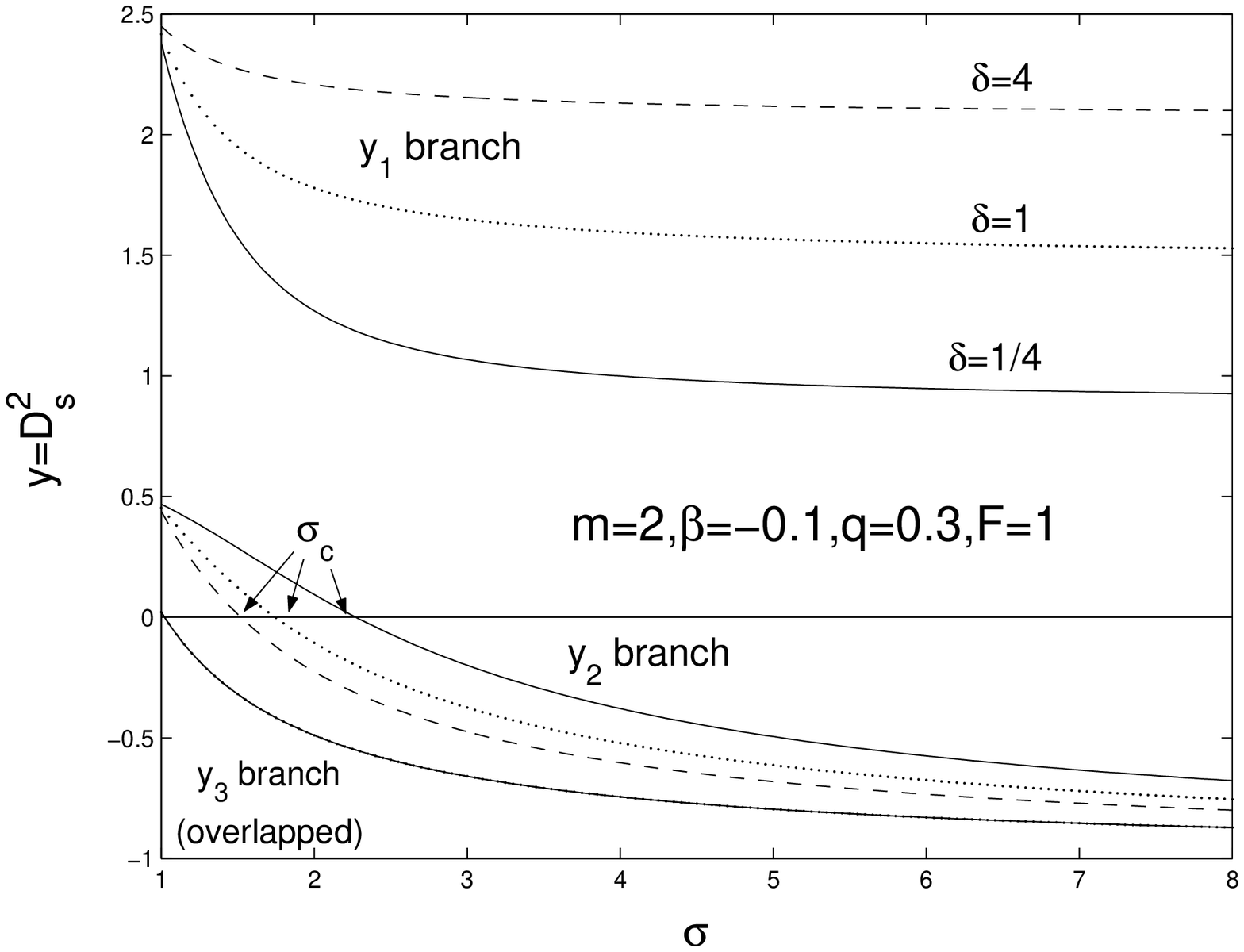}}
    \hspace{1in}
    \subfigure[$m=2,\ \beta=-0.1,\ q=1,\ F=1$]{
      \label{fig:2b}
      \includegraphics[width=75mm,height=52mm]{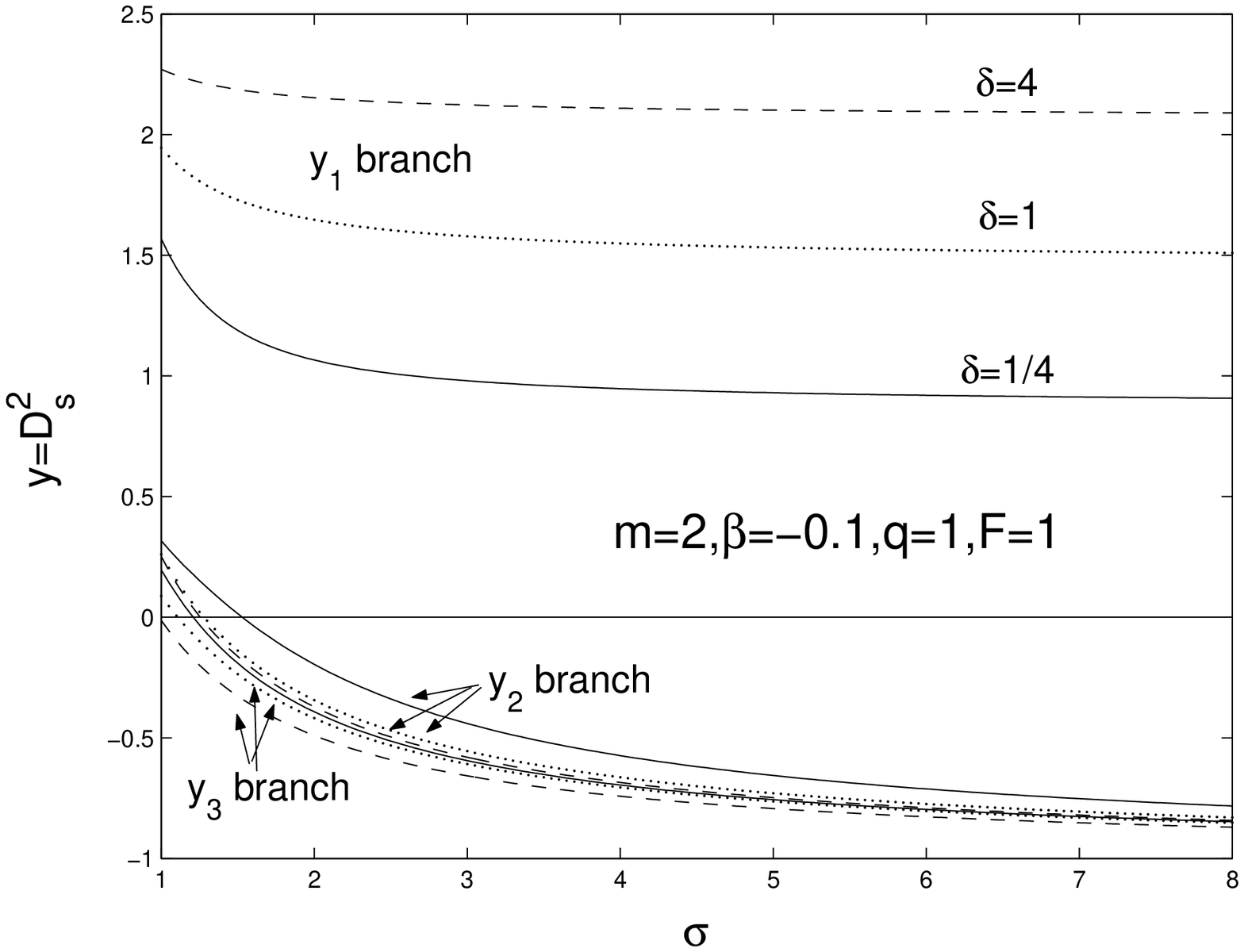}}
    \hspace{1in}
    \subfigure[$m=2,\ \beta=-0.1,\ q=3,\ F=1$]{
      \label{fig:2c}
      \includegraphics[width=75mm,height=52mm]{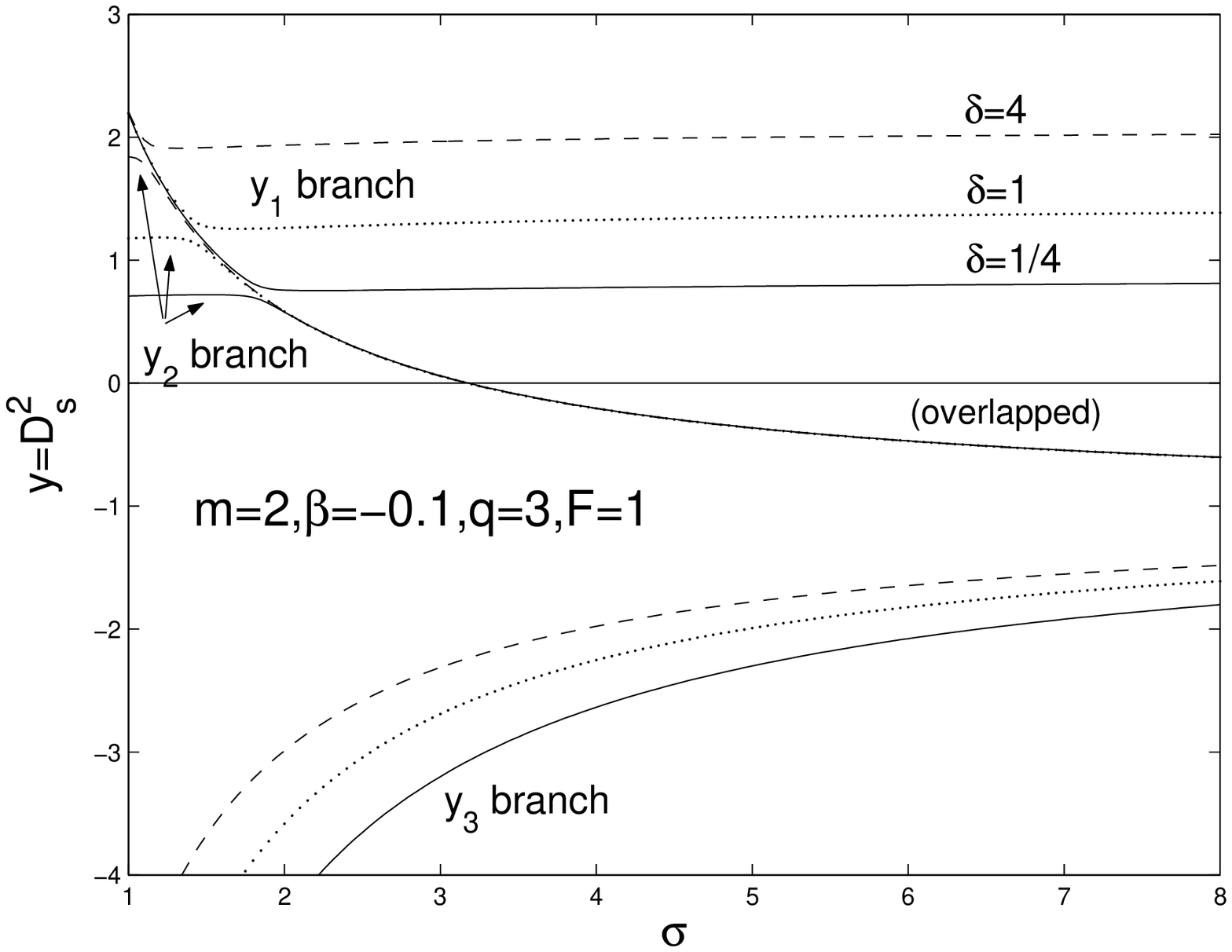}}
  \caption{The solution curves of $D_s^2$ versus $\sigma$ for aligned
  $m=2$ (bar-like) full discs with $F=1$, $\beta=-0.1$, $\delta=1/4$
  (solid curve),
  $1$ (dotted curve) and $4$ (dashed curve) and $q=0.3$, $1$ and $2.5$
  in panels (a), (b) and (c), respectively. In panel (a), curves for
  the three different $\delta$ values of the $y_3$ branch almost overlap
  with each other. The same thing happens in panel (c) for large
  $\sigma\gsim 1.9$ in the $y_2$ branch and for small $\sigma$ in the
  $y_1$ branch; the dashed curve of $y_2$ branch almost coincides with
  the solid curve of $y_1$ branch for small $\sigma\lsim 1.9$.
  In panel (b) with $q=1$, the $y_2$ and $y_3$ branches are
  close to each other. The three curves of the $y_3$ branch
  are lower than the three curves of the $y_2$ branch.
  }\label{fig:2}
\end{figure}

As expected, the magnetic field has a very slight influence on the
composite disc system when it becomes weak enough (e.g., $q=0.3$).
For example, we compare our panel (b) of Fig.\ \ref{fig:3} with
panel (b) of Fig.\ $3$ in \citet{b10} and see that the $y_1$ and
$y_2$ branches are strikingly similar to their mutual counterparts.
A point in common is that when $\sigma$ becomes large (i.e., small
$\eta$ corresponding to a situation where the velocity dispersion
in the stellar disc is much higher than the sound speed in the
magnetized gas disc), the curves tend to be more or less flat. As
$q$ becomes larger, or equivalently, the Alfv\'en speed is
comparable to or even higher than the gas sound speed, the magnetic
field effect can be more prominent, especially for small $\sigma$.
As $\beta$ becomes very small (e.g., $\beta=-0.24$), the magnetic
field provides an attraction force, a stronger magnetic field should
thus enhance the angular velocity of the magnetized gas disc (i.e.,
a larger $D_g^2$). Meanwhile, we see from Fig.\ \ref{fig:1} that it
reduces the angular velocity for the stellar disc. Since here
$D_g>D_s$, so the existence of a magnetic field enlarges the angular
speed difference between the gaseous and stellar discs [see also
equation (\ref{eq:22})]. As $\beta$ becomes slightly larger (e.g.,
$\beta=-0.1$), $D_g$ still increases when $q$ grows with a greater
attractive Lorentz force as expected (not shown in figures); for
$D_s$ solution, the situation is more complicated.
Here in the $\beta=-0.1$ cases with a small $\sigma$, the increase
of magnetic field strength first lowers the $y_1$ and $y_2$ branches
and then raises them upward when $q$ becomes sufficiently large. One
interesting phenomenon is that for large $q$, the $y_1$ branch is flat
for large $\sigma$ and the curve suddenly rises when $\sigma$ reduces
to a certain value; meanwhile, the $y_2$ branch catches up and
stretches that flat curve at small $\sigma$ [see Fig.\ \ref{fig:2c},
Fig.\ \ref{fig:3c}, Fig.\ \ref{fig:4c}]. For large $\beta$ (e.g.,
$\beta=1/4$ and $\beta=0.4$), a larger $q$ directly results in an
increase in all three $D_s^2$ solution branches.

We now examine the effect of varying the surface mass density
ratio $\delta$. We observe a higher $y_1$ branch for a larger
$\delta\equiv\Sigma_0^g/\Sigma_0^s$ and a higher $y_2$ branch
for a smaller $\delta$ when $\beta$ and $q$ are small [see Fig.\
\ref{fig:1a} and Fig.\ \ref{fig:2a}]. This is consistent with
the hydrodynamic results of \citet{b10} where magnetic field
is absent. In the presence of magnetic field, we notice a new feature
that as $\beta$ and $q$ becomes sufficiently large, there emerges a
`convergent point' where for different $\delta$ values, some
branches converge and then their relative positions are changed [see
Fig.\ \ref{fig:3b} and Fig.\ \ref{fig:3c}].
For different branches, the convergent point varies.
In addition, we see from equation (\ref{eq:69}) that since
$\mathcal{X}$ is independent of $\delta$, the convergent point for
$D_g^2$ solution is the same as that for the $D_s^2$ solution.

It is of some interest to consider the case in Fig.\ \ref{fig:4}
for $\beta=0.4$. Different from other cases, here $D_g$ may
become smaller than $D_s$ in the presence of magnetic field. This
happens when $\sigma$ approaches 1 and $q$ becomes fairly large.
Specifically in Fig.\ \ref{fig:4c}, $D_g^2$ solution becomes
negative and thus unphysical even though $D_s^2$ remains positive.

\begin{figure}
  \centering
  \subfigure[$m=1,\ \beta=-0.24,\ \delta=1,\ q=0,\ F=1$]{
      \label{fig:5a}
      \includegraphics[width=75mm,height=55mm]{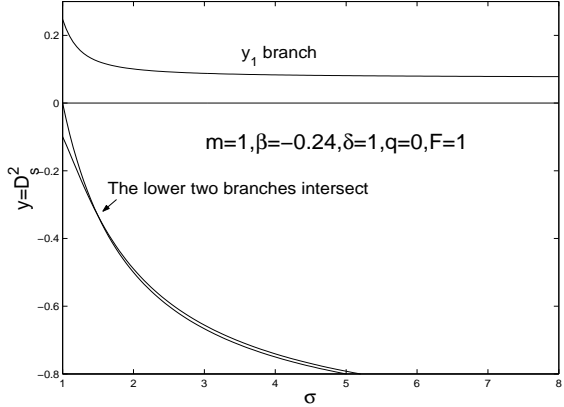}}
    \hspace{1in}
    \subfigure[$m=1,\ \beta=-0.24,\ \delta=1,\ q=1,\ F=1$]{
      \label{fig:5b}
      \includegraphics[width=75mm,height=55mm]{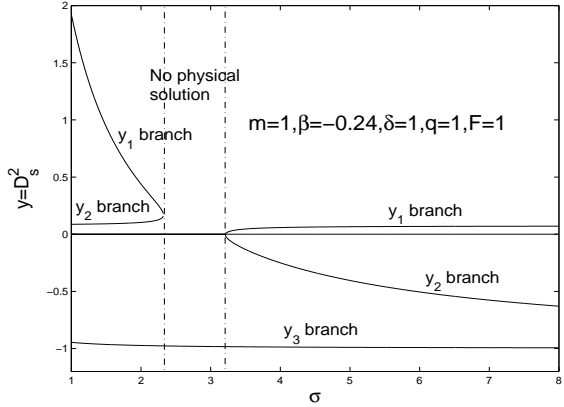}}
  \caption{The $D_s^2$ solution curves versus $\sigma$ for
  aligned $m=1$ full discs with $\beta=-0.24$, $\delta=1$, $F=1$,
  and $q=0$ and 1 shown in panels (a) and (b), respectively. For
  $q=0$, the lower two $D_s^2$ solution branches actually intersect
  with each other. For $q=1$ and $2.3<\sigma<3.2$, there is no
  physical $D_s^2$ solutions.
   }\label{fig:5}
\end{figure}

Regarding the dependence on $m$, there are not many notable features
except for $m=1$. In this case, it is possible that the solution
branches intersect and also there can be no physical $D_s^2$
solutions. We show a case in Fig.\ \ref{fig:5} with $\beta=-0.24$.
We note that in panel (b) of Fig.\ \ref{fig:5}, unlike previous $m=2$
cases, here the $y_2$ branch is the newly emerged branch due to the
magnetic field. It intersects with $y_3$ branch at first [i.e., $q=0$
in Fig.\ \ref{fig:5a}]. With the increase of magnetic field strength,
it moves up and separates from the $y_3$ branch and then meets the
$y_1$ branch. The result is that complex conjugate roots appear and
the only real root lies along the $y_3$ branch which is negative and
thus unphysical. Therefore, the parameter regime between the two
meeting points of $y_1$ and $y_2$ branches (approximately
$2.3<\sigma<3.2$) allows no stationary global MHD perturbation modes.

\subsubsection{Phase Relationships among Perturbation Variables}

Phase relationships among aligned perturbation variables reflect
their connections on large scales and provide models in interpreting
the results from optical and synchrotron radio observations of barred
or barred-spiral galaxies. In the aligned cases, most variable
quantities are real (except for the $r$-component magnetic
perturbation $R$) and thus they are qualitatively either `in phase'
or `out of phase'. Our main
interest is to examine the phase relationships between surface mass
density for the two discs and their relation to the magnetic field.
To derive a quantitative expression, we start from equations
$(\ref{eq:58})-(\ref{eq:60})$ and obtain

\begin{equation}
\begin{split}
&{S^g}/{S^s}=-1-{\mathcal{M}_A}/
{(2\pi G\Sigma_0^sr\mathcal{P}_m)}\ ,\\
&\frac{\mathrm{i}R}{S^g}
=\frac{mB_0}{\Sigma_0^g}\bigg[\Theta_A
\bigg(1+\frac{S^s}{S^g}\bigg)+\Xi_A\bigg]\ ,\\
&Z={(4\beta-1)}\mathrm{i}R/{(2m)}\ .
\end{split}\label{eq:72}
\end{equation}
There is another version of expression for
$\mathrm{i}R/S^g$ obtained by a combination of
equations (\ref{eq:55}) and (\ref{eq:56}), namely
\begin{equation}
\frac{\mathrm{i}R}{S^g}=\frac{mB_0}{\Sigma_0^g}
\bigg[\frac{4\pi G\beta\Sigma_0^gr\mathcal{P}_m}
{\mathcal{K}_A}\bigg(1+\frac{S^s}{S^g}\bigg)
-\frac{\mathcal{L}_A}{\mathcal{K}_A}\bigg]\ .\label{eq:73}
\end{equation}
Equations (\ref{eq:72}) and (\ref{eq:73}) for $\mathrm{i}R/S^g$ are
exactly the same for one can check that their combination yields the
dispersion relation (\ref{eq:64}). From equation (\ref{eq:73}), it
would be easier to check this consistency. Setting $\beta=0$ for the
isothermal case, equation (\ref{eq:73}) reduces to equation (68) of
\citet{b7} (they considered the ratio $S^g/Z$) as expected. Next is
to evaluate these equations to express them in physical parameters.
Note that $\mathrm{i}R/S^g$ is no longer dimensionless, to evaluate
this relationship, we take the proportional factor contained in the
brackets (i.e., removing $B_0/\Sigma_0^g$) of equation (\ref{eq:72})
for calculations.
We show these results below.
\begin{equation}
\begin{split}
&\frac{S^g}{S^s}=-1-\frac{(\mathcal{B}_mD_s^2
-\mathcal{A}^A_m)}{(D_s^2+1)}
\frac{\mathcal{C}(1+\delta)}{\mathcal{P}_m\mathcal{A}^A_m}\ ,\\
&\frac{\mathrm{i}R}{S^g}\propto\frac{m}{(1+2\beta)^2D_g^2
-(1/2-2\beta)q^2}\bigg[(1+2\beta)D_g^2-1
\\ &
\ \ +\frac{[D_g^2+1+{(4\beta-1)q^2}/{(4\beta+2)}]
\mathcal{P}_m\delta}{\mathcal{C}(1+\delta)}
\bigg(1+\frac{S^s}{S^g}\bigg)\bigg]\ .
\end{split}\label{eq:74}
\end{equation}
The first expression in equation (\ref{eq:74}) is the same as
equation (42) of \citet{b10} where a composite disc system of
gravitationally coupled unmagnetized discs is analyzed.

In equations (\ref{eq:72}) and (\ref{eq:73}), there is an imaginary
unit $\mathrm{i}$ on $R$ indicating that the radial component of the
magnetic field perturbation lags a phase $\pi/2$ to the mass density
perturbation. In the third
expression of equation (\ref{eq:72}), the factor $4\beta-1$ means that
the azimuthal component of the magnetic field perturbation is ahead of
or lags behind the radial component of the magnetic field perturbation
by $\pi/2$ for $\beta>1/4$ or $\beta<1/4$, respectively. In the special
case of $\beta=1/4$, we have $Z=0$ and there is no azimuthal component
of magnetic field perturbation.
\begin{figure}
    \centering
    \subfigure[$m=2,\ \beta=-0.24,\ q=0.3,\ F=1$]{
      \label{fig:6a}
      \includegraphics[width=75mm,height=55mm]{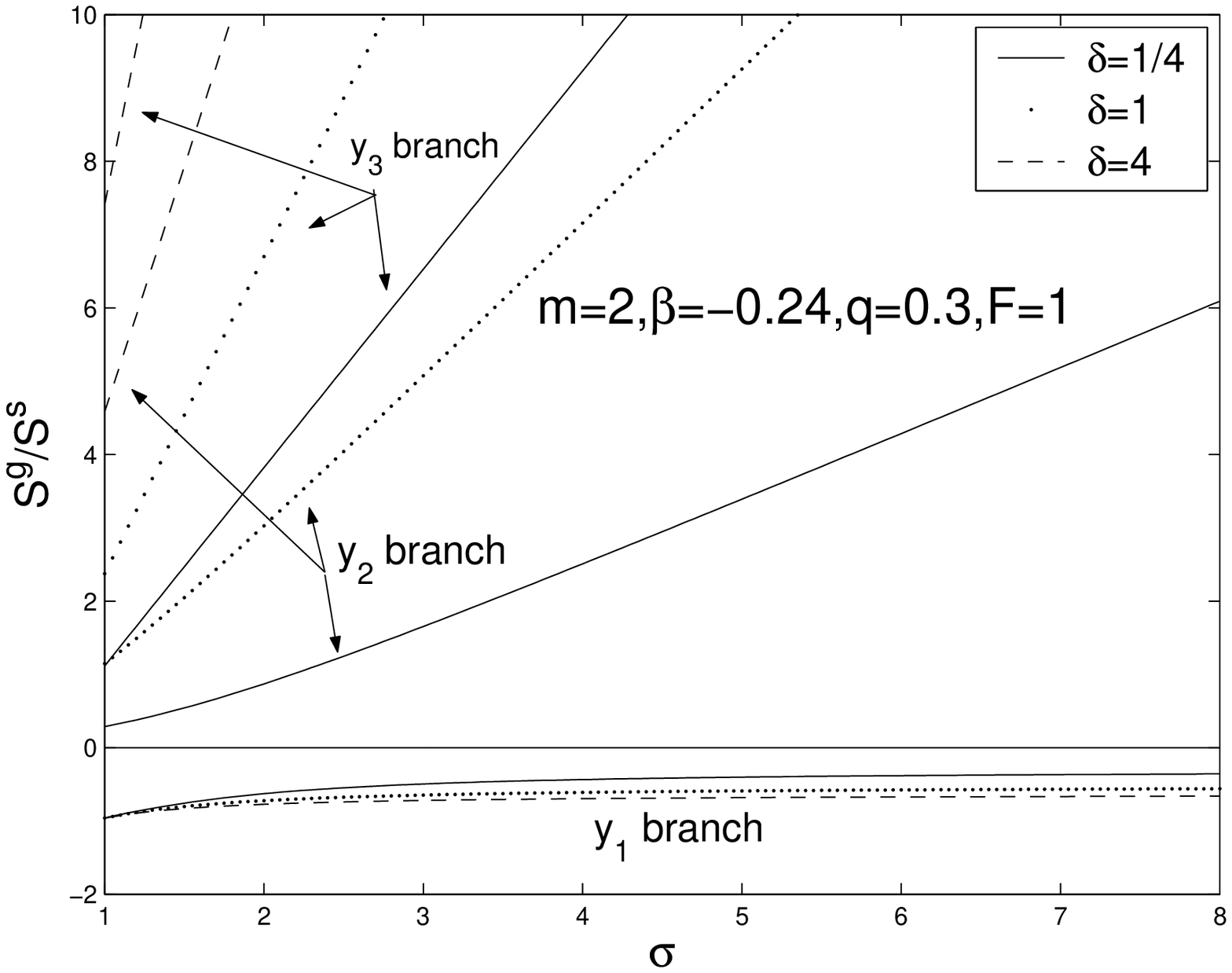}}
    \hspace{1in}
    \subfigure[$m=2,\ \beta=-0.24,\ q=1,\ F=1$]{
      \label{fig:6b}
      \includegraphics[width=75mm,height=55mm]{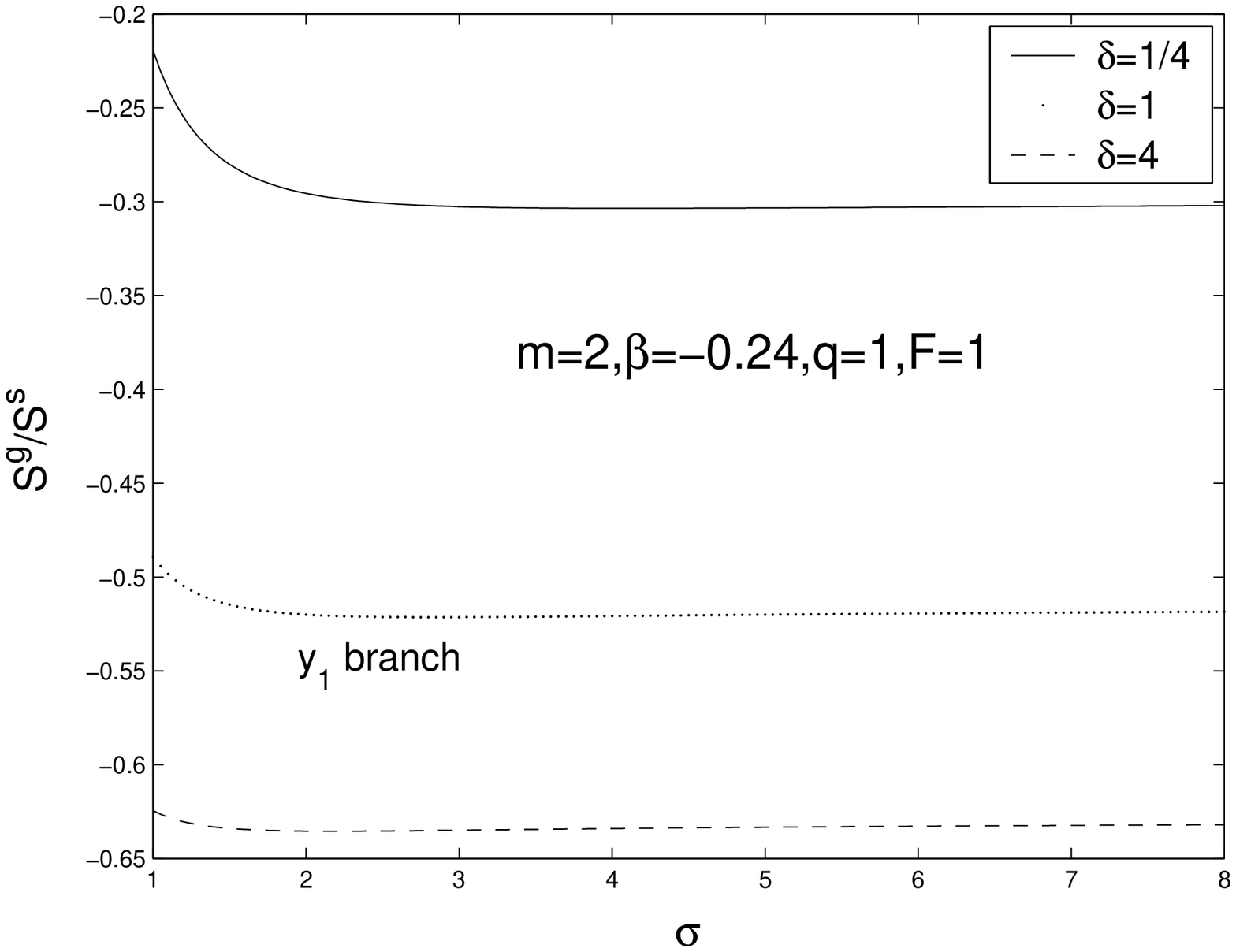}}
    \hspace{1in}
    \subfigure[$m=2,\ \beta=-0.24,\ q=3,\ F=1$]{
      \label{fig:6c}
      \includegraphics[width=75mm,height=55mm]{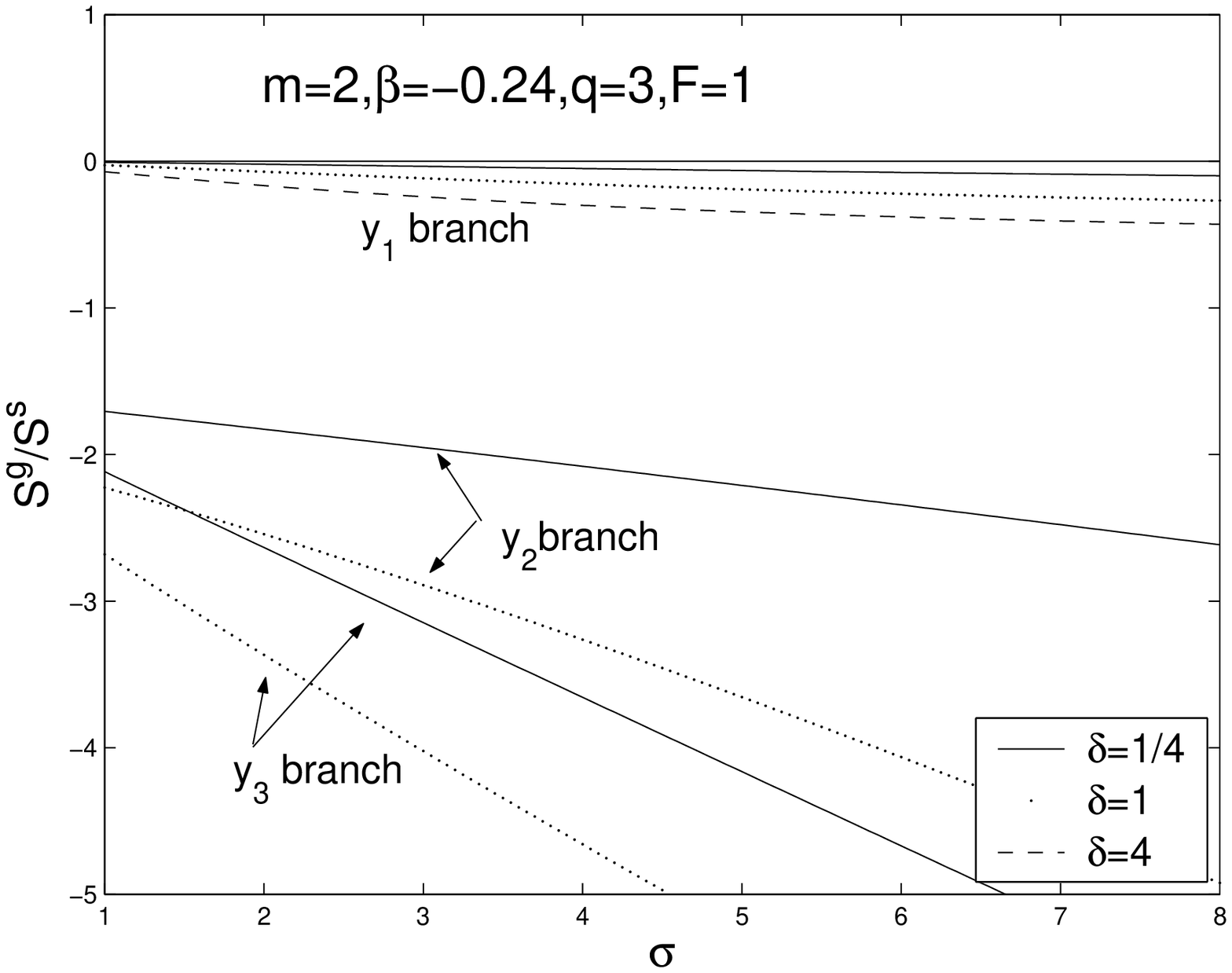}}
  \caption{Phase relation curves as represented by the ratio $S^g/S^s$
  for surface mass density perturbations versus $\sigma$ in the $m=2$
  cases (bar-like) for full discs with $F=1$. Taking $\beta=-0.24$,
  we show curves with $\delta=-1/4,\ 1,\ 4$ and $q=0.3,\ 1,\ 3$ in
  panels (a), (b), (c), respectively. These parameters are the same
  as those in Fig.\ \ref{fig:1} for the convenience of reference. In
  panel (a), the upper, middle, and lower curves with corresponding
  linetypes are the $y_3$, $y_2$, and $y_1$ branches, respectively.
  In panel (b), only the $y_1$ branch is valid and physical. In panel
  (c), the curves from up to down
  with distinct linetypes correspond to the $y_1$, $y_2$, and $y_3$
  branches, respectively. For $\delta=4$, there is only one real branch.
   }\label{fig:6}
\end{figure}
\begin{figure}
    \centering
    \subfigure[$m=2,\ \beta=1/4,\ q=0.3,\ F=1$]{
      \label{fig:7a}
      \includegraphics[width=75mm,height=55mm]{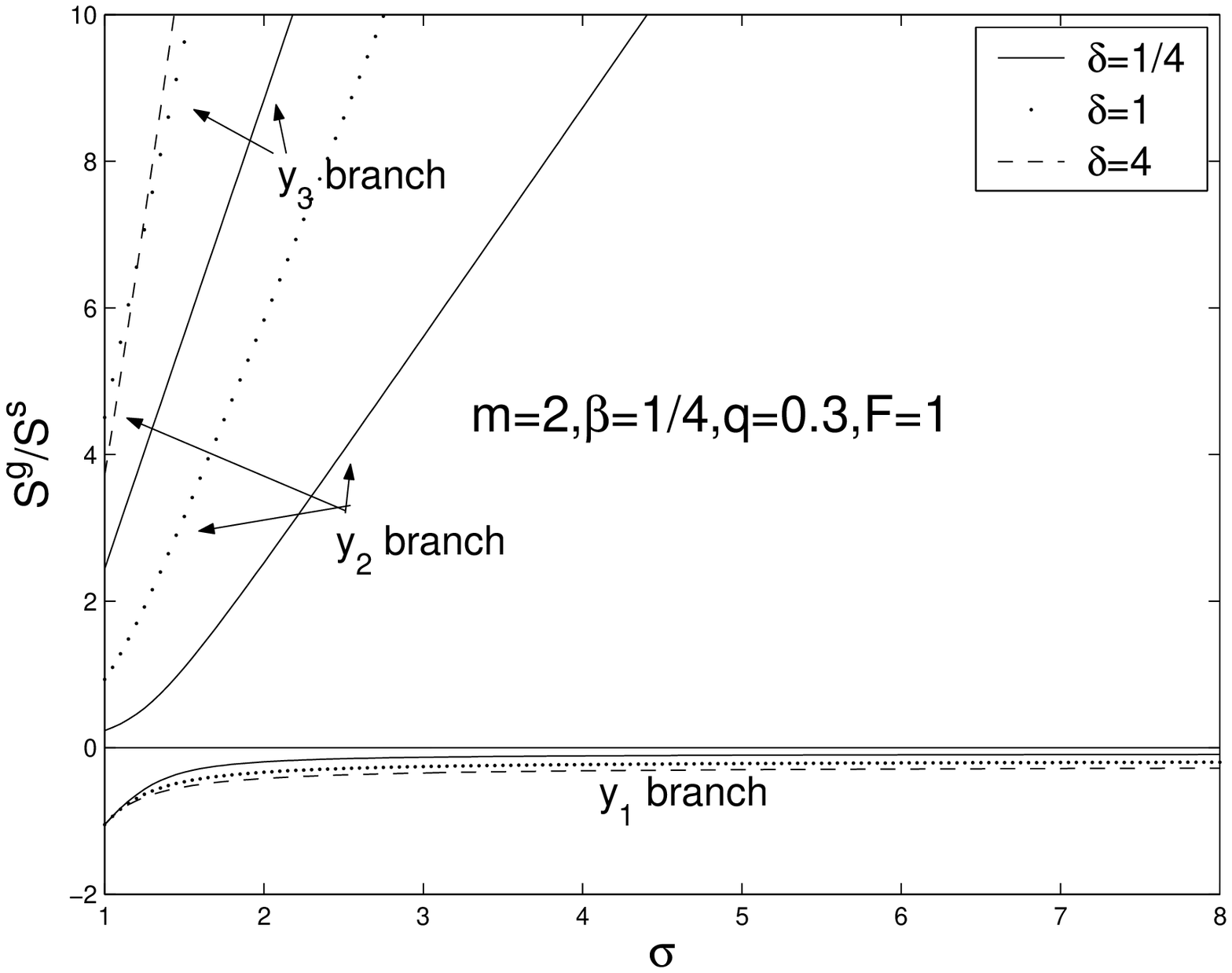}}
    \hspace{1in}
    \subfigure[$m=2,\ \beta=1/4,\ q=3,\ F=1$]{
      \label{fig:7c}
      \includegraphics[width=75mm,height=55mm]{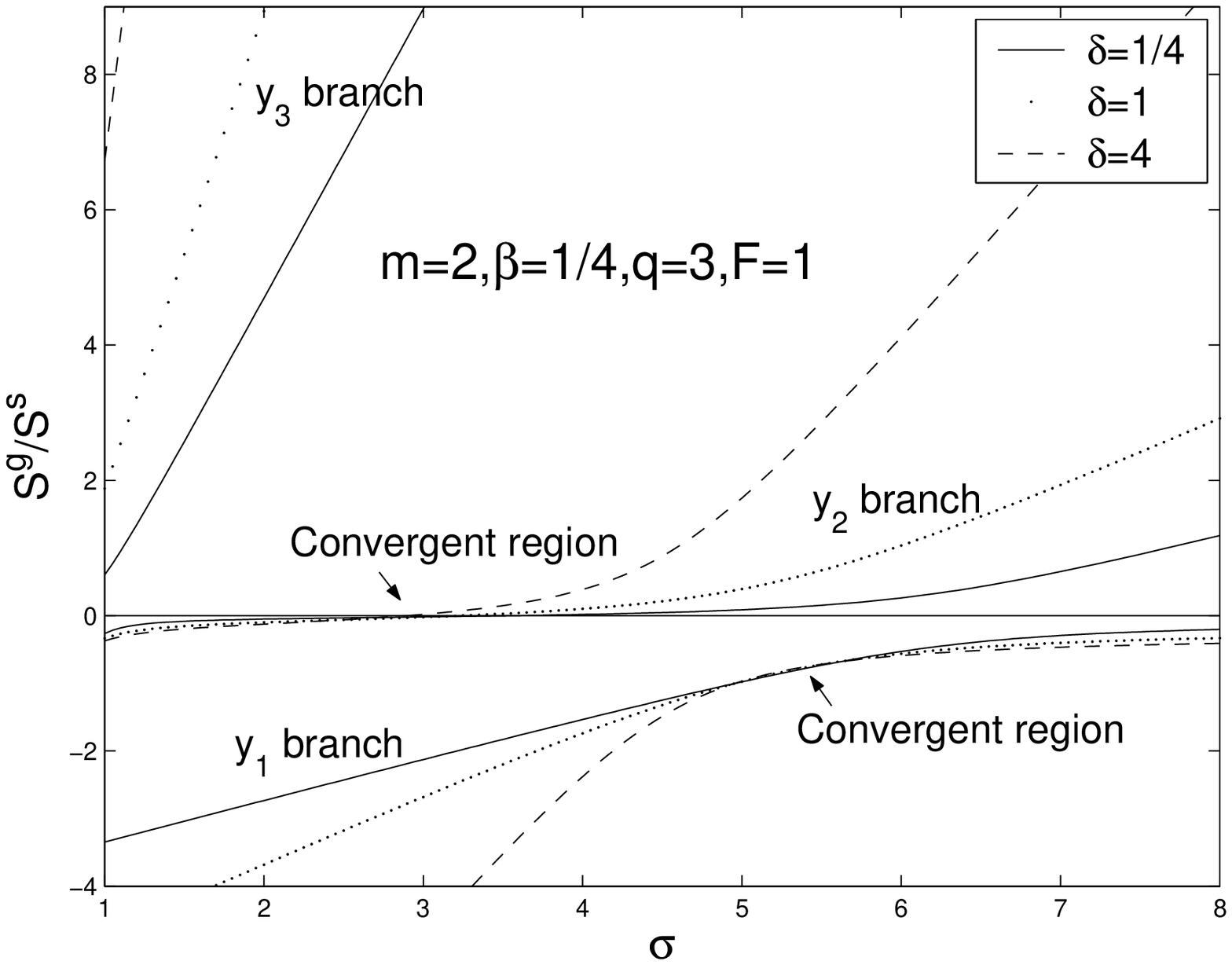}}
  \caption{Curves of $S^g/S^s$ ratio versus $\sigma$ for $m=2$ full
  discs (bar-like) with $F=1$, $\beta=1/4$, $\delta=1/4,\ 1,\ 4$ and
  $q=0.3,\ 3$ shown in panels (a) and (b), respectively. We adopt the
  same parameters as used in Fig.\ \ref{fig:3} for reference and
  comparison. In panel (a), the curves with different linetypes from
  up to down correspond to $y_3$, $y_2$, $y_1$ branches, respectively,
  and for convenience, we just label the corresponding solid curves.
  In panel (b), there appear `convergent regions' for a larger $q$.
   }\label{fig:7}
\end{figure}
\begin{figure}
  \includegraphics[width=75mm,height=55mm]{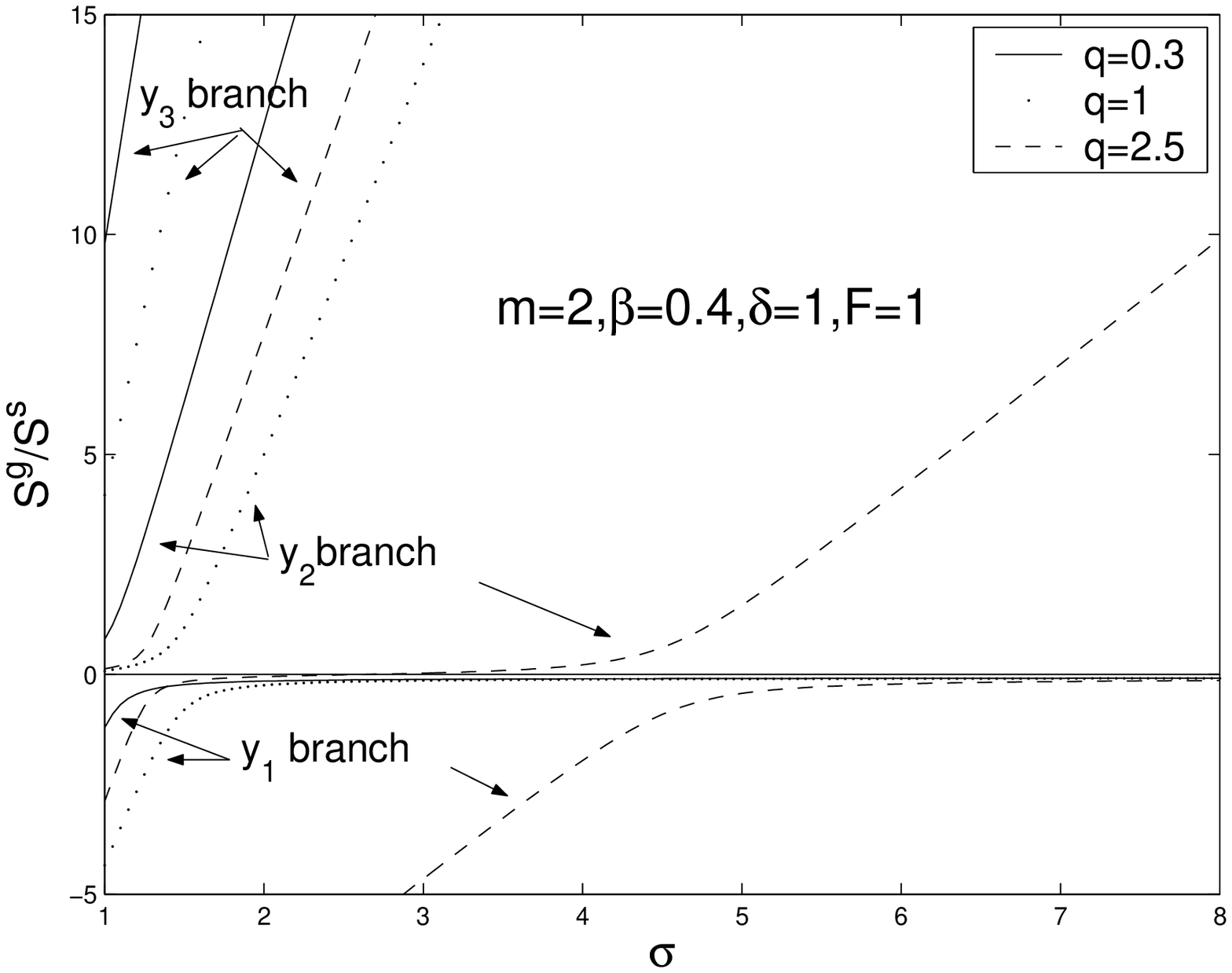}
  \caption{Curves of $S^g/S^s$ ratio versus $\sigma$ for $m=2$ full
  discs with $F=1$, $\beta=0.4$, $\delta=1$ and $q=0.3,\ 1,\ 2.5$,
  respectively. The curves from up to down correspond to $y_3$, $y_2$,
  and $y_1$ branches, respectively, and for convenience, we just label
  the solid curve (i.e., $q=0.3$).
  }\label{fig:8}
\end{figure}

The phase relationships for surface mass densities of aligned
perturbations are displayed in Fig.\ \ref{fig:6}$-$Fig.\ \ref{fig:8}.
In these three figures, we adopt the same parameters as used in
computing the dispersion relations shown in Fig.\ \ref{fig:1}$-$Fig.\
\ref{fig:4} for complete information.
For the convenience of statement, we still refer to the phase relation
curves as $y_1$, $y_2$, and $y_3$ branches corresponding to the $y_1$,
$y_2$ and $y_3$ of $D_s^2$ solutions in the dispersion relation.

The magnetic field influences the
perturbation mass density ratio.
We demonstrate a few examples in Fig.\ \ref{fig:7} and
Fig.\ \ref{fig:8}. In the case of a small $q$ (e.g., $q=0.3$), the $y_1$
branch is negative (i.e., out of phase) while $y_2$ and $y_3$ branches
are positive (i.e., in phase). However, as the magnetic field increases,
the $y_2$ branch becomes negative for small $\sigma$. This means that
for aligned stationary MHD density waves, the density phase relationship
between the two scale-free discs are out of phase for super fast MHD
density waves and in phase for fast MHD density waves. For the middle
one, the mass density ratios are in phase in the regime of a weaker
magnetic field but are out of phase in the regime of a stronger
magnetic field.

We explore phase relationships of surface mass density perturbations
with several different $\delta$ values in Fig.\ \ref{fig:6} and Fig.
\ \ref{fig:7}. One might expect that large $\delta$ leads to higher
$S^g/S^s$ ratio; however, this is not always the case, especially when
both $\beta$ and $q$ are large (e.g., Fig.\ \ref{fig:7} and Fig.\
\ref{fig:8}). We notice `convergent regions' where the phase relation
curves with different $\delta$ ratios seem to converge. By carefully
examining these phase curves, they do not actually converge but just
become very close to each other. This phenomenon results from a complex
interaction between the magnetic field and the two-fluid disc system and
is a new feature in our model as compared with previous works [e.g.,
\citet{b6,b7,b10}; Lou \& Zou 2006].

\begin{figure}
    \centering
    \subfigure[$m=2,\ \beta=-0.24,\ q=0.3,\ F=1$]{
      \label{fig:9a}
      \includegraphics[width=75mm,height=55mm]{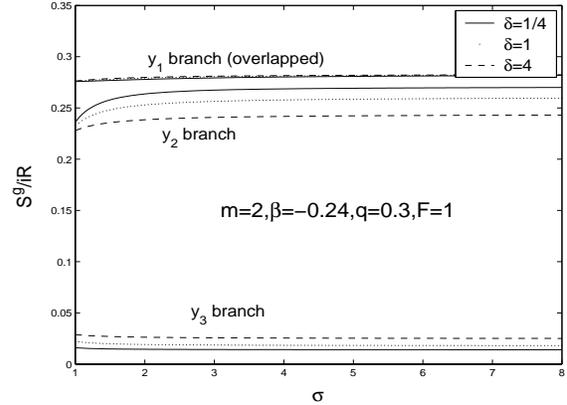}}
    \hspace{1in}
    \subfigure[$m=2,\ \beta=-0.24,\ q=1,\ F=1$]{
      \label{fig:9b}
      \includegraphics[width=75mm,height=55mm]{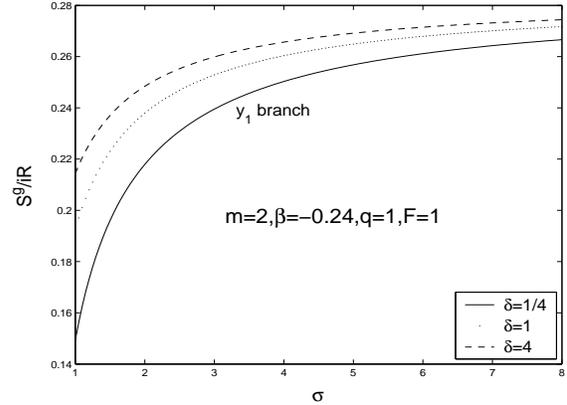}}
    \hspace{1in}
    \subfigure[$m=2,\ \beta=-0.24,\ q=3,\ F=1$]{
      \label{fig:9c}
      \includegraphics[width=75mm,height=55mm]{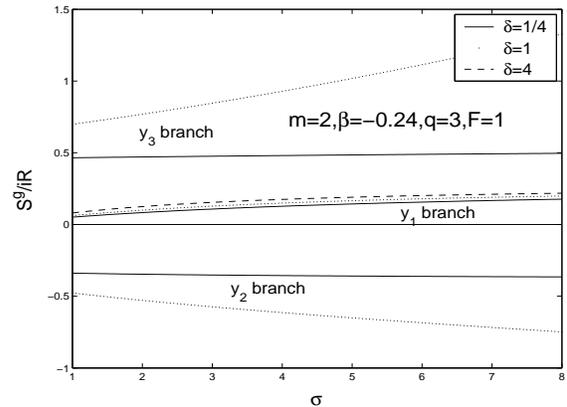}}
  \caption{Perturbation phase relation curves for gas surface
  mass density and radial component magnetic field in the $m=2$
  (bar-like) cases for full discs. Taking $\beta=-0.24$, we
  show curves with $\delta=1/4,\ 1,\ 4$ and $q=0.3,\ 1,\ 3$
  in panels (a), (b), and (c), respectively.
  From relation (73), we see that $S^g/(\mathrm{i} R)$ and
  $S^g/Z$ are proportional to each other with a sign change
  depending on whether $\beta<1/4$ or not.}
  \label{fig:9}
\end{figure}
\begin{figure}
  \includegraphics[width=75mm,height=55mm]{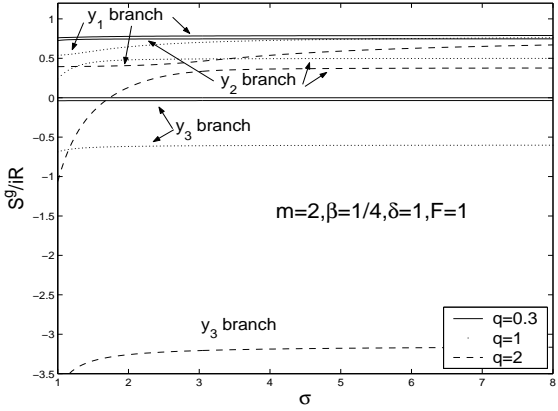}
  \caption{Curves of $S^g/(\mathrm{i}R)$ ratio versus $\sigma$ for
  $m=2$ full discs with $F=1$, $\beta=1/4$, $\delta=1$ and $q=0.3,\
  1,\ 2$. The upper, middle and lower branches with different linetypes
  correspond to $y_1$, $y_2$ and $y_3$ branches, respectively.
  For the solid curves with $q=0.3$, the upper two branches
  $y_1$ and $y_2$ are quite close to each other and the negative
  $y_3$ branch approaches zero.
  }\label{fig:10}
\end{figure}

We now examine phase relationships between gas mass density
perturbation and the radial component of magnetic field perturbation
(Fig.\ \ref{fig:9} and Fig.\ \ref{fig:10}) with $m=2$ (bar-like).
From equation (\ref{eq:73}), we know that $S^g/Z$ is proportional
to $S^g/(\mathrm{i}R)$ with a factor $2m/(4\beta-1)$. When
$\beta\neq1/4$ the phase curves regarding $S^g/Z$ and
$S^g/(\mathrm{i}R)$ are the same in shape. For $\beta >1/4$, $S^g/Z$
and $S^g/(\mathrm{i}R)$ are of the same sign, while for $\beta <1/4$,
$S^g/Z$ and $S^g/(\mathrm{i}R)$ are of the opposite sign. For
$\beta=1/4$, we simply have $Z=0$. For these reasons we just discuss
$S^g/(\mathrm{i}R)$ for convenience; the information of $S^g/Z$ can
be readily derived accordingly. In the case of $\beta=-0.24$, we
already know that the $y_1$ branch is always physical. Here in Fig.\
\ref{fig:9}, we see that $\mathrm{i}R$ and $S^g$ remain always in phase,
and therefore from equation (\ref{eq:72}), we know that $Z$ and $S^g$
are out of phase. The enhancement of the magnetic field lowers the
ratio $S^g/(\mathrm{i}R)$, as a stronger magnetic field tends to
induce a larger magnetic perturbation. In the case of larger $\beta$,
the dependence of this ratio to the magnetic field is somehow more
sensitive.
We take the case of
$\beta=1/4$ as an example of illustration. In this case, the
azimuthal magnetic perturbation vanishes, with a nonzero
$r-$component magnetic field perturbation. We see in Fig.\
\ref{fig:10} that when $q$ is small, the $y_1$ and $y_2$ branches
are positive and are very close to each other. The $y_3$ branch is
negative and borders on zero (recall that $y_2$ and $y_3$ branches
can be physical when $\sigma<\sigma_{c2}$ or $\sigma<\sigma_{c3}$).
As $q$ increases, all three branches lower and the $y_2$ branch
intersects zero indicating a phase relation reversal across the
critical point. When $q$ is increased further, the entire $y_2$
branch becomes negative.

\subsubsection{Partial Discs Embedded in a Dark Matter Halo}

\begin{figure}
    \centering
    \subfigure[$m=1,\ \beta=1/4,\ \delta=1,\ \sigma=2$]{
      \label{fig:11a}
      \includegraphics[width=75mm,height=55mm]{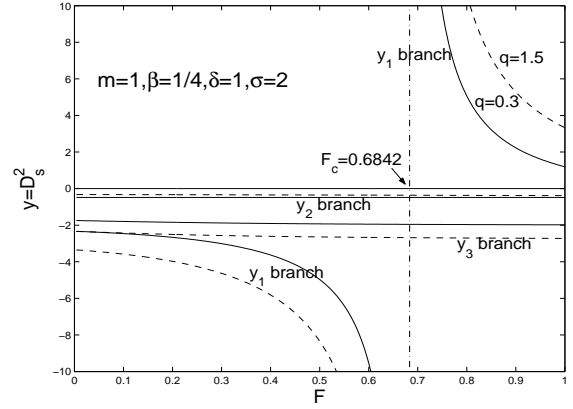}}
    \hspace{1in}
    \subfigure[$m=1,\ \beta=-0.24,\ \delta=1,\ \sigma=1.2,\ q=0.4$]{
      \label{fig:11b}
      \includegraphics[width=75mm,height=55mm]{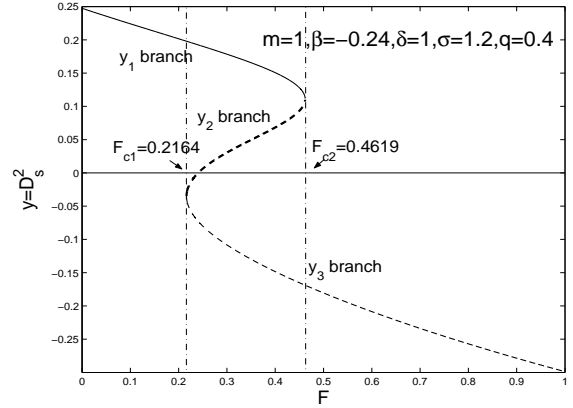}}
    \hspace{1in}
    \subfigure[$m=2,\ \beta=1/4,\ \delta=1,\ \sigma=2$]{
      \label{fig:11c}
      \includegraphics[width=75mm,height=55mm]{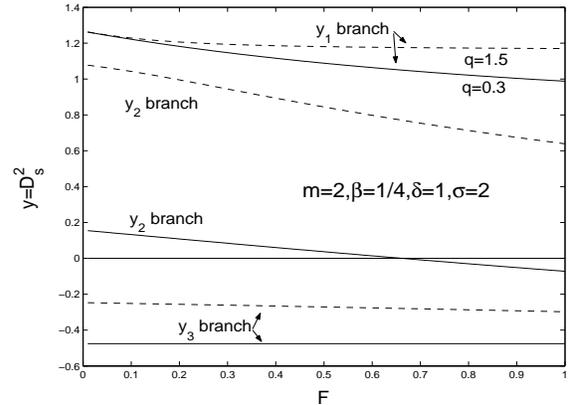}}
  \caption{Dispersion relationships of $D_s^2$ versus $F$
  variation for a composite system of partial scale-free discs.
  There are a few interesting
  features in the case $m=1$. There is only one real $D_s^2$ root
  in panel (b), while the other two roots are complex conjugates.
  It happens that the three $D_s^2$ branches are connected smoothly.
  In panel (c) for $m=2$, $y_1$ branch represents the upper two
  curves ($q=1.5$ for dashed curve and $q=0.3$ for solid curve).
  The $y_2$ branch is labeled separately and the $y_3$ branch
  represents the lower two curves. }\label{fig:11}
\end{figure}

Coplanar aligned MHD perturbation structures in full scale-free discs
are extensively discussed above. For partial discs with factor $F<1$,
we expect that the introduction of an axisymmetric dark matter halo
may lead to some novel features. In fact, through numerical
computations we see that new features manifest mainly through the
$m=1$ case as shown in Fig.\ \ref{fig:11}.

We first note that one of the $D_s^2$ or $D_g^2$ solutions diverges
for the cubic coefficient $C^A_3=0$. As our $C^A_3$ is proportional
to $C_2$ of equation (48) in \citet{b11}, we expect to have a
similar result in this aspect. This phenomenon appears at $m=1$. Let
$\beta=1/4$ and we see in Fig.\ \ref{fig:11a} the divergent point
lies at $F_c\simeq0.6842$ which is exactly the same as that in the
analysis of \citet{b11}, and we have an additional $y_3$ branch
(although negative and thus unphysical). Another difference is that
here the $y_2$ branch is also negative and therefore the only
physical solution lies in the positive portion of $y_1$ solutions.
The direct consequence is that for $0<F\leq F_c$, there can be no
physical solutions.

We also evaluate the $m=1$ cases with different $\beta$ values. For a
sufficiently small $\beta$ (e.g., $\beta=-0.24$), there appears a kind
of discontinuity seen in Fig.\ \ref{fig:11b}. Actually, the range of
$F$ is divided into three regimes. For $0<F<F_{c1}$, only $y_1$ branch
exists and the other two $D_s^2$ roots are complex conjugates. For
$F_{c1}\leq F\leq F_{c2}$, all three branches are real. For $F_{c2}
<F\leq1$, only the $y_3$ branch exists. Since $y_3$ branch is negative,
there is no physical solution for $F_{c2}<F\leq1$. Immediately at
$F=F_{c2}$, there emerges suddenly a new solution (double roots), and
this solution then bifurcates into two branches as $F$ becomes smaller.
The $y_2$ branch turns negative before $F$ encounters $F_{c1}$ and
from then on there is only one physical solution for smaller $F$.

For $m=2$ bar-like configurations, we take $\beta=1/4$, $\delta=1$,
$\sigma=2$
with $q=0.3$ and $1.5$ respectively to obtain solutions in panel
(c) of Fig.\ \ref{fig:11}. For this set of parameters, the $y_1$
branch remains
always positive while the $y_3$ branch remains always negative. Whether
the $y_2$ branch is positive or not depends on the choice of parameters
$F$ and $q$. It is clear that an increase of magnetic field raises all
three branches. This raises the $y_1$ branch only slightly but greatly
lifts the lower two branches. Especially for $y_2$ branch, only smaller
$F$ values can ensure a positive $y_2$ when $q$ is sufficiently small,
while a sufficiently large $q$ parameter actually makes $y_2$ positive
for all $0<F\leq1$. For other values of $m$, the global aligned MHD
perturbation patterns and trends of variation are qualitatively similar.

\section[]{Unaligned  MHD Configurations \\
\qquad of Global Logarithmic Spirals}

In this section, we study the unaligned MHD perturbations
by introducing a special yet useful potential-density pair
referred to as the logarithmic spirals.
We first derive the MHD dispersion relation for the composite
system of two coupled scale-free discs. We then focus on
computations of $m\geq1$ cases. Finally, we study the $m=0$
case which leads to the discussion of the axisymmetric marginal
stability.

\subsection{Stationary MHD Dispersion Relation\\
\ \qquad for Global Logarithmic Spirals}

To construct global logarithmic spiral MHD
perturbations, we take the potential-density
pair in the form of
\begin{equation}
\begin{split}
&S^i(r)=\epsilon^i r^{-3/2}\mathrm{exp}(\mathrm{i}
\xi\mathrm{ln}r)\ ,\\
&V^i(r)=-2\pi GrS^i\mathcal{N}_m(\xi)\ ,\label{eq:76}
\end{split}
\end{equation}
where $\epsilon^{i}$ is a small constant amplitude coefficient,
superscript $i$ denotes either $g$ or $s$ for either gas
or stellar discs, $\xi$ represents the radial wavenumber and
the Kalnajs function $\mathcal{N}_m(\xi)$ is defined by
\begin{equation}
\mathcal{N}_m(\xi)=\frac{\Gamma(m/2+\mathrm{i}\xi/2+1/4)
\Gamma(m/2-\mathrm{i}\xi/2+1/4)}{2\Gamma(m/2+\mathrm{i}\xi/2+3/4)
\Gamma(m/2-\mathrm{i}\xi/2+3/4)}\ \label{eq:77}
\end{equation}
\citep{b3}. In potential-density pair (\ref{eq:76}), the perturbed
surface mass density is complex, with an amplitude radial scaling
of $r^{-3/2}$ and a phase factor $r^{\mathrm{i}\xi}$ for radial
oscillations. The relevant radial variations are no longer the
same as those of the background equilibrium unless $\beta=1/4$.
In fact, there can be more general potential-density pairs as
noted in \citet{b10}. For the analysis of the marginal stability
problem, expression (\ref{eq:76}) is used such that the dispersion
relation (\ref{eq:98}) derived later is real on both sides, and
this choice of scale factor satisfies the scale-free condition in
the perturbation equations. The phase factor is the key to establish
global logarithmic spiral patterns. We call it logarithmic spirals
because the dependence of the phase factor on $r$ follows the
logarithmic function. Kalnajs function (\ref{eq:77}) is an even
function in $\xi$ and thus a consideration of $\xi\geq0$ (i.e.,
leading MHD density waves) suffices.

We now derive the MHD dispersion relation for global
logarithmic spirals with $m\geq1$ cases using equations
$(\ref{eq:41})-(\ref{eq:43})$, (\ref{eq:47}) together
with (\ref{eq:76}). With radial scalings of
$S\propto r^{-3/2+\mathrm{i}\xi}$, $V\propto r^{-1/2+\mathrm{i}\xi}$,
$U\propto r^{-1/2+\beta+\mathrm{i}\xi}$ and
$J\propto r^{1/2+\beta+\mathrm{i}\xi}$, all
quantities in equations $(\ref{eq:41})-(\ref{eq:43})$ and (\ref{eq:47})
are scale-free; these equations become
\begin{equation}
m\Omega_g
S^g-(1/2-\mathrm{i}\xi+\beta)\frac{\Sigma_0^g}{r}\mathrm{i}U^g
+\frac{m\Sigma_0^g}{r^2}J^g=0\ ,\label{eq:78}
\end{equation}

\begin{equation}
\begin{split}
&\bigg\{m\Omega_g-\frac{C_A^2[2m^2-\mathrm{i}\xi(1-4\beta)+2\xi^2]}
{2m\Omega_g r^2}\bigg\}
\mathrm{i}U^g=\\
&\bigg[(1-2\mathrm{i}\xi)\pi G\mathcal{N}_m-\frac{(1-2\mathrm{i}\xi)
a_g^2+(1-4\beta)C_A^2}{2r\Sigma_0^g}\bigg]S^g\\
&\quad\qquad\qquad\qquad+(1-2\mathrm{i}\xi)\pi
G\mathcal{N}_mS^s-\frac{2\Omega_g}{r}J^g\ ,\label{eq:79}
\end{split}
\end{equation}
\begin{equation}
\begin{split}
m\Omega_g &J^g+\bigg[(1-\beta)r\Omega_g-\frac{C_A^2(1-4\beta)}
{2r\Omega_g}\bigg]\mathrm{i}U^g=\\ &\quad\qquad mr\bigg(2\pi
G\mathcal{N}_m-\frac{a_g^2}{r\Sigma_0^g}\bigg)S^g +2\pi
Gmr\mathcal{N}_mS^s\ ,
\end{split}\label{eq:80}
\end{equation}
\begin{equation}
Z=-\mathrm{i}\xi\frac{B_0\mathrm{i}U^g}{m\Omega_g r}\ ,
\qquad\qquad\qquad
R=\frac{B_0U^g}{r\Omega_g}\label{eq:81}
\end{equation}
for the magnetized gas disc and
\begin{equation}
\begin{split}
&\bigg[\frac{(m^2-2+2\beta)\Omega_s^2r^2}{m^2+\xi^2+2\beta+1/4}
-a_s^2+2\pi Gr\Sigma_0^s\mathcal{N}_m\bigg]S^s
\\ &\qquad\qquad\qquad\qquad\qquad
+2\pi Gr\Sigma_0^s\mathcal{N}_mS^g=0\label{eq:82}
\end{split}
\end{equation}
for the stellar disc. Equations $(\ref{eq:78})-(\ref{eq:81})$ are
the same as equation (65) in \citet{b11} for a single magnetized
scale-free disc by setting $S^s=0$ and equation (\ref{eq:82}) is
exactly the same as the first equation of (68) in \citet{b10}. A
combination of equations $(\ref{eq:78})-(\ref{eq:80})$ and
(\ref{eq:82}) gives the full solution for the stationary
dispersion relation of the gravitational coupled MHD discs
for logarithmic spirals. Similar to the aligned cases,
these are homogeneous linear equations in terms of
$(S^g,\ S^s,\ U^g,\ J^g)$. For non-trivial solutions,
the determinant of the coefficients of equations
$(\ref{eq:78})-(\ref{eq:80})$ and (\ref{eq:82}) should
vanish. The result is the stationary MHD dispersion relation.
As necessary requirements, we expect that in special cases,
this dispersion relation will reduce to those discussed by
\citet{b11} or \citet{b10}. Meanwhile, we solve the above
equations directly to reveal physical contents.

A combination of equations (\ref{eq:78}) and (\ref{eq:80})
leads to two expressions of $\mathrm{i}U^g$ and $J^g$ in
terms of $S^g$ and $S^s$, namely
\begin{equation}
\begin{split}
&\mathrm{i}U^g={m\Omega_gr}
[(\Theta_S+\Xi_S)S^g+\Theta_S S^s]/\Sigma_0^g\ ,\\
&J^g={\Omega_gr^2}
\{[-1+(1/2-\mathrm{i}\xi+\beta)(\Theta_S+\Xi_S)]S^g\\
&\qquad\qquad\qquad\qquad
+(1/2-\mathrm{i}\xi+\beta)\Theta_S S^s\}/\Sigma_0^g\ ,
\end{split}\label{eq:83}
\end{equation}
where, for notational simplicity, we denote
\begin{equation}
\begin{split}
&\Theta_S\equiv {2\pi Gr\mathcal{N}_m\Sigma_0^g}/{\Delta_S}\ ,\\
&\Xi_S\equiv {(-a_g^2+\Omega_g^2r^2)}/{\Delta_S}\ ,\\
&\Delta_S\equiv(3/2-\mathrm{i}\xi)\Omega_g^2r^2+(2\beta-1/2)C_A^2\ .
\end{split}\label{eq:84}
\end{equation}
We use the subscript $_S$ to indicate the logarithmic `spiral cases'.
Note that $\Theta_S$ and $\Xi_S$ are two dimensionless constant
parameters. A substitution of equation (\ref{eq:83}) into equation
(\ref{eq:79}) and a further combination with equation (\ref{eq:82})
lead to
\begin{equation}
\begin{split}
[\mathcal{K}_S(\Theta_S+\Xi_S)+\mathcal{L}_S-(1-2\mathrm{i}\xi)\pi
G\Sigma_0^gr\mathcal{N}_m]&S^g
\\ \quad
+[\mathcal{K}_S\Theta_S-(1-2\mathrm{i}\xi)\pi G
\Sigma_0^gr\mathcal{P}_m]&S^s=0\ ,\\
2\pi G\Sigma_0^sr\mathcal{N}_mS^g+(\mathcal{M}_S +2\pi
G\Sigma_0^sr\mathcal{N}_m)&S^s=0\ ,
\end{split}\label{eq:85}
\end{equation}
where, for notational simplicity, we further define
\begin{equation}
\begin{split}
&\mathcal{K}_S\equiv(m^2+2\beta+1-2\mathrm{i}\xi)\Omega_g^2r^2\\
&\qquad\ \qquad -[m^2-\mathrm{i}\xi(1/2-2\beta)+\xi^2]C_A^2\ ,\\
&\mathcal{L}_S\equiv(1/2-2\beta)C_A^2-2\Omega_g^2r^2+(1/2-\mathrm{i}\xi)a_g^2\ ,\\
&\mathcal{M}_S\equiv \frac{(m^2+2\beta-2)\Omega_s^2r^2}
{(m^2+\xi^2+2\beta+1/4)}-a_s^2\ .
\end{split}\label{eq:86}
\end{equation}
We can derive the stationary MHD dispersion
relation for global logarithmic spirals by calculating the
coefficient determinant of equations (\ref{eq:82}) and (\ref{eq:85}).
After manipulations and rearrangements, we obtain
\begin{equation}
\begin{split}
&({\Delta_S}/\mathcal{K}_S)[\mathcal{K}_S(\Theta_S+\Xi_S)
+\mathcal{L}_S-(1-2\mathrm{i}\xi)\pi
G\Sigma_0^gr\mathcal{N}_m]\\
&\quad\qquad \times(\mathcal{M}_S+2\pi G\Sigma_0^sr\mathcal{N}_m)\\
&\ \ \qquad\quad =4\pi^2G^2\Sigma_0^g\Sigma_0^sr^2\mathcal{N}_m^2
[1-(1/2-\mathrm{i}\xi){\Delta_S}/{\mathcal{K}_S}]\ .
\end{split}\label{eq:87}
\end{equation}
In equation (\ref{eq:87}), the left-hand side consists of two factors.
One can show that the factor in the left brackets is exactly the
dispersion relation in a single coplanar magnetized disc discussed
by \citet{b11} and the second factor in parentheses denotes the
dispersion relation of a single scale-free stellar disc. The
right-hand side denotes the mutual gravitational coupling between
the two scale-free discs. By setting $\beta=0$ for the isothermal
case, equation (\ref{eq:87}) reduces to equation (106) in \citet{b7}
where isothermal fluid-magnetofluid discs are investigated.

In all calculations of this subsection, the equations for
logarithmic spirals are strikingly similar to those of the
aligned perturbations. If we replace $\mathcal{P}_m$ by
$\mathcal{N}_m$ and partly replace $2\beta$ by $1/2-\mathrm{i}\xi$,
this similarity arises in the expression of the perturbation
equations. We assert that the similarity in structure must lead
to the similarity in concrete expressions of the dispersion
relation between the two cases. We show this fact in the next
subsection.

\subsection{Stationary Logarithmic Spiral Configurations}

\subsubsection{$D_s^2$ and $D_g^2$ Solutions of the Dispersion Relation}

We now evaluate the spiral dispersion relation for $m\geq1$ cases
numerically to show the dependence of $D_g^2$ or $D_s^2$ on other
dimensionless parameters (i.e., $m,\ \beta,\ \eta,\ \delta,\ q,\ F$).
Similar to the aligned cases, we choose to calculate the coefficient
determinant of equations $(\ref{eq:78})-(\ref{eq:80})$ and (\ref{eq:82}).
We leave details of onerous calculations to Appendix C where the
simplified determinant is shown explicitly. After careful calculations
and simplifications, we obtain the MHD dispersion relation as cubic
polynomial equation in terms of $D_g^2$. It is satisfying to see that
this dispersion relation can be expressed in the exactly same form
as that of the aligned cases merely by a substitution of notations.
We write
\begin{equation}
\begin{split}
\mathcal{A}^S_m(\beta,\ \xi)&\equiv m^2+\xi^2+2\beta+1/4\ ,\\
\mathcal{H}^S_m(\beta,\ \xi)&\equiv {\mathcal{A}^S_m\mathcal{N}_m}
/{\mathcal{C}}+\mathcal{B}_m\ .\\
\end{split}\label{eq:88}
\end{equation}
We still use the notations $\mathcal{B}_m(\beta)$ and
$\mathcal{C}(\beta)$ as defined in equation (\ref{eq:65}).
Now the stationary MHD dispersion relation can be
written in terms of $y\equiv D_g^2$, namely
\begin{equation}
C^S_3y^3+C^S_2y^2+C^S_1y+C^S_0=0\ .\label{eq:89}
\end{equation}
The four coefficients $C^S_3$, $C^S_2$, $C^S_1$,
$C^S_0$ are functions of dimensionless parameters
$m,\ \beta,\ \delta,\ \eta,$ and $q^2$ defined by
\begin{equation}
\begin{split}
&C^S_3\equiv\mathcal{B}_m\mathcal{H}^S_m\eta\ ,\\
&C^S_2\equiv \bigg[(\mathcal{B}_m-\mathcal{A}^S_m)\mathcal{H}^S_m
+\frac{(\mathcal{A}^S_m+\mathcal{B}_m)
(\mathcal{H}^S_m-\mathcal{B}_m)\delta}{(1+\delta)}\bigg]\eta\\
&\qquad\qquad -\frac{(\mathcal{A}^S_m+\mathcal{B}_m)
(\mathcal{H}^S_m\delta+\mathcal{B}_m)}{(1+\delta)}+C^S_{2Q}q^2\ ,\\
&C^S_1\equiv\bigg[-\mathcal{A}^S_m\mathcal{H}^S_m
+\frac{(\mathcal{A}^S_m+\mathcal{B}_m)
(\mathcal{H}^S_m-\mathcal{B}_m)\delta}{(1+\delta)}\bigg]\eta\\
&+(\mathcal{A}^S_m+\mathcal{B}_m)^2
-\frac{(\mathcal{A}^S_m+\mathcal{B}_m)
(\mathcal{H}^S_m\delta+\mathcal{B}_m)}{(1+\delta)}+C^S_{1Q}q^2\ ,\\
&C^S_0\equiv C^S_{0Q3}q^6+C^S_{0Q2}q^4+C^S_{0Q1}q^2\ ,
\end{split}\label{eq:90}
\end{equation}
where the following five coefficients are further defined by
\begin{equation*}
\begin{split}
&C^S_{2Q}\equiv\frac{m^2(4\beta-1)-2\mathcal{A}^S_m
+8\beta^2-20\beta+6}{2(1+\delta)}(\mathcal{H}^S_m
+\mathcal{B}_m\delta)\eta
\\ &\qquad\qquad
-\frac{4(1-2\beta)\mathcal{A}^S_m-(1+2\beta)}
{(4\beta+2)(1+\delta)}\frac{\mathcal{B}_m
\mathcal{N}_m}{\mathcal{C}}\delta\eta\ ,\\
\end{split}
\end{equation*}
\begin{equation*}
\begin{split}
&C^S_{1Q}\equiv -\frac{\mathcal{A}^S_m+5\beta-2}{(1+\delta)}
(\mathcal{H}^S_m+\mathcal{B}_m\delta)\eta
\\ &\qquad\quad
-\frac{\mathcal{A}^S_m(4\beta-3)+2\beta+1}
{(4\beta+2)(1+\delta)}\bigg(\mathcal{H}^S_m +\mathcal{B}_m\delta
-\frac{2\mathcal{B}_m\mathcal{N}_m\delta}{\mathcal{C}}\bigg)\eta
\\ &\qquad\quad
-\frac{\mathcal{A}^S_m(4\beta-3)+2\beta+1}
{(4\beta+2)(1+\delta)}\frac{\mathcal{N}_m}
{\mathcal{C}}(\mathcal{A}^S_m+\mathcal{B}_m)\delta
\\ &\qquad\quad
+(\mathcal{A}^S_m+\mathcal{B}_m)(\mathcal{A}^S_m+5\beta-2)\ ,\\
\end{split}
\end{equation*}
\begin{equation*}
\begin{split}
&C^S_{0Q3}\equiv -\frac{(4\beta-1)^3
(\mathcal{H}^S_m+\mathcal{B}_m\delta)\eta}
{2(4\beta+2)^2(1+\delta)} \\ &\qquad\ \ \
-\frac{(4\beta-1)^2(2\mathcal{A}^S_m-2\beta-1)}{(4\beta+2)^3(1+\delta)}
\frac{\mathcal{B}_m\mathcal{N}_m}{\mathcal{C}}\delta\eta\ ,\\
\end{split}
\end{equation*}
\begin{equation*}
\begin{split}
&C^S_{0Q2}\equiv\frac{(4\beta-1)}{(4\beta+2)}
\bigg[\frac{(4\beta-1)}{2}
+\frac{2\mathcal{A}^S_m-2\beta-1}{(4\beta+2)(1+\delta)}
\frac{\mathcal{N}_m}{\mathcal{C}}\delta\bigg]
\\ &\qquad\quad
\times (\mathcal{A}^S_m+\mathcal{B}_m)
-\frac{(4\beta-1)^2(\mathcal{H}^S_m+\mathcal{B}_m\delta)\eta}
{2(4\beta+2)(1+\delta)}  \\ &\quad
+\frac{(4\beta-1)(2\mathcal{A}^S_m-2\beta-1)}
{(4\beta+2)^2(1+\delta)}\bigg(\mathcal{H}^S_m
+\mathcal{B}_m\delta-\frac{2\mathcal{B}_m\mathcal{N}_m\delta}
{\mathcal{C}}\bigg)\eta\ ,\\
\end{split}
\end{equation*}
\begin{equation}
\begin{split}
&C^S_{0Q1}\equiv\frac{(2\mathcal{A}^S_m-2\beta-1)}
{(4\beta+2)(1+\delta)}\bigg(\mathcal{H}^S_m+\mathcal{B}_m\delta
-\frac{\mathcal{B}_m\mathcal{N}_m\delta}{\mathcal{C}}\bigg)\eta
\\ &\qquad
-\frac{(2\mathcal{A}^S_m-2\beta-1)}{(4\beta+2)(1+\delta)}
\bigg(1+\delta-\frac{\mathcal{N}_m}{\mathcal{C}}\delta\bigg)
(\mathcal{A}^S_m+\mathcal{B}_m)\ .
\end{split}\label{eq:91}
\end{equation}
Before solving dispersion relation (\ref{eq:89}), we see qualitatively
the dependence of $y\equiv D_g^2$ on other dimensionless parameters.
Similar to the aligned cases, we have $C^S_0=0$ for $q=0$ and meanwhile,
$C^S_3,\ C^S_2,\ C^S_1$ are exactly the same as $C_2,\ C_1,\ C_0$
respectively in equation (72) of \citet{b10}. By setting $\eta=0$
and $\delta\rightarrow\infty$, we have $C_3=0$ and the coefficients
actually reduce to the situation of a single MHD scale-free disc
analyzed by \citet{b11} [see their equations (70)]. We also point
out that by setting $\beta=1/4$ and $\xi=0$, we have
$\mathcal{N}_m(1/4,\ 0)=\mathcal{P}_m(1/4)$ and
$\mathcal{A}_m(1/4,\ 0)=\mathcal{A}^S_m(0)$, this logarithmic spiral
case goes back to the aligned $\beta=1/4$ case as expected.

Equation (\ref{eq:89}) is cubic in $y\equiv D_g^2$. We can also
readily cast equation (\ref{eq:89}) in terms of $D_s^2$ instead
of $D_g^2$. This is exactly parallel to what we have done in the
analysis of the aligned cases and we would not repeat the steps
here. For notational clarity, we should add $'$ to the
coefficients of the cubic equation for $D_s^2$ such as $C^{S'}_0$
and so forth, similar to the aligned cases. However, since $D_s^2$
will be frequently used, we denote $y=D_s^2$ instead of $y'$ from
now on unless otherwise stated.

\subsubsection{Dependence of $D_s^2$ Solutions with\\
\quad\qquad Other Dimensionless Parameters}

\begin{figure}
    \centering
    \subfigure[$m=2,\ \beta=-0.24,\ \sigma=1,\ F=1$]{
      \label{fig:12a}
      \includegraphics[width=75mm,height=48mm]{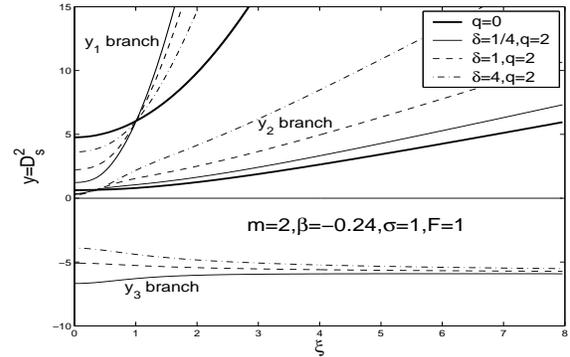}}
    \hspace{1in}
    \subfigure[$m=2,\ \beta=-0.24,\ \sigma=5,\ F=1$]{
      \label{fig:12b}
      \includegraphics[width=75mm,height=48mm]{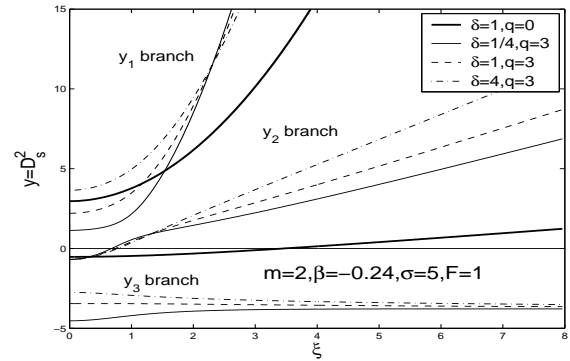}}
    \hspace{1in}
    \subfigure[$m=2,\ \beta=-0.24,\ \sigma=1,\ F=0.5$]{
      \label{fig:12c}
      \includegraphics[width=75mm,height=48mm]{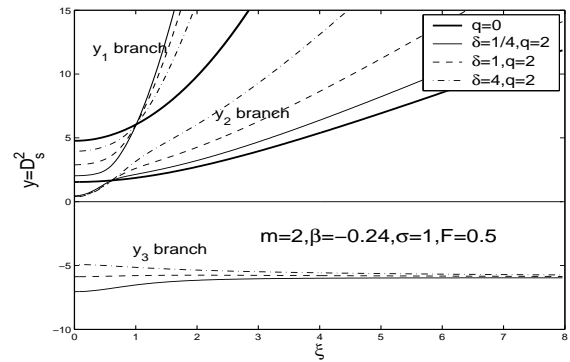}}
  \caption{$D_s^2$ solution branches of equations (\ref{eq:89})
  and (\ref{eq:22}) as functions of radial wave number $\xi$ with
  specified parameters $m=2$ and $\beta=-0.24$. When $q=0$,
  our cases naturally reduce to the unmagnetized coupled discs analyzed
  by \citet{b10} where there are two roots corresponding to faster and
  slower density waves (Lou \& Fan 1998b). For the convenience of
  comparison, we draw this case of $q=0$ in boldface solid curves.
  When $\sigma=1$ and $q=0$, the solution is independent of $\delta$
  \citep{b10}. Setting $q=2$, we then show the solutions with
  $\delta=1/4,\ 1$ and 4 which are displayed in solid, dashed
  and dash-dotted curves, respectively. We examine full discs with
  $\sigma=1$ and $\sigma=5$ shown in panels (a) and (b), while
  for a partial disc, we show a case with $\sigma=1$ and $F=0.5$.
   }\label{fig:12}
\end{figure}

\begin{figure}
    \centering
    \subfigure[$m=2,\ \beta=1/4,\ \sigma=1,\ F=1$]{
      \label{fig:14a}
      \includegraphics[width=75mm,height=55mm]{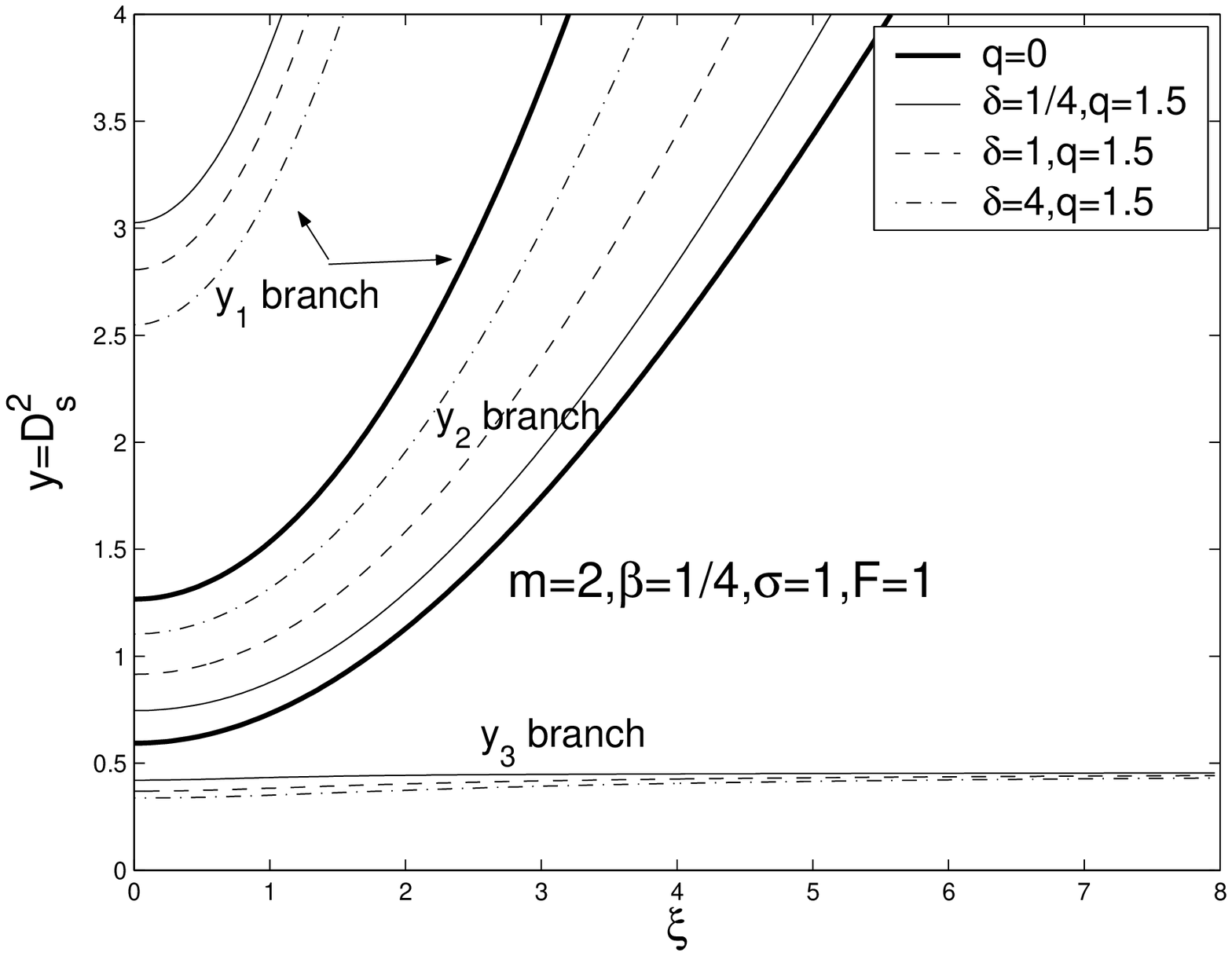}}
    \hspace{1in}
    \subfigure[$m=2,\ \beta=1/4,\ \sigma=5,\ F=1$]{
      \label{fig:14b}
      \includegraphics[width=75mm,height=55mm]{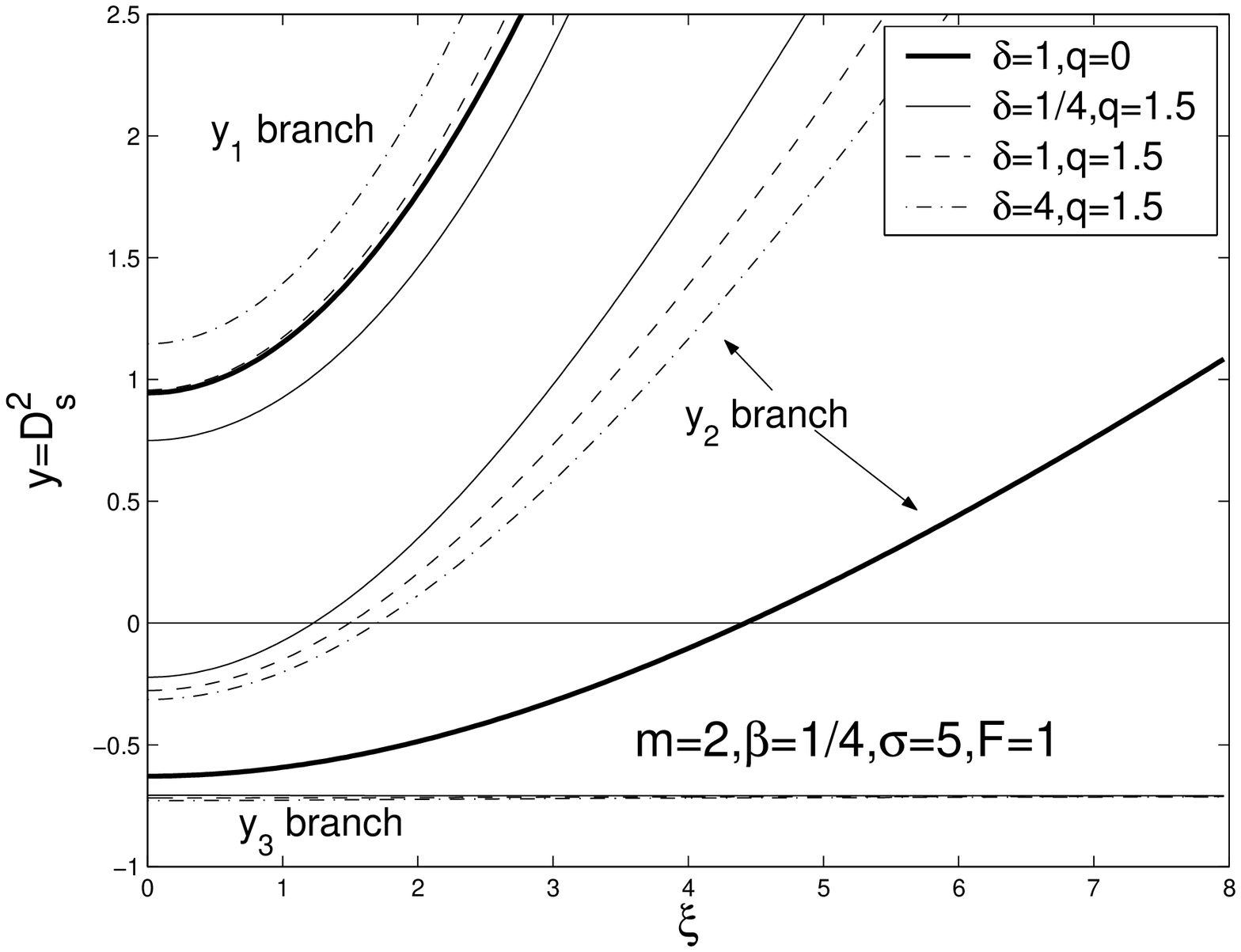}}
    \hspace{1in}
    \subfigure[$m=2,\ \beta=1/4,\ \sigma=1,\ F=0.5$]{
      \label{fig:14c}
      \includegraphics[width=75mm,height=55mm]{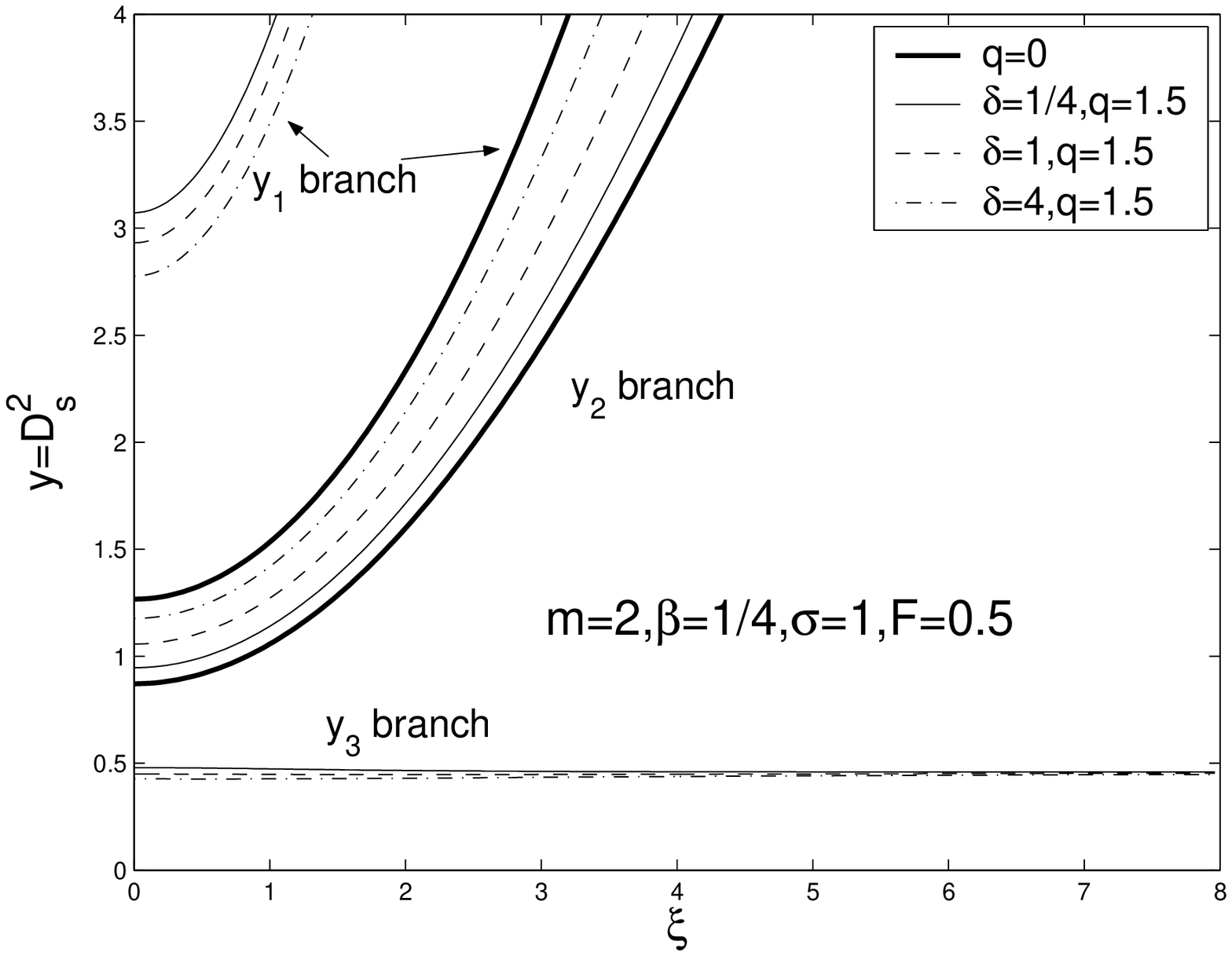}}
  \caption{$D_s^2$ solution curves versus $\xi$ for a
  logarithmic spiral with $m=2$ (two-armed spirals) and
  $\beta=1/4$. By setting $q=1.5$, we show the solutions with
  $\delta=1/4,\ 1$ and 4, respectively. The cases for full
  discs with $\sigma=1$ and $\sigma=5$ are shown in panels (a)
  and (b), respectively. For a partial disc, we set $\sigma=1$
  and $F=0.5$ and show the results in panel (c). The three
  curves of branch $y_3$ overlap with each other.
  }\label{fig:14}
\end{figure}

\begin{figure}
    \centering
    \subfigure[$m=1,\ \beta=-0.1,\ \sigma=2,\ F=1$]{
      \label{fig:15a}
      \includegraphics[width=75mm,height=55mm]{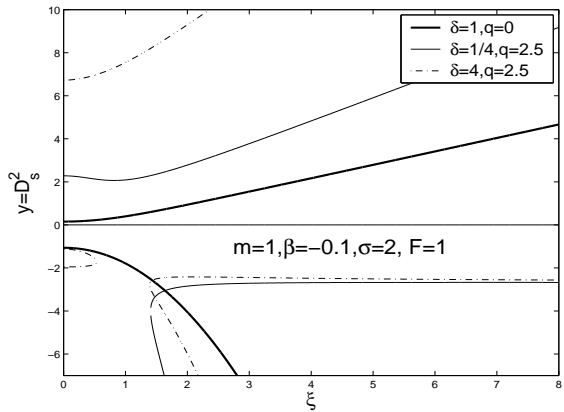}}
    \hspace{1in}
    \subfigure[$m=1,\ \beta=1/4,\ \sigma=2,\ F=1$]{
      \label{fig:15b}
      \includegraphics[width=75mm,height=55mm]{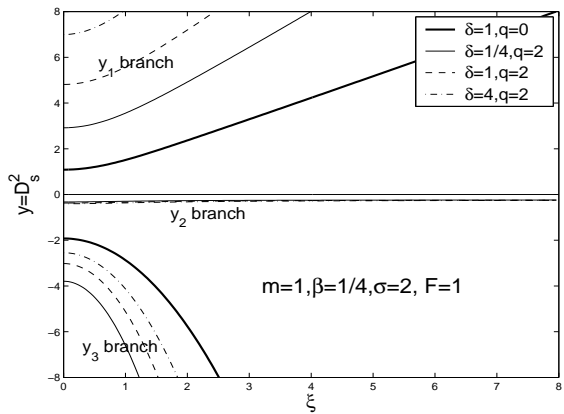}}
  \caption{The $D_s^2$ solution curves versus $\xi$ for a logarithmic
  spiral with $m=1$. For $\beta=-0.1$ in panel (a), there can be only
  one real root and for $\beta=1/4$ in panel (b) the curves are
  similar to the unmagnetized case. In panel (b), the three curves
  of $y_2$ branch almost overlap with each other.
  }\label{fig:15}
\end{figure}

\begin{figure}
    \centering
    \subfigure[$m=1,\ \beta=1/4,\ \delta=1,\ \sigma=2,\ \xi=1$,
     $q=0.3$ (solid curves) and $1.5$ (dashed curve) ]{
      \label{fig:16a}
      \includegraphics[width=75mm,height=55mm]{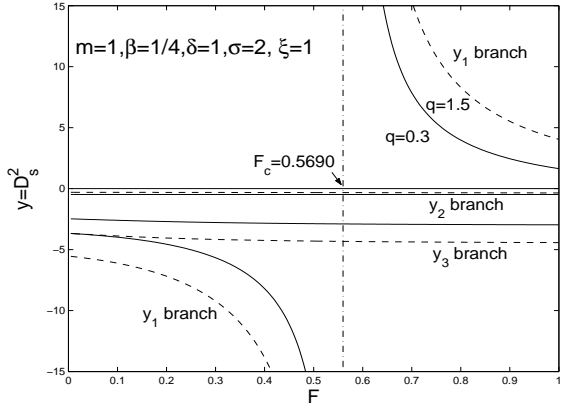}}
    \hspace{1in}
    \subfigure[Critical value of $F_c$ versus $\xi$ for the case $m=1$
    with three different $\beta$ values. ]{
      \label{fig:16b}
      \includegraphics[width=75mm,height=55mm]{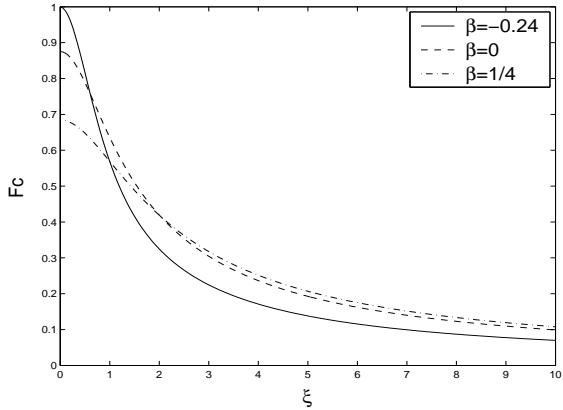}}
    \hspace{1in}
    \subfigure[$m=2,\ \beta=1/4,\ \delta=1,\ \sigma=2,\ \xi=2$,
    $q=0.3$ (solid curves) and $1.5$ (dashed curve) ]{
      \label{fig:16c}
      \includegraphics[width=75mm,height=55mm]{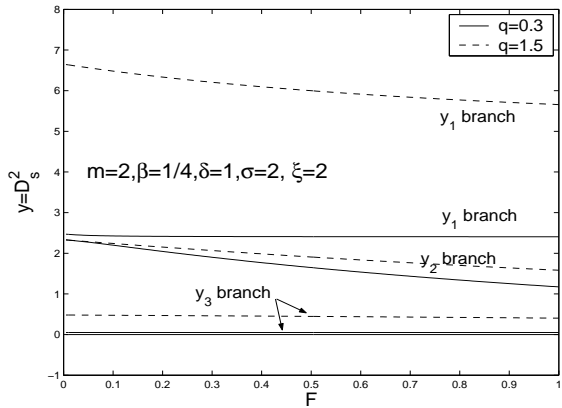}}
  \caption{Solutions for a composite system of partial scale-free
  discs. For $m=1$ cases, there exists $F_c$ such that the $y_1$
  solution diverges [see panel (a)]. This $F_c$ depends only on
  $\beta$ and $\xi$ as shown in panel (b). For $m\geq2$ cases,
  the ratio $F$ does not have much influence on the behaviour of
  $D_s^2$ solutions.
  }\label{fig:16}
\end{figure}

As we have shown before, the dispersion relation for global logarithmic
spirals is expressed in the same form as that for aligned perturbations,
indicating that the behaviours of the two cases bear certain
similarities. Compared with the aligned cases, the differences should
arise only from the substitutions of $\mathcal{A}^A_m(\beta)$
by $\mathcal{A}^S_m(\beta,\ \xi)$ and $\mathcal{P}_m(\beta)$ by
$\mathcal{N}_m(\xi)$, respectively. For logarithmic spiral perturbations,
there is one more parameter $\xi$ corresponding to the radial wavenumber.
This parameter is important for forming large-scale structures in spiral
galaxies. On the one hand, we are interested in the influence of $\xi$
on behaviours of $D_s^2$ solution. On the other hand, by analogy we can
imagine that the dependence of $D_s^2$ solution on parameter $\eta$
should be similar to that of the aligned cases. For these two reasons,
we present our figures in $\xi-y$ plane where $y=D_s^2$. We focus on
the $m=2$ case (viz., two-armed spirals). For cases of different $\beta$
values, we will show in Fig.\ \ref{fig:12}$-$Fig.\ \ref{fig:14}. The
$m=1$ case is shown in Fig.\ \ref{fig:15} briefly. We also show results
for partial discs in Fig.\ \ref{fig:16}.

\begin{figure}
    \centering
    \subfigure[$m=2,\ \beta=0,\ \sigma=1,\ F=1$]{
      \label{fig:13a}
      \includegraphics[width=75mm,height=55mm]{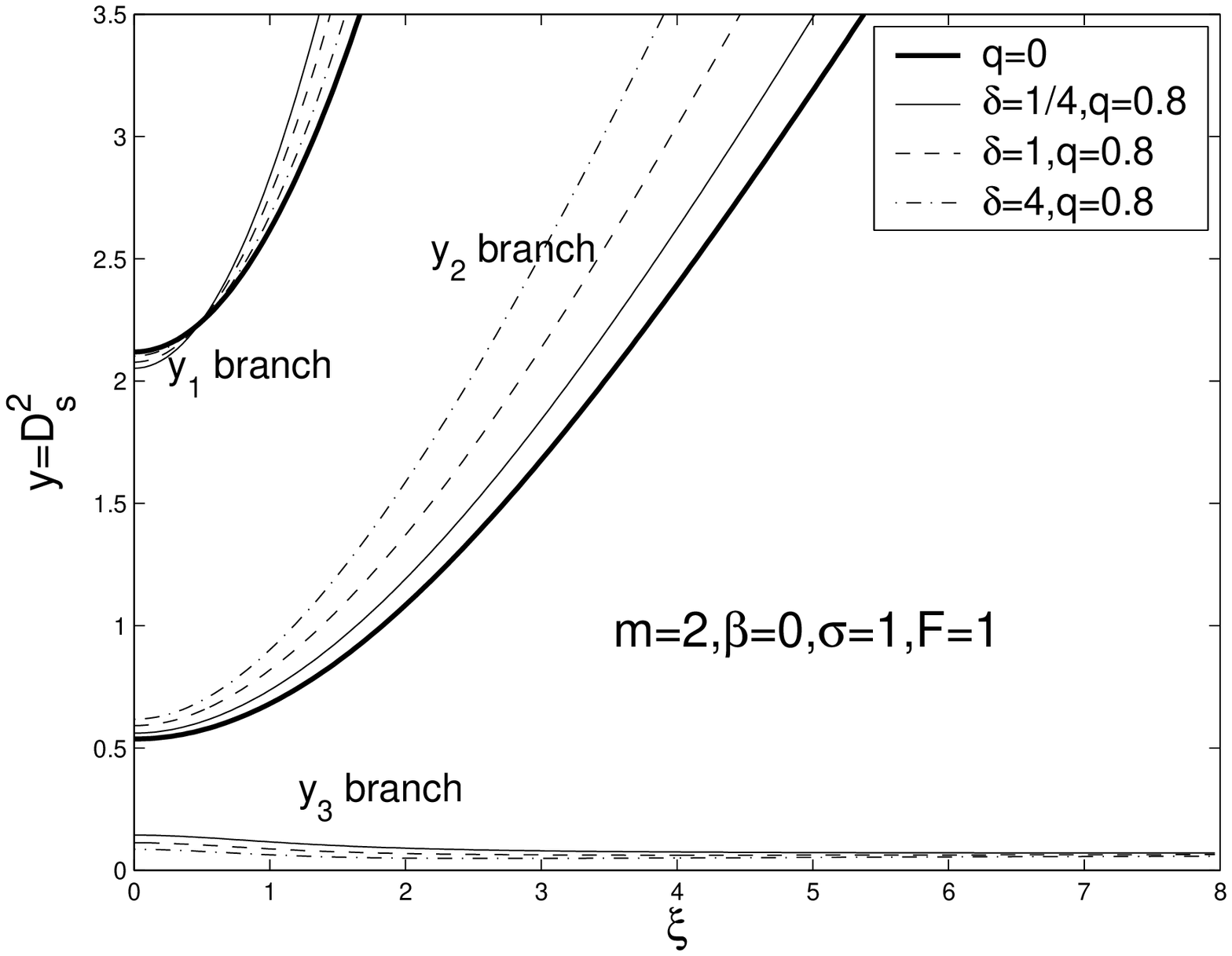}}
    \hspace{1in}
    \subfigure[$m=2,\ \beta=0,\ \sigma=1,\ F=1$]{
      \label{fig:13b}
      \includegraphics[width=75mm,height=55mm]{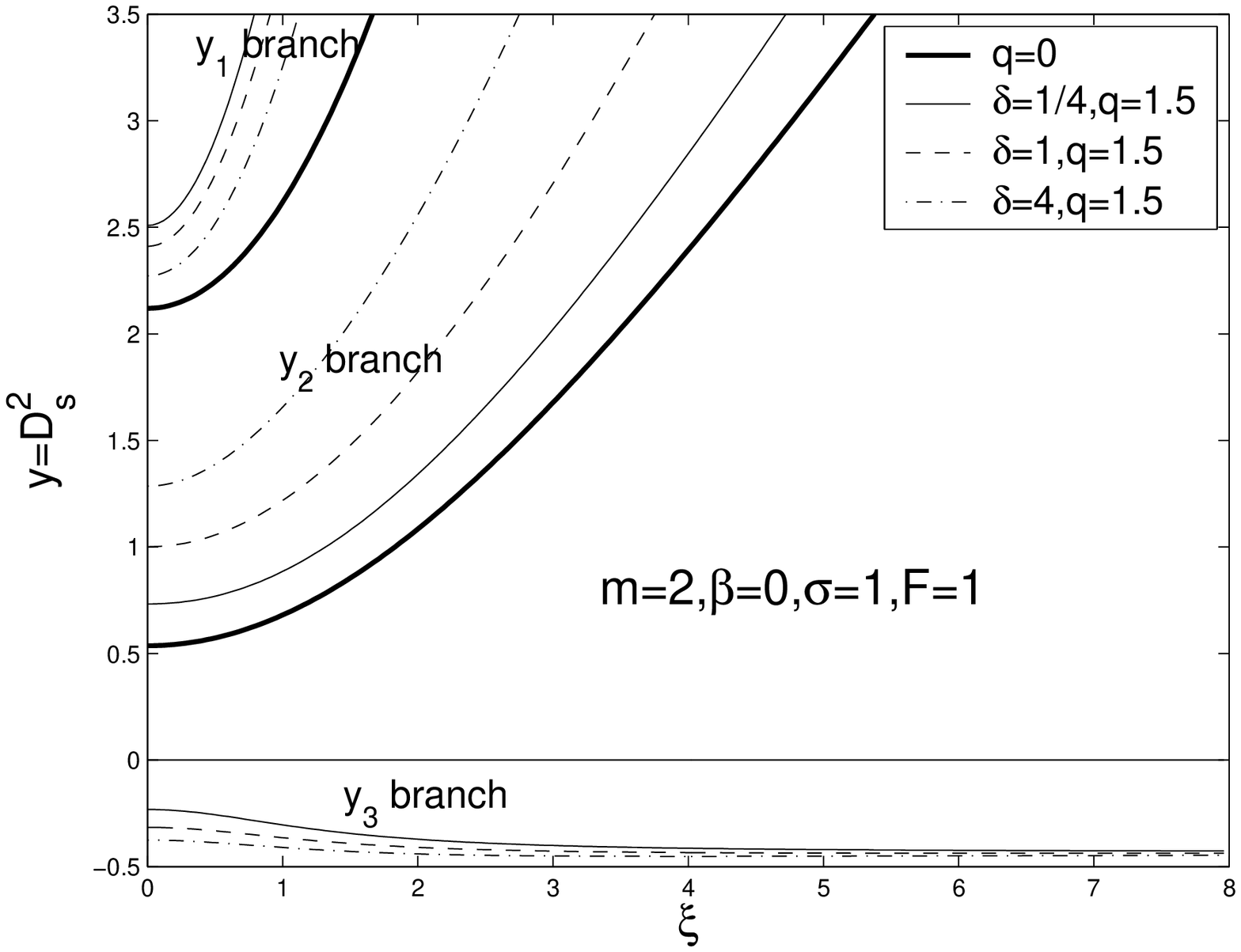}}
    \hspace{1in}
    \subfigure[$m=2,\ \beta=0,\ \sigma=5,\ F=1$]{
      \label{fig:13c}
      \includegraphics[width=75mm,height=55mm]{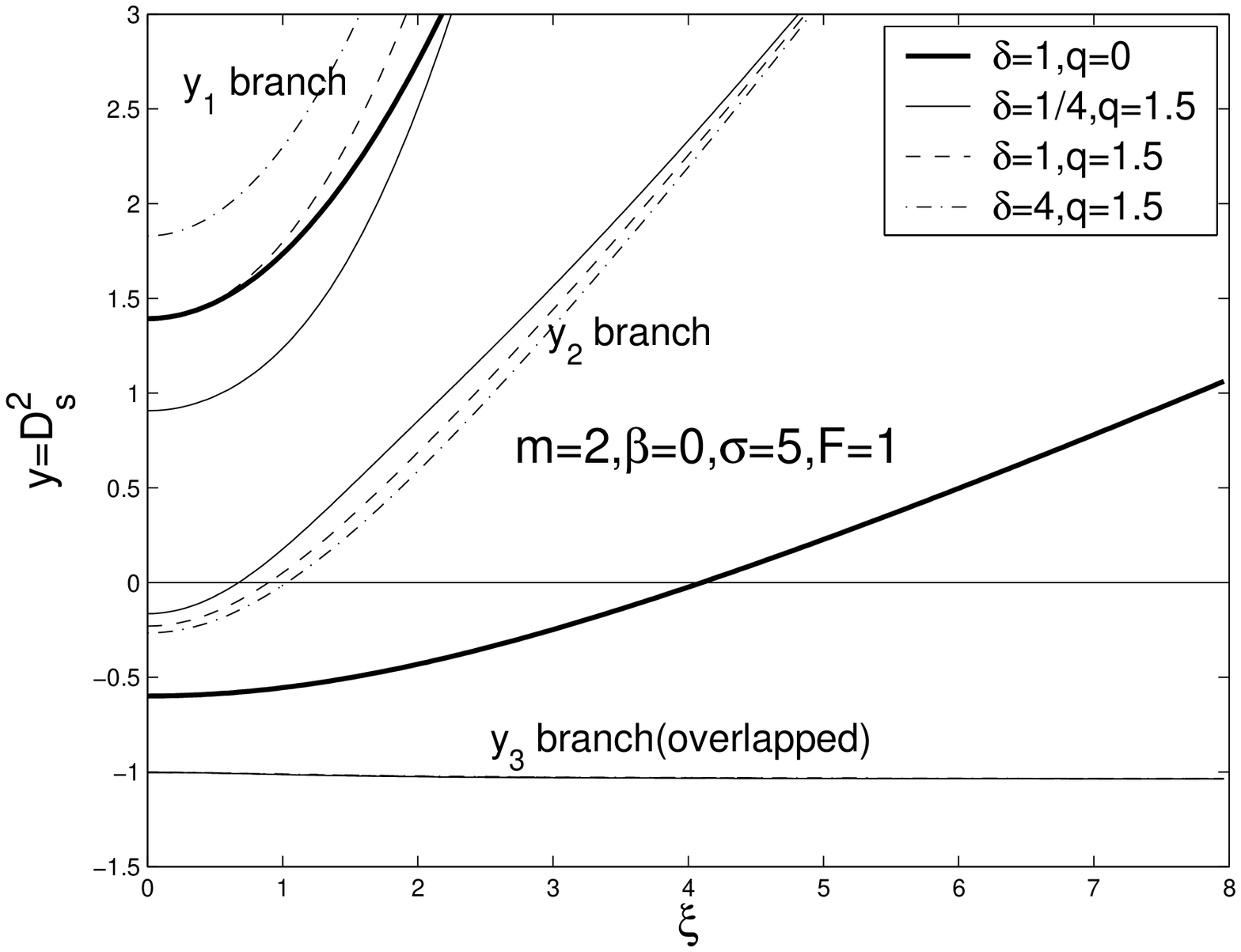}}
  \caption{$D_s^2$ solution curves versus $\xi$ for a
  logarithmic spiral case with $m=2,\ \beta=0,\ F=1$.
  Magnetic field has a considerable influence on the disc system
  [see panels (a) and (b)]. A weak magnetic field results in a
  physical $y_3$ solution but when the magnetic field strength
  becomes stronger, this $y_3$ solution falls negative. In panel
  (c), we show the influence of $\sigma$, where the three curves
  of $y_3$ branch pact together.
  }\label{fig:13}
\end{figure}

Similar to the aligned cases, the $y_1$ and $y_2$ branches correspond
to the unmagnetized solutions of Lou \& Fan (1998b) and \citet{b10}
and $y_3$ is an extra solution (i.e., SMDWs) due to the very existence
of magnetic field. Meanwhile, the upper two branches are monotonically
increasing with the increase of $\xi$, while the $y_3$ branch remains
always flat as $\xi$ varies. The $y_3$ branch is sensitive to the $q$
variation.
This is a sign that the $y_3$ solution is the direct result
of the magnetic field.
Generally, the $y_1$ branch remains always positive, the
$y_2$ branch can be either positive or negative and the lowest $y_3$
branch is mostly negative and thus unphysical. There are critical
values of $\xi$ at which these branches change signs. These critical
values are determined by $C^{S'}_0=0$. As they cannot be evaluated
analytically, we simply show their presence in the figures. In each
figure, we choose $\sigma=1$ and $\sigma=5$ and see that all three
branches are lowered for larger values of $\sigma$ as expected. This
trend of variation is similar to that in the aligned cases.

For logarithmic spiral cases with $m=2$, the `convergent point' is
a common phenomenon where curves with different $\delta$ values
tend to converge at a certain point. For a small $\beta$ (e.g.,
$\beta=-0.24$), there is no convergent point in the aligned cases
but here the variation of $\xi$ produces this convergent point.
For larger $\beta$ values (e.g., $\beta=0,\ 1/4$), the situation
becomes similar to the aligned cases where variations of $\sigma$
lead to the existence of convergent point. For example, in Fig.\
\ref{fig:13} and Fig.\ \ref{fig:14} the sequence of curves with
different $\delta$ values show a sign of convergence when $\sigma=5$.

A strong coplanar magnetic field generally raises all three branches
[see Fig.\ \ref{fig:13b} and Fig.\ \ref{fig:14}], in accordance with
the aligned cases. Nevertheless, this is not always the case. For a
small $\beta$ (e.g., $\beta=-0.24$), a larger $q$ tends to lower the
solution curve as the radial wavenumber $\xi$ becomes small (see
Fig.\ \ref{fig:12}). In the $\beta=0$ case corresponding a composite
system of two coupled SIDs (Lou \& Zou 2004, 2006), the increase of
magnetic field strength first lowers the $y_1$ and $y_2$ branches
for small $\xi$ and then turns to raise them. Meanwhile, $y_3$ is
positive when $q$ is not too large, but as $q$ grows, $y_3$ is
lowered down to become negative and thus unphysical (see Fig.\
\ref{fig:13}). We made a few comments about the $y_3$ branch. We
already see that it bear little relation to the spiral pattern.
Since it is the new branch originated from the magnetic field, we
have $y_3=0$ when $q=0$ and its dependence on $q$ is easily seen
from our figures. Generally speaking, the $y_3$ branch is sensitive
to both $\sigma$ and $q$.

The $m=1$ case reveals some interesting features to note. We show a
few cases in Fig.\ \ref{fig:15}. Here the unmagnetized two solutions
of \citet{b10} correspond to our $y_1$ and $y_3$ branches. The $y_3$
branch is monotonically decreasing with $\xi$. For a small $\beta$,
there can be only one real root and the other two are complex
conjugates [see Fig.\ \ref{fig:15a}]. We also see that both $q$ and
$\delta$ have a strong influence on the solution behaviours. For the
physical $y_1$ solution branch, a strong magnetic field as well as
a high $\delta$ greatly raises the solution curves.

For a composite system of two partial scale-free discs embedded in
an axisymmetric dark matter halo, we distinguish $m=1$ and $m>1$
cases. In the $m=1$ case, there always exists the critical value
$F_c$ for the potential ratio $F$ parameter such that the $y_1$
solution branch diverges from either side of $F_c$. Mathematically,
this critical value $F_c$ is determined by $C^S_3=0$ and is quite
similar to that in the aligned cases. For $F\leq F_c$,
there is no physical solution, while for $F>F_c$ the $y_1$ branch
is physical and sensitive to $F$ variation; the existence of an
axisymmetric dark matter halo can significantly influence global
configurations in a composite disc system with $m=1$ MHD
perturbations. Here $F_c$ depends only on $\beta$ and $\xi$ and we
demonstrate this dependence using examples in Fig.\ \ref{fig:16b}.
The $F_c$ curves with different $\beta$ decrease monotonically as
$\xi$ increases. For a small radial wavenumber $\xi\leq1$, the
critical value $F_c$ approaches 1 as seen in panels (a) and (b) of
Fig.\ \ref{fig:16}. Since in reality the dark matter halos play
significant roles in disc galaxies (i.e., $F$ is small and not close
to 1), we propose that a spiral galaxy with only one spiral arm with
a relatively large proportion of dark matter halo must have a
relatively large radial wavenumber (i.e., tightly wound) to support
a global stationary pattern. In principle, this may be tested by
observations.

For spiral perturbations with $m>1$, however, the dark matter seems to be
less important in determining the stationary dispersion relation. Because
in the entire range of $0<F\leq1$, there are only slight variations in
the $D_s^2$ solutions. We show such an example in Fig.\ \ref{fig:16c}.
With a radial wavenumber $\xi=2$, all three solution branches $y_1$,
$y_2$ and $y_3$ remain positive and rise only slightly as $F$ decreases.
We can also compare Fig.\ \ref{fig:12a} and Fig.\ \ref{fig:12c}, Fig.\
\ref{fig:14a} and Fig.\ \ref{fig:14c} and see that in $m=2$ cases,
the dark matter halo has little influence on solutions. For cases of
$m\geq3$, the situations are qualitatively similar. Nevertheless, a
dark matter halo should have a strong stabilizing effect in general.

\subsubsection{Phase Relationships among Perturbation Variables}

For logarithmic spiral cases, the phase relationships among coplanar
MHD perturbation variables have more features to explore. The
introduction of radial oscillations with radial wavenumber $\xi$
brings an imaginary part into our equations which describes the
phase difference between perturbed surface mass density and the
magnetic field perturbation as well as other perturbation variables.
To describe these phase relations more specifically, we begin with
equations $(\ref{eq:82})-(\ref{eq:84})$ to obtain
\begin{equation}
\begin{split}
&{S^g}/{S^s}=-1-\mathcal{M}_S/(2\pi G\Sigma_0^sr\mathcal{N}_m)\ ,\\
&\frac{\mathrm{i}R}{S^g}=\frac{mB_0}
{\Sigma_0^g}\bigg[\Theta_S\bigg(1+\frac{S^s}{S^g}\bigg)+\Xi_S\bigg]\ ,\\
&Z=\xi\mathrm{i}R/m\ .
\end{split}\label{eq:92}
\end{equation}
There is another independent expression of $\mathrm{i}R/S^g$ obtained
by a combination of equations (\ref{eq:78}) and (\ref{eq:79}), namely
\begin{equation}
\frac{\mathrm{i}R}{S^g}=\frac{mB_0}{\Sigma_0^g}
\bigg[\frac{(1-2\mathrm{i}\xi)\pi
G\Sigma_0^gr\mathcal{N}_m}{\mathcal{K}_S}
\bigg(1+\frac{S^s}{S^g}\bigg)
-\frac{\mathcal{L}_S}{\mathcal{K}_S}\bigg]\ .\label{eq:93}
\end{equation}
It is no wonder that these expressions are quite similar
to the aligned phase relationships. We next express these
results in physical parameters. As $\mathrm{i}R/S^g$ is
no longer dimensionless, we take the proportional factor
in brackets of equation (\ref{eq:92}).
The results are
\begin{equation}
\begin{split}
&\frac{S^g}{S^s}=-1-\frac{(\mathcal{B}_mD_s^2-\mathcal{A}^S_m)}
{(D_s^2+1)}
\frac{\mathcal{C}(1+\delta)}{\mathcal{N}_m\mathcal{A}^S_m}\ ,\\
&\frac{\mathrm{i}R}{S^g}\propto\frac{m}
{(1+2\beta)(3/2-\mathrm{i}\xi)D_g^2-(1/2-2\beta)q^2}
\bigg[(1+2\beta)D_g^2 \\ &
+\frac{D_g^2+1+{(4\beta-1)q^2}/{(4\beta+2)}}
{\mathcal{C}(1+\delta)}\bigg(1+\frac{S^s}{S^g}\bigg)
\mathcal{N}_m\delta -1\bigg]\ .\label{eq:94}
\end{split}
\end{equation}
Since $\mathrm{i}R/S^g$ contains imaginary
part, the phase difference between surface perturbation mass
density and the perturbed magnetic field becomes complicated. For
this reason, we show such phase relations in special figures.

\begin{figure}
    \centering
    \subfigure[$m=2,\ \beta=-0.24,\ \sigma=1,\ F=1$]{
      \label{fig:17a}
      \includegraphics[width=75mm,height=50mm]{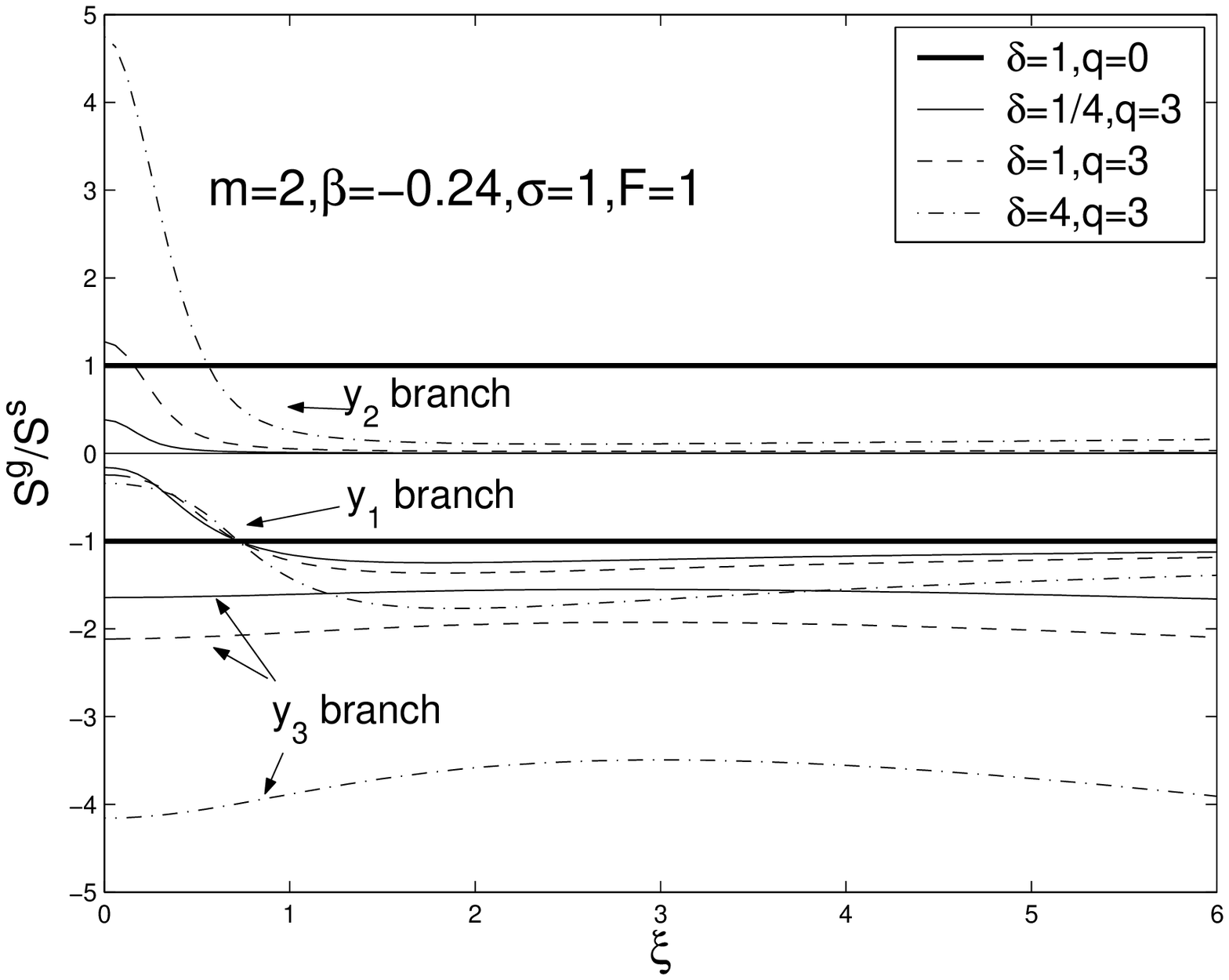}}
    \hspace{1in}
    \subfigure[$m=2,\ \beta=-0.24,\ \sigma=5,\ F=1$]{
      \label{fig:17b}
      \includegraphics[width=75mm,height=50mm]{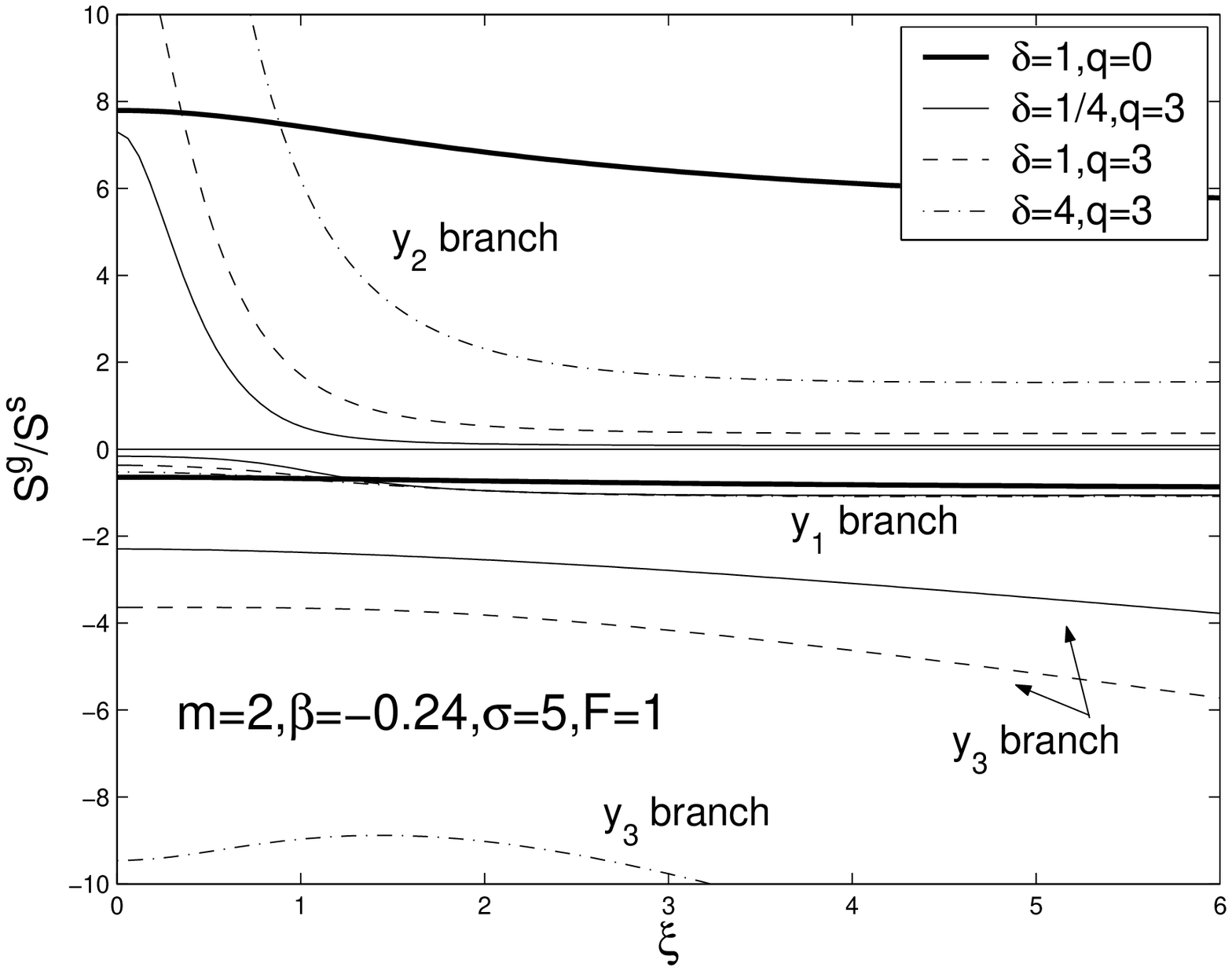}}
    \hspace{1in}
    \subfigure[$m=2,\ \beta=-0.24,\ \sigma=1,\ F=0.5$]{
      \label{fig:17c}
      \includegraphics[width=75mm,height=50mm]{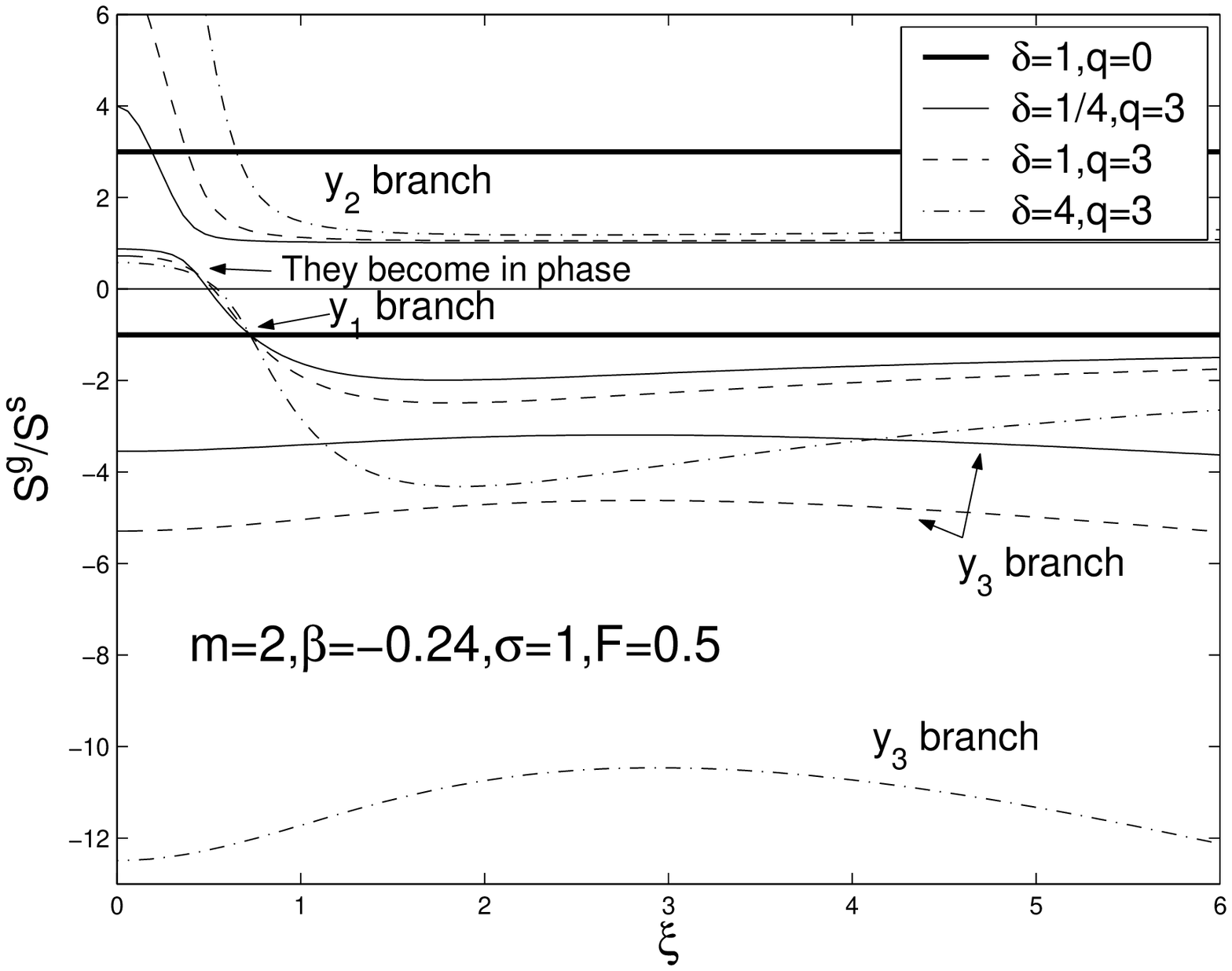}}
  \caption{Phase relation curves for the ratio $S^g/S^s$ of
  surface mass density perturbation versus the radial wave number
  $\xi$ for the case of $m=2$ and $\beta=-0.24$. The solutions with
  $q=3,\ \delta=1/4,\ 1$ and $4$ are shown in light solid curve,
  dashed curve and dash-dotted curve, respectively. We examine full
  discs ($F=1$) with $\sigma=1$ and $\sigma=5$ in panels (a) and (b),
  respectively. For a partial disc with $F=0.5$, we show solutions
  with $\sigma=1$ in panel (c). In all panels, the solution branches
  $y_1$, $y_2$, and $y_3$ can be readily identified. Each branch
  contains three distinct solutions with $\delta=1/4$ (solid curve),
  1 (dashed curve) and 4 (dash-dotted curve), respectively.
  In each panel, the two boldface curves for $y_1$ and $y_2$
  branches correspond to $q=0$ (the $y_3$ branch is trivial).
  }\label{fig:17}
\end{figure}

\begin{figure}
    \centering
    \subfigure[$m=2,\ \beta=1/4,\ \sigma=1,\ F=1$]{
      \label{fig:18a}
      \includegraphics[width=75mm,height=55mm]{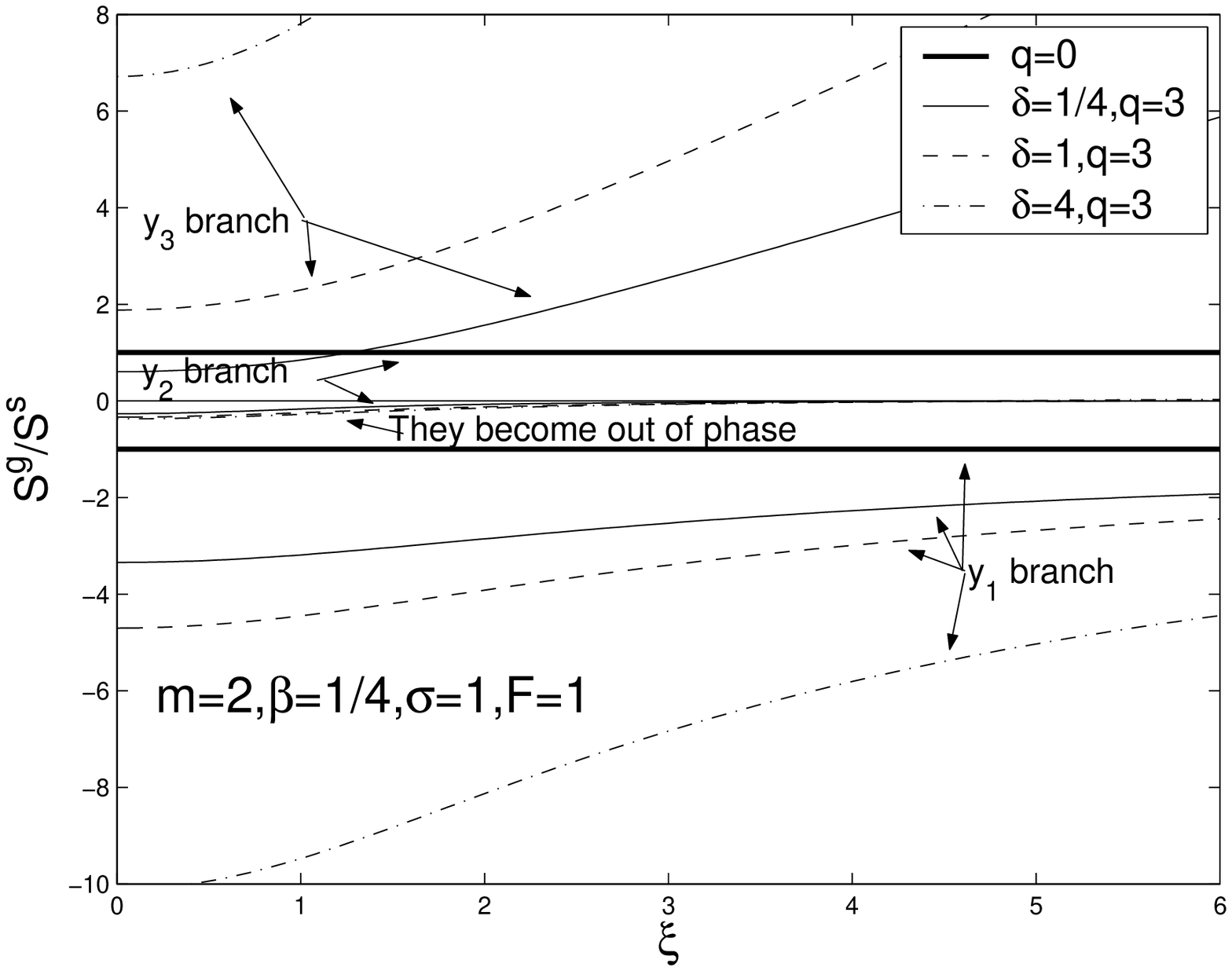}}
    \hspace{1in}
    \subfigure[$m=2,\ \beta=1/4,\ \sigma=3,\ F=1$]{
      \label{fig:18b}
      \includegraphics[width=75mm,height=55mm]{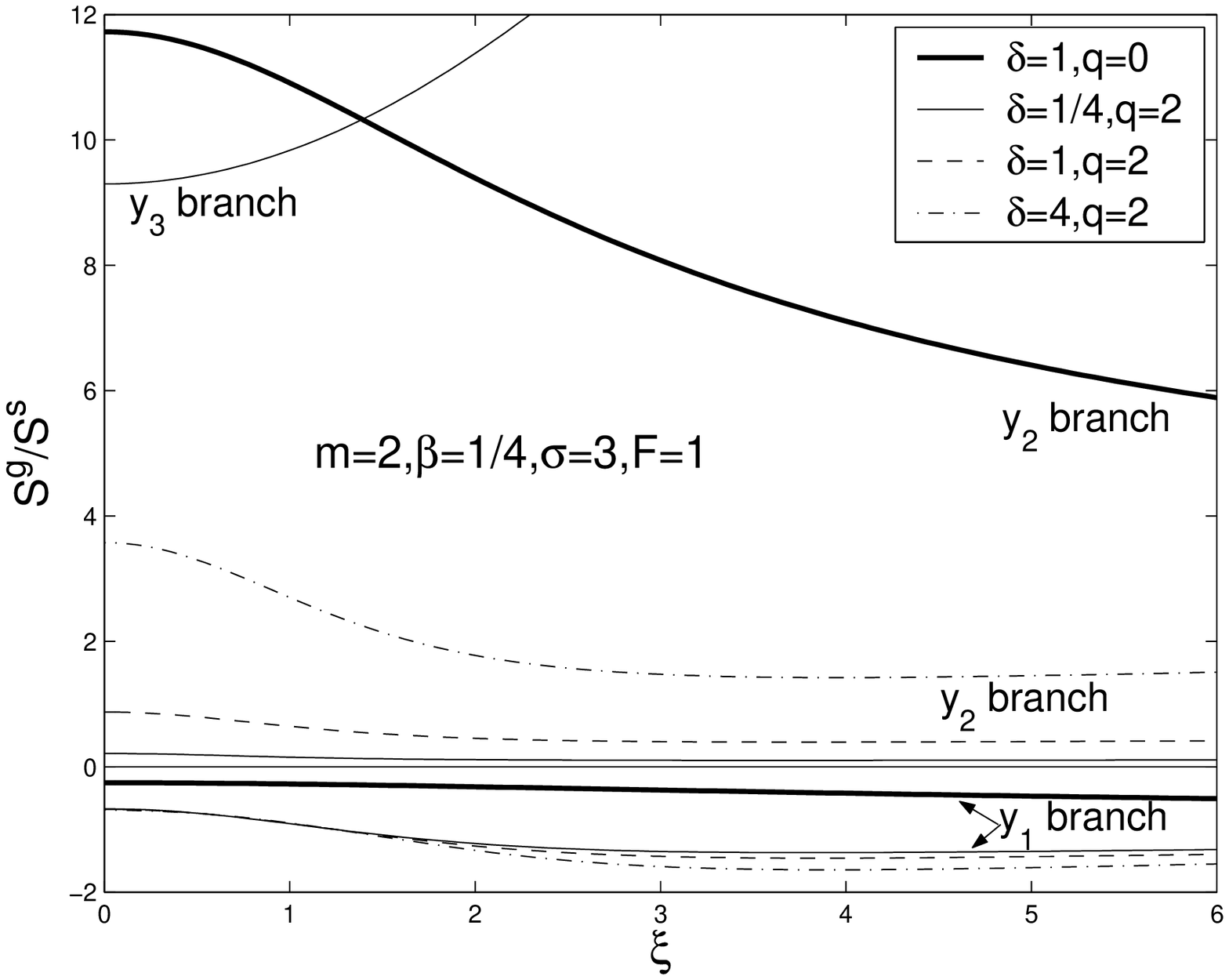}}
    \hspace{1in}
    \subfigure[$m=2,\ \beta=1/4,\ \sigma=1,\ F=0.5$]{
      \label{fig:18c}
      \includegraphics[width=75mm,height=55mm]{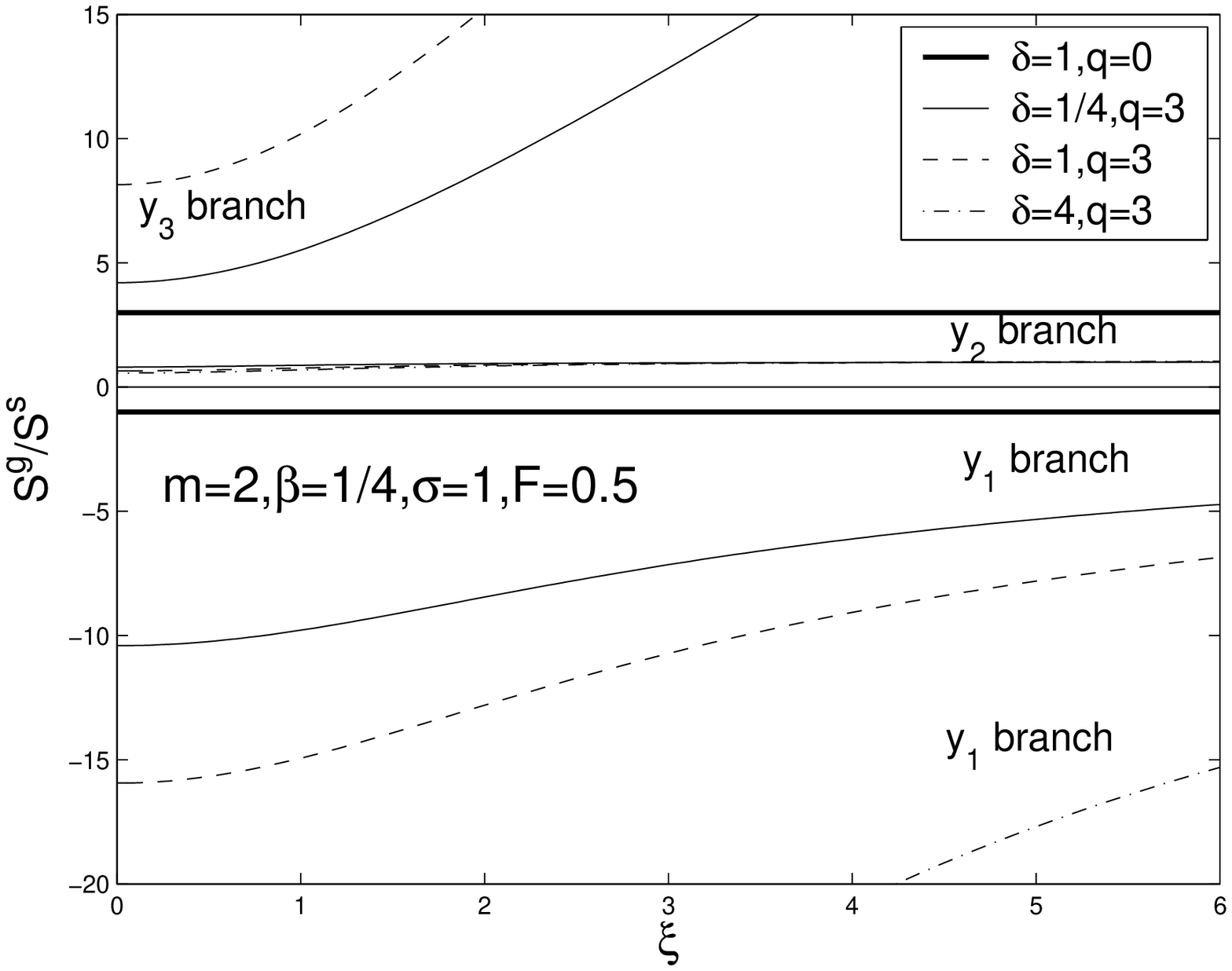}}
  \caption{The phase relation curves for the ratio $S^g/S^s$ of
  surface mass density perturbations versus $\xi$ for a logarithmic
  spiral case with $m=2$ and $\beta=1/4$. We show separately phase
  curves for full discs ($F=1$) with $\sigma=1$ and $\sigma=3$ in
  panels (a) and (b), respectively. For a partial disc of $F=0.5$
  with $\sigma=1$, we show solution branches in panel (c). Note that
  in the case $q=0$ (boldface curves) and $\sigma=1$, the two solution
  branches have $S^g/S^s=\pm1$ [see equation (73) of \citet{b10}].
   }\label{fig:18}
\end{figure}

We first quickly note that when $q=0$ (unmagnetized) and $\sigma=1$
(the same `sound speeds'), the ratio $S^g/S^s=\pm 1$ for the $y_1$
and $y_2$ solution branches, respectively. We can prove this claim
by using the fact that $y_1=\mathcal{A}_m/\mathcal{B}_m$ and $y_2=
(1-\mathcal{N}_m/\mathcal{C})\mathcal{A}_m/\mathcal{H}_m$ for the
two $D_s^2$ solutions \citep{b10} and by substituting these explicit
$y_1$ and $y_2$ expressions into equation (\ref{eq:94}). For fairly
arbitrary parameters, the tendency of the phase relation curves for
the ratio of the perturbed surface mass densities in the two
scale-free discs are shown in Fig. \ref{fig:17} and Fig.
\ref{fig:18}. We would emphasize that the relative order of the
three branches varies in the phase relation curves. For example, for
a small $\beta=-0.24$, the $y_3$ branch is the lowest, while for a
larger $\beta=1/4$, it becomes the upper one.

\clearpage

\begin{figure}
    \centering
    \subfigure[$m=2,\ \beta=-0.24,\ \delta=1,\ \sigma=1,\ F=1$]{
      \label{fig:19a}
      \includegraphics[width=75mm,height=53mm]{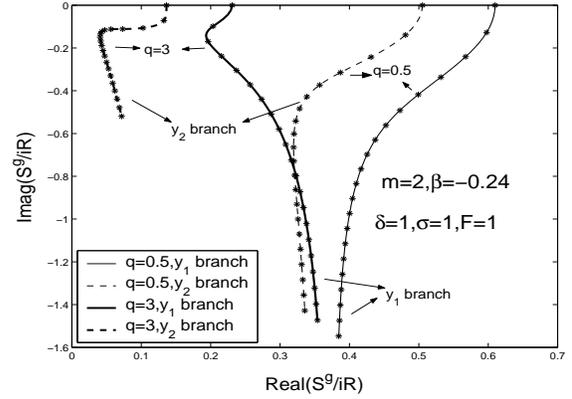}}
    \hspace{1in}
    \subfigure[$m=2,\ \beta=-0.24,\ \delta=1,\ \sigma=5,\ F=1$]{
      \label{fig:19b}
      \includegraphics[width=75mm,height=53mm]{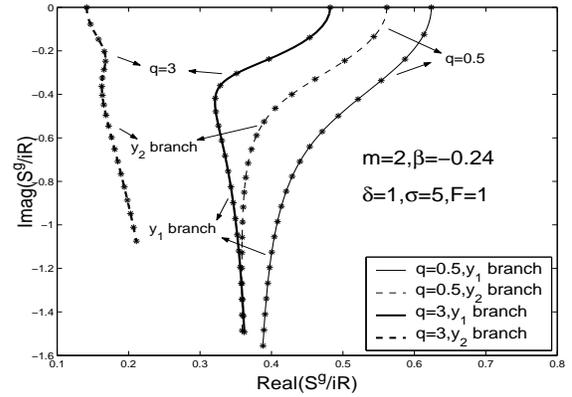}}
    \hspace{1in}
    \subfigure[$m=2,\ \beta=-0.24,\ \delta=1,\ \sigma=1,\ F=0.5$]{
      \label{fig:19c}
      \includegraphics[width=75mm,height=53mm]{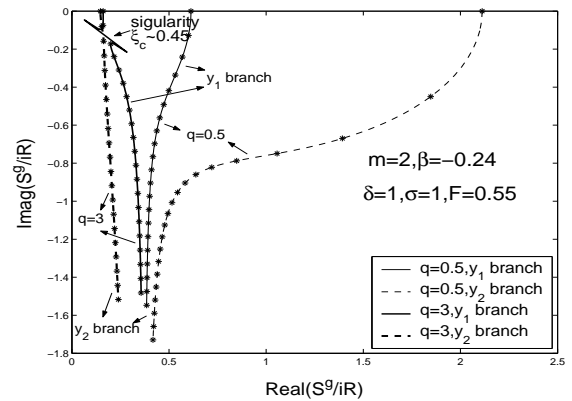}}
  \caption{The phase relation curves for the perturbed radial magnetic
  field and gas surface mass density versus $\xi$ for the case of $m=2$
  and $\beta=-0.24$. The ratio $S^g/(\mathrm{i}R)$ is generally complex
  and is shown here in the complex plane with real and imaginary axes. We
  take the range of $\xi=0$ to 6. For $\xi=0$, the imaginary part is zero
  according to equation (\ref{eq:94}). As $\xi$ grows, the imaginary part
  emerges and becomes more significant. To demonstrate the relation with
  $\xi$, we show $*$ to mark an interval of 0.3 in the variation of $\xi$
  from $0$ to $6$. In this case, only the $y_1$ and $y_2$ solution
  branches are physical and the unphysical $y_3$ branch is ignored.
  }\label{fig:19}
\end{figure}

In the case of $\beta=-0.24$ and for full discs, the phase relation
of the perturbed surface mass densities in the two scale-free discs
is out of phase for the $y_1$ branch and in phase for the $y_2$
branch ($y_3$ branch is unphysical) even with a strong magnetic
field; this situation differs from the aligned cases. For the $y_2$
branch in full discs, a small $\xi$ leads to a larger $S^g/S^s$ ratio,
while for a large $\xi$, the ratio can be rather small compared with
$\delta$. For the $y_1$ branch in full discs, the $S^g/S^s$ curves with
different $\delta$ values do not vary significantly. For a small $\xi$,
the ratio $|S^g/S^s|$ is fairly small and for a large $\xi$, this ratio
lies between 1 and 2. The increase of $\sigma$ generally raises the two
branches to different levels. The presence of magnetic field gives rise
to variations in the $y_1$ solution curves in two trends: the curves
move above and below the $q=0$ solution curve for small and large
$\xi$, respectively (see Fig. \ref{fig:17}).
For a composite system of two coupled partial scale-free discs,
some of these tendencies does not hold. First, $S^g/S^s$ may
become positive for the $y_1$ branch with a sufficiently small
radial wavenumber $\xi$. It also raises the $y_2$ branch
considerably and separates solutions of $y_1$ branch with
different $\delta$ values.
In other words, the influence of $\delta$ becomes more prominent.

In the case of $\beta=1/4$, the phase relation curves of the
perturbed surface mass densities for the $y_1$ solution branch
remain negative. For small $\sigma$, variations of $\delta$
produce considerable differences in phase curves,
while for large $\sigma$, phase curves with
various $\delta$ values become much closer. For the $y_2$ branch
with a small $\sigma$, the ratio $S^g/S^s$ can become negative at a
certain $\sigma_c$ with $S^g=0$ and the curves with different
$\delta$ values are bunched together. For a large $\sigma$, the
ratio $S^g/S^s$ are positive and the curves with different $\delta$
values are separated. In the case of a small $\sigma=1$, the $y_3$
branch is physical with a positive $S^g/S^s$ ratio growing with
$\xi$. The value of $\delta$ influences the phase relation in this
branch in a significant manner. In a composite system of partial
discs, the presence of a dark matter halo raises the $y_2$ and $y_3$
solution branches and lowers the $y_1$ solution branch in the phase
relation curve. Meanwhile, the $S^g/S^s$ ratio is positive for the
$y_2$ branch, and for the $y_1$ and $y_3$ branches, the ratio
magnitude $|S^g/S^s|$ becomes much larger than that for a composite
system of full discs.

\begin{figure}
    \centering
    \subfigure[$m=2,\ \beta=1/4,\ \delta=1,\ \sigma=1,\ F=1$]{
      \label{fig:20a}
      \includegraphics[width=75mm,height=53mm]{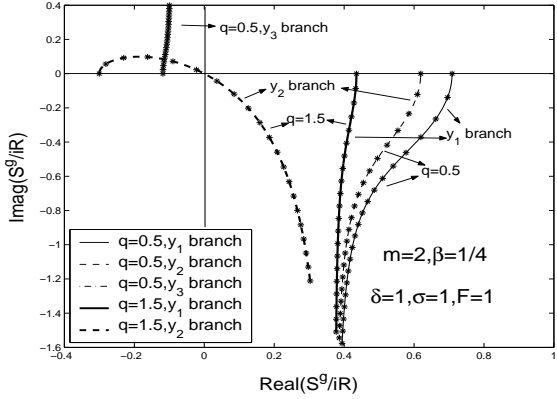}}
    \hspace{1in}
    \subfigure[$m=2,\ \beta=1/4,\ \delta=1,\ \sigma=3,\ F=1$]{
      \label{fig:20b}
      \includegraphics[width=75mm,height=53mm]{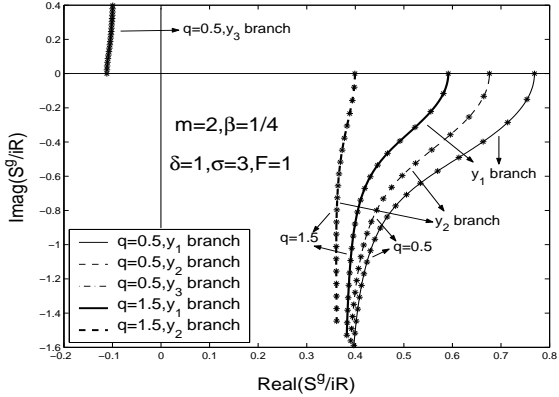}}
    \hspace{1in}
    \subfigure[$m=2,\ \beta=1/4,\ \delta=1,\ \sigma=1,\ F=0.5$]{
      \label{fig:20c}
      \includegraphics[width=75mm,height=53mm]{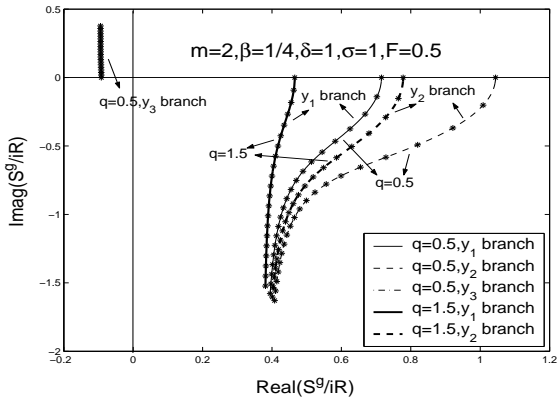}}
  \caption{The phase relation curves for real and imaginary parts
  of $S^g/(\mathrm{i}R)$ between the radial component of the
  perturbation magnetic field and surface mass density versus $\xi$
  for logarithmic spirals with $m=2$ and $\beta=1/4$ in a complex
  plane. We show separately curves of full discs ($F=1$) with
  $\sigma=1$ and $\sigma=3$ in panels (a) and (b), respectively,
  and for a partial disc with $\sigma=1$ in panel (c). We also add
  the phase curve for the physical $y_3$ branch.
  When $q=1.5$, the $y_3$ branch is very far away from the other curves;
  we have omitted this curve for the compactness
  of presentation.
  }\label{fig:20}
\end{figure}

\begin{figure}
    \centering
    \subfigure[$m=2,\ \beta=1/4,\ \delta=1,\ \sigma=1,\ F=1$]{
      \label{fig:sz1a}
      \includegraphics[width=75mm,height=55mm]{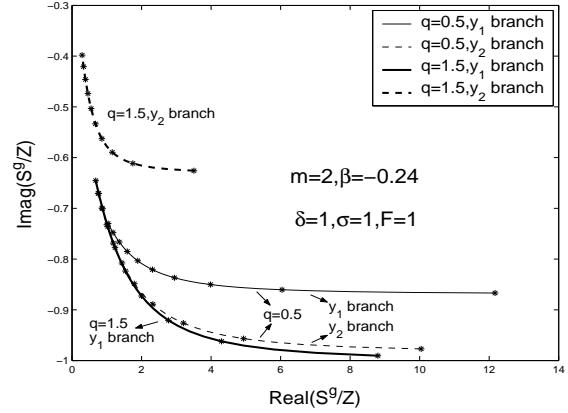}}
    \hspace{1in}
    \subfigure[$m=2,\ \beta=1/4,\ \delta=1,\ \sigma=3,\ F=1$]{
      \label{fig:sz1b}
      \includegraphics[width=75mm,height=55mm]{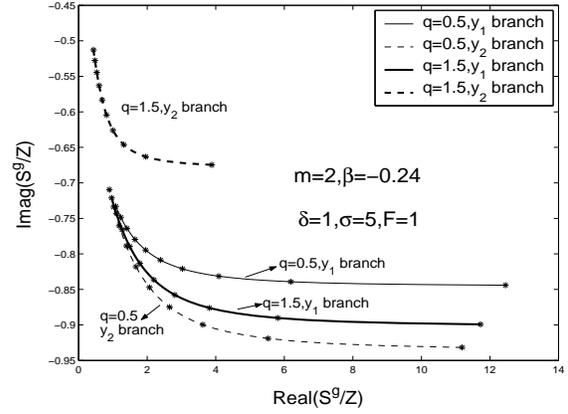}}
  \caption{The phase relation curves between the perturbed azimuthal magnetic
  field and surface mass density versus $\sigma$ for logarithmic spirals
  with $m=2,\ \beta=-0.24$. We show curves of full discs ($F=1$) with $\sigma=1$
  and $\sigma=3$ in panels (a) and (b). We take the range of $\xi=0.4$ to 4.
  Since $S^g/Z\rightarrow\infty$ as $\xi\rightarrow0$, we start from
  $\xi=0.4$ and use $*$ to label an interval of 0.4 in the variation of $\xi$.
  The phase points moves from right to left as $\xi$ increases.}\label{fig:sz1}
\end{figure}
\begin{figure}
    \centering
    \subfigure[$m=2,\ \beta=1/4,\ \delta=1,\ \sigma=1,\ F=1$]{
      \label{fig:sz2a}
      \includegraphics[width=75mm,height=55mm]{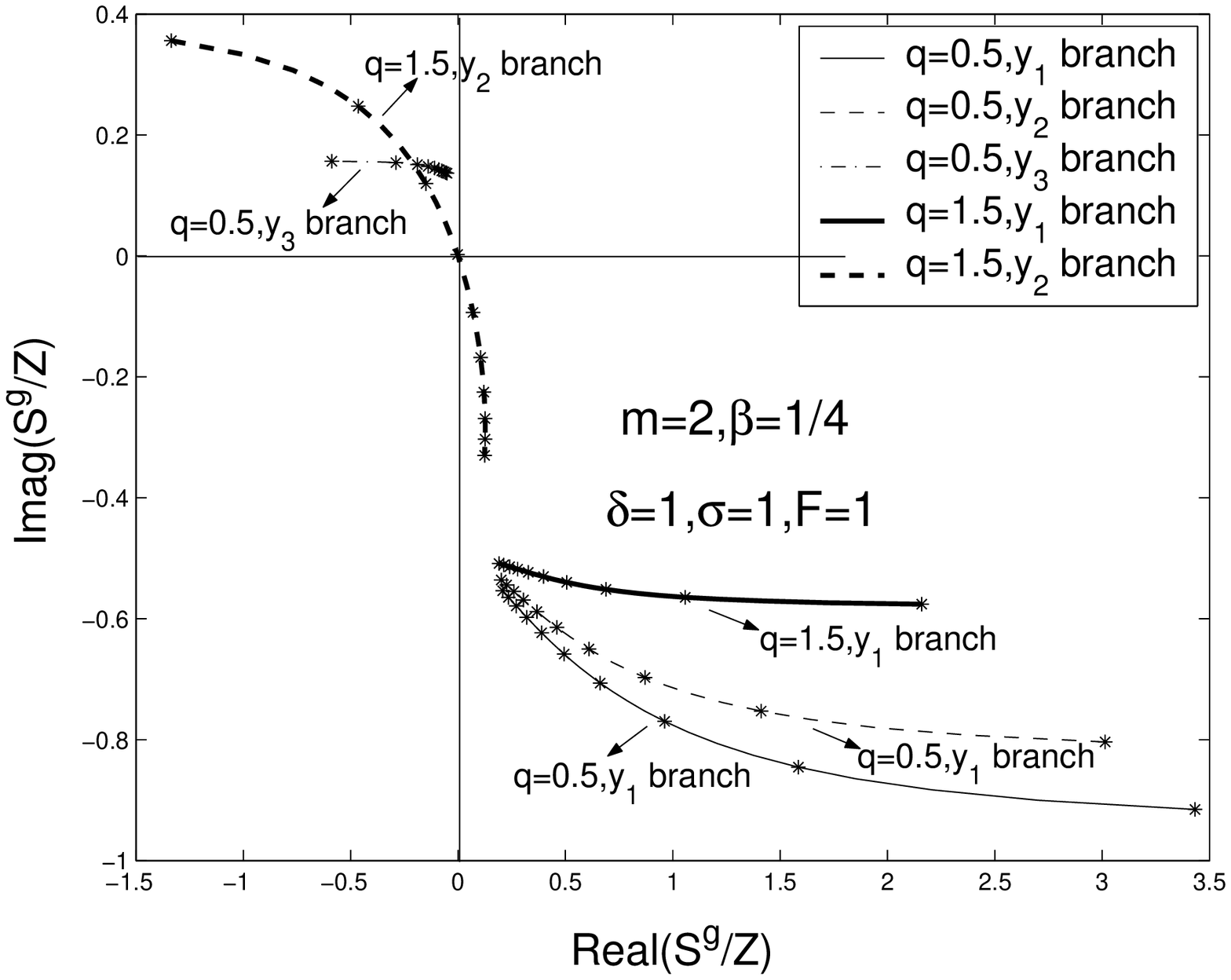}}
    \hspace{1in}
    \subfigure[$m=2,\ \beta=1/4,\ \delta=1,\ \sigma=3,\ F=1$]{
      \label{fig:sz2b}
      \includegraphics[width=75mm,height=55mm]{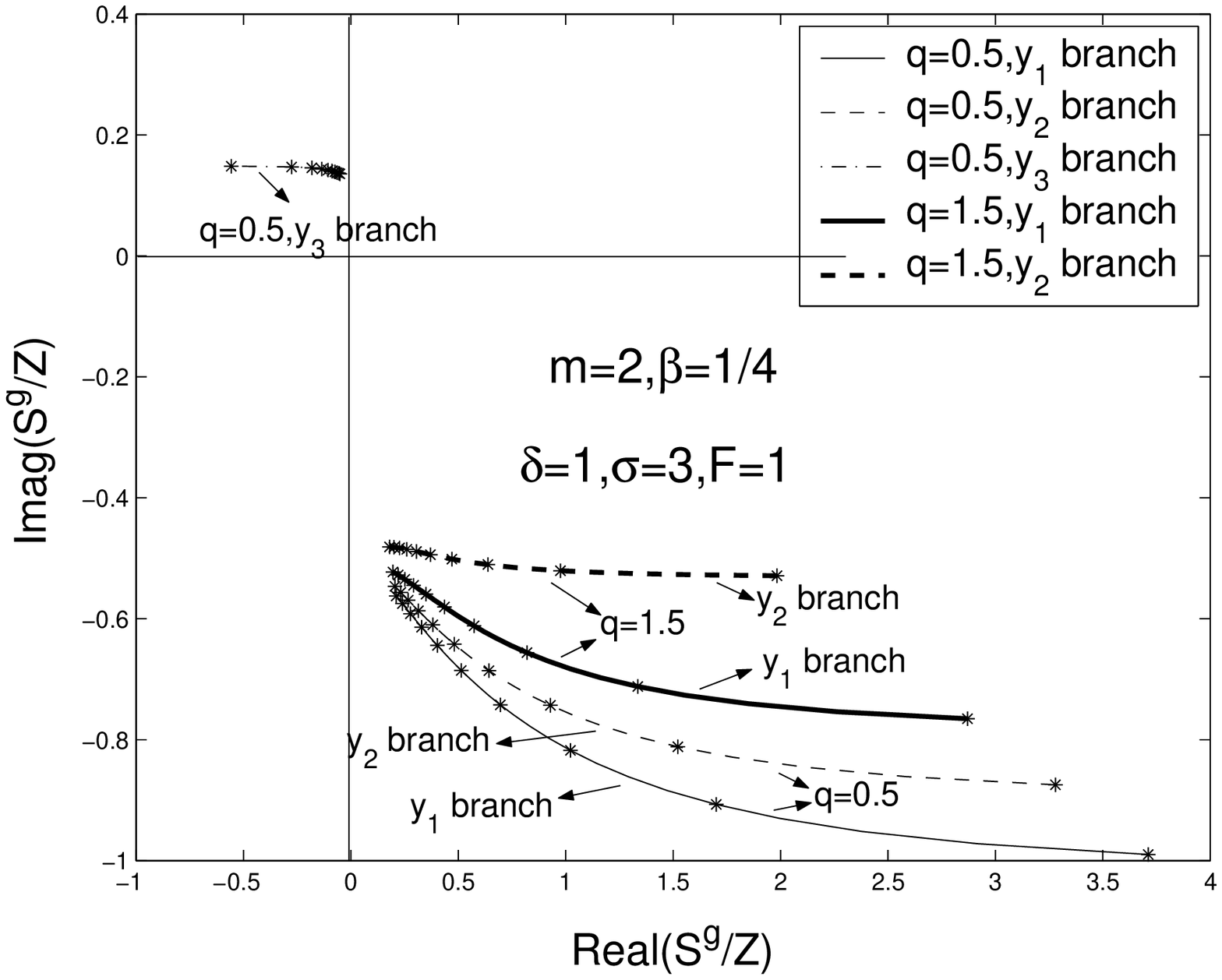}}
  \caption{The phase relation curves between the perturbed azimuthal
  magnetic field and surface mass density versus $\sigma$ for
  logarithmic spirals with $m=2,\ \beta=1/4$. We show curves of full
  discs ($F=1$) with $\sigma=1$ and $\sigma=3$ in panels (a) and (b).
  We also add the physical $y_3$ branch. When $q=1.5$, the $y_3$
  branch is very far away from the other curves and we have omitted
  this curve for the compactness of presentation. We start from
  $\xi=0.4$ and use $*$ to label an interval of 0.4 in the variation
  of $\xi$. The phase points moves from right to left for the $y_1$
  and $y_2$ branches and from left to right for the $y_3$ branch as
  $\xi$ increases.
  }\label{fig:sz2}
\end{figure}

We now examine the phase relationship between the perturbed gas
surface mass density and the perturbed magnetic field as these
aspects may provide useful diagnostics for observations. Since both
$S^g/(\mathrm{i}R)$ and $S^g/Z$ are complex in general, we show such
phase relation in the complex plane. In Fig.\ \ref{fig:19} and
Fig.\ \ref{fig:20} for the complex ratio $S^g/(\mathrm{i}R)$, the
variation of $\xi$ is labelled by an asterisk $*$ with an interval of
$\Delta\xi=0.3$ starting from the real axis. According to equation
(\ref{eq:94}), the $S^g/(\mathrm{i}R)$ ratio is real for $\xi=0$ and
thus the curves start from the real axis. In Fig.\ \ref{fig:sz1} and
Fig.\ \ref{fig:sz2} for $S^g/Z$, the variation of $\xi$ is labelled
by an asterisk $*$ with an interval of $\Delta\xi=0.4$. According to
equation (\ref{eq:92}), $S^g/Z\rightarrow\infty$ as $\xi\rightarrow0$.
Equation (\ref{eq:92}) also tells us that the phase relations of the
azimuthal magnetic field perturbation and the gas surface mass density
are totally opposite for leading ($\xi>0$) and trailing ($\xi<0$)
spiral MHD density waves.

In the case of $\beta=-0.24$, only the $y_1$ and $y_2$ branches are
physical, and we show these two branches in Fig.\ \ref{fig:19} and
Fig.\ \ref{fig:sz1}. The main trend is that with the increase of
the radial wavenumber $\xi$, the real part of $S^g/(\mathrm{i}R)$
is positive and does not vary much as the imaginary part decreases
steadily from zero. The result is that for tight-winding spiral
arms with large $\xi$, the radial component of the magnetic field
perturbation becomes almost in phase with the gas density perturbation
(i.e., $S^g/R\rightarrow$ a positive number), while for a small $\xi$,
the radial component of the magnetic field perturbation lags behind
the perturbed gas density by $\sim\pi/2$, a trend similar to the
aligned cases. The azimuthal component of the magnetic field
perturbation $Z=\xi\mathrm{i}R/m$ remains always ahead of the radial
component of the magnetic field perturbation by $\pi/2$ in phase. We
note that for small $\xi$ the imaginary part of $S^g/Z$
approaches a constant as its real part moves to infinity. Meanwhile,
the azimuthal magnetic field perturbation can be much larger than the
radial magnetic field perturbation for large $\xi$. For a composite
system of partial scale-free discs with a coplanar magnetic field, we
note the following two points.
First, by decreasing the potential parameter $F$, the $y_2$ branch
moves towards right while the $y_1$ branch does not move much.
Secondly, for a large magnetic parameter $q=3$, singularity occurs
in the $y_1$ branch as $F$ decreases. This singularity can be
determined from equation (\ref{eq:94}), namely
\begin{equation}
\begin{split}
\frac{[D_g^2+1+(4\beta-1)q^2/(4\beta+2)]
\mathcal{N}_m\delta}{\mathcal{C}(1+\delta)}
\bigg(1+\frac{S^s}{S^g}\bigg)
\qquad\\ \qquad
-1+(1+2\beta)D_g^2=0\ ,\label{eq:95}
\end{split}
\end{equation}
where $F$ parameter is implicitly contained in coefficient $\mathcal{C}$.

The $y_3$ branch can be positive in the case of $\beta=1/4$ (see
Fig.\ \ref{fig:14} for this SMDW mode); we therefore include the
phase information of the $y_3$ branch in Fig.\ \ref{fig:20} and
Fig.\ \ref{fig:sz2}. Different from the $y_1$ and $y_2$ branches,
the $y_3$ branch has a $S^g/(\mathrm{i}R)$ characterized by an
almost negative real part and an imaginary part increasing
(rather than decreasing) steadily with the increase of $\xi$.
The phase relation of this $y_3$ branch has a phase difference
of $\sim\pi$ relative to the other two branches. For $S^g/Z$, the
$y_3$ branch still lies in the second quadrant and moves from left
to right and tends to the upper imaginary axis. It indicates that
for this branch the azimuthal magnetic perturbation lags behind the
gas surface mass density perturbation by $\sim\pi$ for small $\xi$
and by $\sim\pi/2$ as $\xi$
becomes fairly large. There is yet another salient feature. For a
small $\sigma=1$ and a large $q=1.5$, the $y_2$ branch appears
fairly special. In panel (a) of Fig.\ \ref{fig:20}, starting from a
negative real value as $\xi$ increases, the imaginary part first
increases and then decreases, meanwhile the real part increases with
increasing $\xi$; the curve intersects the real axis again at the
origin exactly. It then keeps this tendency such that the imaginary
part falls down steadily as $\xi$ increases. In terms of phase
difference between $S^g$ and $R$ or $S^g$ and $Z$, this branch is
special in that there is a phase shift of $\pi$ at some $\xi_c$.
This point corresponds to $S^g=0$, also related to the phase shift
of $S^g/S^s$ discussed earlier.
For cases of partial discs with $F=0.5$, such `singularity'
no longer exists and all curves are continuous. Meanwhile,
such $\sim\pi$ phase shift of $y_2$ branch in the presence
of a strong magnetic field no longer exists either.

\subsection{Marginal Stability Curves for
\\ \ \qquad Axisymmetric Disturbances}

The $m=0$ `spiral' case is very special. It is considered as a
purely radial oscillation in a composite MHD disc system and is
related to part of the MHD disc stability problem. In a general
disc problem, the dispersion relation for time-dependent MHD
perturbations involves a $\omega^2$ term. A linearly stable disc
system requires $\omega^2\geq0$. In our analysis, we set $\omega=0$
to obtain stationary patterns and to also meet the requirement
of the scale-free condition. In the $m=0$ case, the stationary
dispersion relation with $\omega=0$ actually corresponds to
the marginal stability curve \citep{b9,b10a,b10,b7,b7a}.
Meanwhile to treat the problem properly, a different limiting
process should be taken similar to the aligned $m=0$ case.
That is, we first set $m=0$ with $\omega\neq 0$ and then
let $\omega\rightarrow0$ in the relevant MHD perturbation
equations. The same as before, we begin with equations
$(\ref{eq:48})-(\ref{eq:52})$. Equation (\ref{eq:48}) requires
$U\propto r^{1/2+2\beta+\mathrm{i}\xi}$ and then equation
(\ref{eq:50}) gives $J\propto r^{1/2+\beta+\mathrm{i}\xi}$.
Substituting these two expressions into equation (\ref{eq:49}),
we see that the scale-free condition cannot be met unless
$\omega=0$ where $U=0$ with $U/\omega\neq0$. In the process of
taking the limit of $\omega\rightarrow0$, we also remove terms
that are not scale free. We show the result of this limiting
process below.
\begin{equation}
\begin{split}
&r^2(\omega^2-\kappa_g^2)S^g=(\xi^2+1/4)\Sigma_0^g\Psi^g
+(\xi^2+1/4-\beta)C_A^2S^g\ ,\\
&r^2(\omega^2-\kappa_s^2)S^g=(\xi^2+1/4)\Sigma_0^s\Psi^s\ .
\end{split}\label{eq:96}
\end{equation}
A substitution of equations (\ref{eq:35}) and (\ref{eq:76})
into equation (\ref{eq:96}) with $\omega=0$ yields
\begin{equation}
\begin{split}
&[\mathcal{A}^{S'}_0a_g^2+\kappa_g^2r^2+(\mathcal{A}^{S'}_0-\beta)C_A^2
-2\pi Gr\mathcal{A}^{S'}_0\mathcal{N}_0\Sigma_0^g]S^g\\
&\qquad\qquad\qquad\qquad\qquad\qquad
-2\pi Gr\mathcal{A}^{S'}_0\mathcal{N}_0\Sigma_0^gS^s=0\ ,\\
&[\mathcal{A}^{S'}_0a_s^2+\kappa_s^2r^2-2\pi
Gr\mathcal{A}^{S'}_0\mathcal{N}_0\Sigma_0^s]S^s\\
&\qquad\qquad\qquad\qquad\qquad\qquad
-2\pi Gr\mathcal{A}^{S'}_0\mathcal{N}_0\Sigma_0^sS^g=0\ ,\\
\end{split}\label{eq:97}
\end{equation}
where we define
\begin{equation*}
\mathcal{A}^{S'}_0\equiv\xi^2+1/4
\end{equation*}
and the corresponding notation
\begin{equation*}
\mathcal{H}^{S'}_0\equiv {\mathcal{A}^{S'}_0
\mathcal{N}_0}/{\mathcal{C}}+\mathcal{B}_0\ .
\end{equation*}
The difference between $\mathcal{A}^S_0$ and $\mathcal{A}^{S'}_0$
is that the latter does not contain the term $2\beta$; this is
exactly the result of different limiting processes.

The two relations in equation (\ref{eq:97})
cannot be satisfied unless the coefficient
determinant vanishes, which leads to
\begin{equation}
\begin{split}
&[\mathcal{A}^{S'}_0a_g^2+\kappa_g^2r^2+(\mathcal{A}^{S'}_0
-\beta)C_A^2-2\pi Gr\mathcal{A}^{S'}_0\mathcal{N}_0\Sigma_0^g]
\\ & \qquad\qquad
\times[\mathcal{A}^{S'}_0a_s^2+\kappa_s^2r^2-2\pi
Gr\mathcal{A}^{S'}_0\mathcal{N}_0\Sigma_0^s]
\\ & \qquad\qquad\qquad\qquad\qquad\qquad
=(2\pi Gr\mathcal{A}^{S'}_0\mathcal{N}_0)^2\Sigma_0^g\Sigma_0^s\ .
\end{split}\label{eq:98}
\end{equation}
Substitutions of equations $(\ref{eq:20})-(\ref{eq:23})$ lead
to the explicit expression of this stationary dispersion relation
(\ref{eq:98}) in terms of physical parameters in the form of
\begin{equation}
\begin{split}
&\bigg\{\frac{\mathcal{A}^{S'}_0}{(1+2\beta )}+2(1-\beta)D_g^2
+\frac{q^2(\mathcal{A}^{S'}_0-\beta)}{(1+2\beta )}\\
&-\mathcal{A}^{S'}_0\mathcal{N}_0F
\frac{[D_g^2+1+(4\beta-1)q^2/(4\beta+2)]\delta}
{2\beta\mathcal{P}_0(1+\delta)}\bigg\}\\
&\times\bigg[\frac{\mathcal{A}^{S'}_0}{1+2\beta}+2(1-\beta)D_s^2
-\frac{\mathcal{A}^{S'}_0\mathcal{N}_0F(D_s^2+1)}
{2\beta\mathcal{P}_0(1+\delta)}\bigg]\\
&=\bigg[\frac{\mathcal{A}^{S'}_0\mathcal{N}_0F}
{2\beta\mathcal{P}_0(1+\delta)}\bigg]^2
\bigg[D_g^2+1+\frac{(4\beta-1)q^2}{(4\beta+2)}\bigg](D_s^2+1)\delta\ .
\end{split}\label{eq:99}
\end{equation}
By rearranging equation (\ref{eq:99}), we obtain a quadratic
equation for the marginal stability in terms of $y=D_s^2$, namely
\begin{equation}
C^M_2y^2+C^M_1y+C^M_0=0\ ,\label{eq:100}
\end{equation}
where
\begin{equation}
\begin{split}
&C^M_2\equiv\mathcal{B}_0\mathcal{H}^{S'}_0\sigma\ ,\\
&C^M_1\equiv
\bigg[(\mathcal{B}_0-\mathcal{A}^{S'}_0)\mathcal{H}^{S'}_0
+\frac{(\mathcal{A}^{S'}_0+\mathcal{B}_0)
(\mathcal{H}^{S'}_0-\mathcal{B}_0)}{(1+\delta)}\bigg]\sigma
\\ &\qquad
-\frac{(\mathcal{A}^{S'}_0+\mathcal{B}_0)
(\mathcal{H}^{S'}_0+\mathcal{B}_0\delta)}{(1+\delta)}
\\ &\qquad
-\frac{(\mathcal{A}^{S'}_0+\mathcal{B}_0-4\beta+3)
(\mathcal{H}^{S'}_0\delta+\mathcal{B}_0)}{(1+\delta)}q^2\ ,
\\ &
C^M_0\equiv\bigg[-\mathcal{A}^{S'}_0\mathcal{H}^{S'}_0
+\frac{(\mathcal{A}^{S'}_0+\mathcal{B}_0)
(\mathcal{H}^{S'}_0-\mathcal{B}_0)}{(1+\delta)}\bigg]\sigma
\\ &\qquad
+(\mathcal{A}^{S'}_0+\mathcal{B}_0)\bigg[\mathcal{A}^{S'}_0
+\mathcal{B}_0-\frac{\mathcal{H}^{S'}_0
+\mathcal{B}_0\delta}{1+\delta}\bigg]
\\ &
+(\mathcal{A}^{S'}_0+\mathcal{B}_0-4\beta+3)
\bigg[\mathcal{A}^{S'}_0+\mathcal{B}_0
-\frac{\mathcal{H}^{S'}_0+\mathcal{B}_0\delta}{1+\delta}\bigg]q^2\ .\\
\end{split}\label{eq:101}
\end{equation}

\begin{figure}
    \centering
    \subfigure[$m=0,\ \beta=-1/8,\ \delta=1,\ \sigma=5,\ F=1$]{
      \label{fig:21a}
      \includegraphics[width=75mm,height=55mm]{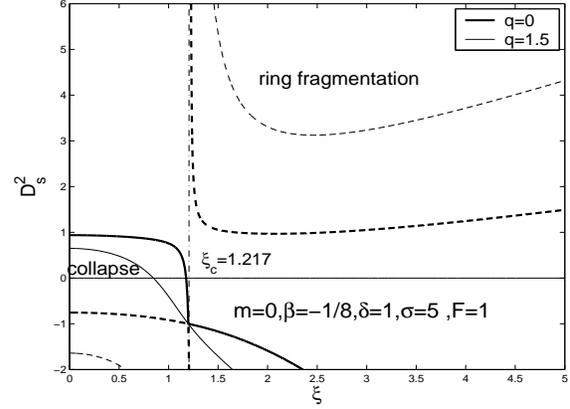}}
    \hspace{1in}
    \subfigure[$m=0,\ \beta=-1/8,\ \delta=1,\ \sigma=5,\ F=0.6$]{
      \label{fig:21b}
      \includegraphics[width=75mm,height=55mm]{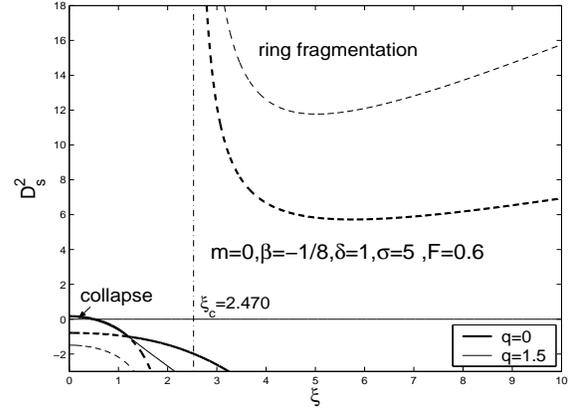}}
  \caption{The marginal stability curves for the case of $m=0$,
  $\beta=-1/8$, $\delta=1$, $\sigma=5$.
  For comparison, we show the cases $q=0$ (boldface curves) and
  $q=1.5$ (light curves) in one figure. Note that for a
  full disc ($F=1$) and a partial disc ($F=0.6$), the marginal
  stability curves behave quite differently.}\label{fig:21}
\end{figure}

\begin{figure}
    \centering
    \subfigure[$m=0,\ \beta=1/4,\ \delta=1,\ \sigma=5,\ F=1$]{
      \label{fig:21a}
      \includegraphics[width=75mm,height=55mm]{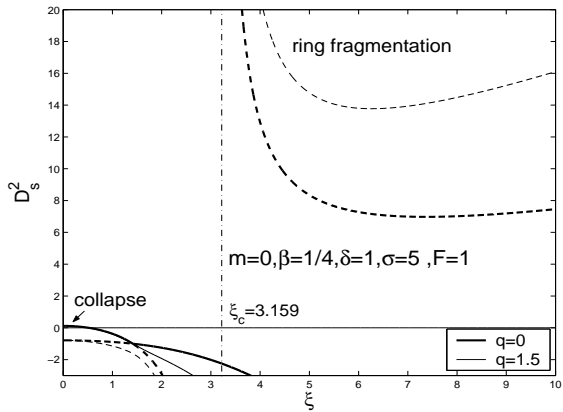}}
    \hspace{1in}
    \subfigure[$m=0,\ \beta=1/4,\ \delta=1,\ \sigma=5,\ F=0.6$]{
      \label{fig:21b}
      \includegraphics[width=75mm,height=55mm]{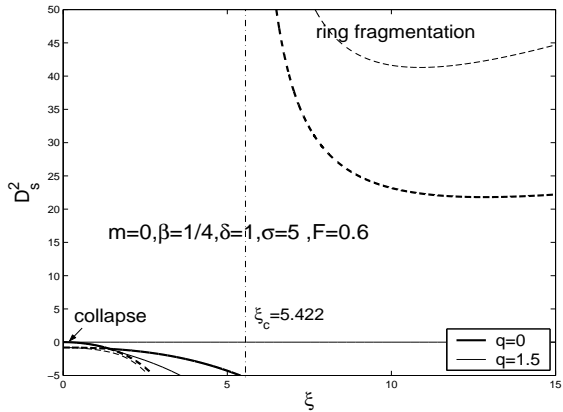}}
  \caption{The marginal stability curves for the case $m=0$,
  $\beta=1/4$, $\delta=1$, $\sigma=5$. For both $F=0.6$ and $F=1$,
  the two cases of $q=0$ and $q=1.5$ are computed. Compared with
  a smaller $\beta$, the ring fragmentation regions rise up while
  the MHD Jeans collapse regions shrink.}\label{fig:22}
\end{figure}

One necessary check for a consistent derivation is to set $q=0$
and the coefficients (\ref{eq:101}) reduce to those in the case
of unmagnetized coupled scale-free discs \citep{b10}. Equation
(\ref{eq:100}) is quadratic in $D_s^2$ and has two solution
branches corresponding to the marginal stability curves. For
several chosen sets of parameters, we demonstrate these $D_s^2$
curves in Fig.\ \ref{fig:21} and Fig.\ \ref{fig:22}. The unstable
regime consists of two regions. In the upper-right corner, the two
coupled discs rotate too fast to be stable and are susceptible to
the MHD ring fragmentation instability, while in the lower-left
corner the two coupled discs rotate too slow to resist large-scale
Jeans collapse modified by rotation and magnetic field
\citep{b3b,b13,b12,b5,b5c,b6,b6a,b10,b11,b7,b7a}. The marginal curves
represent precise global treatments of axisymmetric linear stability.
In the WKBJ or tight-winding regime, one can readily derive the local
dispersion relation for a single MHD disc and the local dispersion
relation for a composite system of gravitationally coupled discs in the
WKBJ regime can be expressed in a parallel form as equation (\ref{eq:98}).
As discussed for a single scale-free MHD disc by \citet{b11}, the WKBJ
approximation is well justified for short wavelength perturbations
(i.e., large $\xi$) but not as good for long wavelength perturbations.

We comment on the MHD ring fragmentation instability here. 
In fact, this is the MHD generalization of the well-known Toomre 
instability characterized by the so-called $Q$ parameter 
\citep{b8a,b13a}. Besides in contexts of single discs, there have 
been numerous studies to identify such an effective $Q$ parameter 
for a composite disc system \citep{b1z,b2c,b5b} and in a single 
coplanar magnetized SID \citep{b5a}. \citet{b12} noted that in 
a single SID, the minimum of the ring fragmentation branch is 
effectively related to the Toomre $Q$ parameter. For a coplanar
magnetized SID, \citet{b5} and \citet{b5c} emphasized that the 
MHD ring fragmentation branch is also closely related to the
generalized $Q_M$ parameter \citep{b5a}. The conceptual 
connection of two instabilities comes from their dependence 
on radial wavenumber. Physically, \citet{b9} suggested a 
straightforward $D$ criterion to determine the axisymmetric 
stability against ring fragmentation. In our composite
scale-free discs with a coplanar magnetic field, we therefore
interpret this MHD ring fragmentation as an extension of the
Toomre instability, with $D_s$ parameter being used for the 
stability criterion. In addition, this $D_s$ parameter 
describes a global MHD instability criterion, while the $Q$ 
parameter is only defined locally.

When $C_2^M=0$ occurs for $\xi=\xi_{c}$, one branch of the stability
curves diverges at $\xi=\xi_{c}$. For a specified $\beta$ value, the
divergent point $\xi_{c}$ depends only on $F$ parameter contained in
the coefficient $\mathcal{C}$.

For hydrodynamic perturbations in coupled scale-free discs without
magnetic field, \citet{b10} have discussed the results thoroughly
and their figures clearly show the influence of the two parameters
$\sigma$ and $\delta$ (see their Fig.\ 4 to Fig.\ 8), for example,
the increase of $\sigma$ or $\delta$ lowers both the collapse and
ring fragmentation regions.
In other words, the increase of $\sigma$ or $\delta$ reduces the
danger of the Jeans collapse instability but makes the composite
system more vulnerable to the ring fragmentation instability. By
our exploration in the presence of magnetic field, this trend of
variation remains. For this reason, we shall not repeat the analysis
in this regard similar to theirs. We mainly focus on the case of
$\delta=1,\ \sigma=5$ and show the marginal stability curves for
$q=0$ and $q=1.5$ in one figure. Similar to the study of \citet{b11}
where global coplanar MHD perturbation structures in a single
scale-free gaseous disc with a coplanar magnetic field are analyzed,
we arrive at the same conclusion in our model of a composite system
of coupled scale-free discs. The enhancement of a coplanar magnetic
field reduces the danger of ring fragmentation in an apparent manner;
and at the same time, it only suppresses Jeans collapse instabilities
for the $\beta$ range of $-1/4<\beta<1/4$ and tends to aggravate Jeans
collapse instabilities for the $\beta$ range of $1/4<\beta<1/2$. As
has been already mentioned in \citet{b11}, this collapse feature can
be understood from the coupling between the surface mass density and
the coplanar magnetic field of the background rotational MHD equilibrium.

We also examine global MHD perturbation structures in a composite system
of partial scale-free discs with the potential factor $F=0.6$. It is
clear from Fig.\ \ref{fig:21} and Fig.\ \ref{fig:22} that the presence of
an axisymmetric dark matter halo suppresses both the ring fragmentation
and Jeans collapse instabilities. This results from the fact that the
dark matter halo exists as a stationary massive gravitational background.
It can reduce the ring fragmentation because its gravitational force
acts as centripetal force to resist such danger. It can suppress
Jeans collapse because its stationary background supports the
stationary structure of the discs \citep{b7e,b2b2,b7f,b1}.

\section{Summary and discussion}

From several perspectives, this paper represents a
generalization and extension of previous works
[\citet{b13,b12,b3,b5,b5c,b6a,b7,b7a,b10,b11}]. We consider a composite
disc system consisting of gravitationally coupled stellar and gas
scale-free discs with a coplanar magnetic field and construct global
stationary coplanar MHD perturbation structures within such a composite
disc system. More specifically, such discs are treated as razor-thin
and scale-free so that the entire problem is two-dimensional with all
background quantities varying in power laws of radius $r$. For the
background rotational MHD equilibrium, the surface mass density
scales as $\Sigma_0\propto r^{-\alpha}$, the rotation curve scales
as $v_0\propto r^{-\beta}$ and the azimuthal magnetic field scales
as $B_0\propto r^{-\gamma}$. The barotropic equation of state is
expressed as $\Pi=K\Sigma^n$. By the scale-free condition, the
indices $\alpha,\ \beta,\ \gamma,\ n$ are all expressed in terms of
$\beta$ by equation (\ref{eq:19}). The two discs are gravitationally
coupled through the Poisson integral and the scale-free condition
requires that the physical quantities of the two discs should share
the same radial scalings with different proportional coefficients
allowed. Meanwhile, the potential factor $F$ for a composite partial
disc system is included to represent the ratio of the gravitational
potential arising from the composite disc system to that arising
from the entire system including a massive axisymmetric dark matter
halo mass distribution which is assumed unresponsive
to perturbations in the
composite disc system. For convenience and clarity, several
dimensionless constant parameters are introduced. The first one is
the effective rotational Mach number $D$ (we omit either subscript
or superscript `$g$' or `$s$' for simplicity here) defined as the
ratio of the disc rotation speed to the disc sound speed yet with
an additional scaling factor $(1+2\beta)^{-1/2}$. In essence, this
is a measure for the disc rotational velocity. The second one is
$q$ parameter corresponding to the ratio of the Alfv\'en wave speed
to the gas sound speed which provides a measure for the strength of
the coplanar magnetic field. The other two parameters are $\delta$
and $\eta$ which represent the ratio of surface mass densities of
the two discs and the square of the sound speed ratio (the velocity
dispersion in the stellar disc mimics the sound speed in our
formulation) of the two discs, respectively. For the convenience
of figure presentation, we also define $\sigma\equiv1/\eta$. These
parameters completely characterize the behaviour of a composite disc
system for the background equilibrium and are independent of each
other; to a greater extent, they can be freely chosen but within a
certain bound. In our model, there is a wide range of freedom to
choose these parameters.
Our goal is to study stationary coplanar MHD perturbations to
construct global aligned and unaligned configurations described by
the relevant dispersion relations containing all these parameters
and to analyze partially the linear instability problem of such
a composite disc system.

Our model is highly idealized in several aspects, nevertheless,
these simplifications are sensible and useful in making the problem
tractable. First, we adopt the razor-thin approximation for the
composite disc system so that vertical variations across the disc
system are ignored [e.g., \citet{b8,b13,b2a,b12,b5b}]. This is
justifiable so long as the radial scale of a galactic disc is much
larger than the disc thickness. Secondly, we simplify a galactic
system to just three major constituent components: a magnetized gas
disc, a stellar disc (treated as a fluid) and an axisymmetric dark
matter halo. In reality, there exists another component of cosmic
rays, important from the diagnostic perspective for observing
large-scale radio structures as revealed by synchrotron emissions
from relativistic electrons gyrating around the galactic magnetic
field \citep{b5d}.
Technically, we need to deal with subsystems in a
piece-wise manner and different subsystems will
manifest different aspects of the physical problem.
In our model formalism, we mainly focus on the magnetized
interstellar gases and the stars in a galactic disc system
in the presence of an axisymmetric dark matter halo.
This is typical for disc galaxies including our own Milky-Way.
Thirdly, we adopt a simple fluid description for both the stellar and
magnetized gas discs. In principle, a distribution function approach
to the stellar disc would be more accurate and comprehensive.
However, for a large-scale dynamics without resonances etc.,
a fluid approximation suffices \citep{b4,b1,b0d,b6,b7,b10}.
Fourthly, the basic problem is formulated here within the scale-free
framework. Now all quantities are required to vary with power-law
scalings in $r$, and the barotropic equation of state must be
imposed [e.g., \citet{b13,b10,b11}]. The scale-free condition
further requires that the two coupled discs share the same radial
scalings as well as the same barotropic index [e.g., \citet{b10}].
Another major constraint is that such scale-free configurations must
be stationary with $\omega=0$ [e.g.,
\citet{b5,b5c,b6,b6a,b7,b7a,b10,b11}]. Nevertheless, the model of
coupled scale-free discs represents a significant extension of
isothermal discs with $\beta=0$. In particular, global MHD
perturbation structures can be constructed with the relevant MHD
dispersion relations determined analytically. Fifthly, the
background magnetic field is presumed to be azimuthal and bears a
radial power-law scaling $r^{-\gamma}$. As the magnetic field
distribution in disc galaxies is generally complex, either the
isopedic or the coplanar magnetic field geometry represents an
idealization. Model analysis in terms of these two `orthogonal'
geometries provide a theoretical basis for physical intuitions about
magnetic field effects.
Our current model focuses on the case of a coplanar magnetic field,
while for the case of an isopedic magnetic field, the interested
reader is referred to \citet{b12,b6a} and \citet{b13e}.

Our model makes advances in the research in the MHD disc problem.
By including two coupled scale-free discs with a coplanar magnetic
field, our model contains various model ingredients of previous
models of \citet{b13,b12,b5,b5c,b7,b7a,b10,b11}.
In a fluid-magnetofluid composite disc system with a coplanar
magnetic field, we construct both aligned and spiral global MHD
perturbation configurations and they are in accordance with
previous works and meanwhile there are several new features in
our model. With such an analysis, the MHD dispersion relations
are derived analytically without any approximation, meanwhile,
the axisymmetric marginal stability criteria is also obtained
analytically. Our model provides a new perspective for
understanding large-scale MHD structures in disc galaxies. We
now summarize the main results in the following.

(\textrm{i})\emph{ Aligned Stationary MHD Configurations}

For the aligned cases, we derive the MHD dispersion relation and
provide ample numerical examples concerning global MHD perturbation
configurations and the corresponding phase relations among
perturbation variables. In our fluid-magnetofluid model of two
coupled scale-free discs, the MHD dispersion relation is a cubic
equation in terms of $y=D_s^2$ and there are three $y=D_s^2$
solution branches in general. By our analysis, the existence of the
three branches is due to the gravitational coupling of perturbations
in the two discs and the interaction of the magnetic field with the
disc system [see the dispersion relation of equation (\ref{eq:64})]
and they correspond to purely azimuthal propagation of
super fast MHD density waves (sFMDWs), fast MHD density waves (FMDWs)
and slow MHD density waves (SMDWs), respectively. Generally speaking,
the $y_1$ branch (sFMDW) is always positive and the $y_3$ branch
(SMDW) is more often negative thus unphysical.
Our interest lies mainly on the $m=2$ (i.e., bar-like)
case and we discuss the influence of several parameters.

%
%
%
%

The magnetic field serves as a centripetal force when $\beta<1/4$
(i.e., magnetic tension force is stronger than the magnetic pressure
force) and as a centrifugal force when $\beta>1/4$ (i.e., magnetic
tension force is weaker than the magnetic pressure force) for the
background MHD rotational equilibrium state. In the presence of
coplanar MHD perturbations, there is a gravitational interaction between
the two discs and there are also MHD interactions within the magnetized
gas disc. For cases of $q^2\ll 1$, the magnetic field has very weak
influence on the composite disc system. The $y_3$ branch becomes trivial.
Our $y_1$ and $y_2$ branches correspond to solutions of \citet{b10}.
While for $q\geq1$ (i.e., the Alfv\'en wave speed is comparable to
or larger than the sound speed in the gas disc), the behaviour of
solution branches can be totally different from the unmagnetized
cases of \citet{b10} and magnetic field effect can become dominant,
especially for small $\sigma$ (e.g., $\sigma\sim1$). In the presence
of magnetic field, we observe a new feature that as $\beta$ and $q$
become large enough, there appears a `convergent point' where for
different $\delta$ values some solution branches converge and then
their relative positions are changed. Under the assumption that
$\eta\leq1$ (or $\sigma\geq1$, the gas sound speed is less than
the velocity dispersion of the stellar disc) and when $\beta<1/4$, we
have $D_s<D_g$ according to equation (\ref{eq:22}); the magnetized
gas disc rotates faster and the existence of magnetic field enlarges
the rotational speed difference between the gas and stellar discs.
Yet for a relatively strong magnetic field and for
$\beta>1/4$, there are solutions with $D_s>D_g$ and the stellar
disc rotates faster instead (see Fig.\ \ref{fig:4}).

In the aligned cases, the phase relations among perturbation
variables are relatively simple because all quantities are real and
they must be either in phase or out of phase. For the $y_1$ branch,
the perturbation mass densities in the two discs are always out of
phase while for the other two branches $y_2$ and $y_3$, the
perturbation mass densities are generally in phase. However for
the $y_2$ branch with a strong magnetic field, the perturbation
mass densities can also become out of phase. For the radial component
of the magnetic field perturbation, $\mathrm{i}R$ and $S^g$ are always
in phase for the $y_1$ branch; they are in phase for the $y_2$
branch and out of phase for the $y_3$ branch in many cases but on the
whole this phase relation depend on both $q$ and $\sigma$ parameters
(see Fig.\ \ref{fig:10}). Since $Z=(4\beta-1)\mathrm{i}R/2m$, the
phase relation between gas mass density perturbation and the azimuthal
component of the magnetic field perturbation is similar to that of
$S^g$ and $\mathrm{i}R$ when $\beta>1/4$ and has an additional phase
shift $\pi$ for $\beta<1/4$. When $\beta=1/4$ there is no azimuthal
magnetic field perturbation.


We also note that when $\beta$ becomes sufficiently small (e.g.,
$\beta=-0.24$) with a mild magnetic field strength, there exists
only one solution branch and the other two are complex conjugates.

Regarding $m$ dependence, there are not many new features except
for $m=1$. In this case, the solutions are quite different from
$m=2$ modes. In particular, it is now possible that the solution
branches intersect and also there can be no physical $D_s^2$
solutions.

(\textrm{ii})\emph{Stationary Configurations of Logarithmic Spirals}

The MHD dispersion relation for the spiral cases is quite similar
to that for the aligned cases in form. Differences only occur by
a systematic replacement of notations. Therefore, there are many
common features in solution behaviours. Physically, logarithmic
spirals involve both radial and azimuthal propagations of MHD
density waves, while the aligned cases involve purely azimuthal
propagations of MHD density waves \citep{b5}.

In the $m=2$ case, the $y_3$ branch is due to the existence of
magnetic field and seems to bear little relation with the spiral
pattern (i.e., it almost does not vary much for arbitrary $\xi$).
This SMDW is in most cases negative thus unphysical. Those physical
SMDWs in the spiral cases appear when $\beta$ is not too small (i.e.,
$\beta\geq0$) and $\sigma$ is small enough (i.e., $\sigma=1$).
The $y_1$ and $y_2$ solution branches are monotonically increasing
with radial wavenumber $\xi$, suggesting that to maintain a stationary
tight-winding logarithmic spiral pattern, a fast disc rotation speed
is required for sFMDWs or FMDWs. The increase of the magnetic field
strength generally raises the solution branches, while a large
$\sigma$ tends to lower the solution branches. The `convergent point'
appears more often in the logarithmic spiral cases as $\xi$ varies.

We now describe the phase relations among perturbation variables for
logarithmic spirals. The phase relation between perturbation surface
mass densities remain either in phase or out of phase, and the
behaviour is similar to those in the aligned cases, but the phase
relation between surface mass density perturbation and the magnetic
field perturbation becomes fairly involved. The dependence of
$S^g/(\mathrm{i}R)$ and $S^g/Z$ on $\xi$ is described by a complex
expression. Now the azimuthal magnetic field perturbation is
$Z=\xi\mathrm{i}R/m$ which is quite different from that of the
aligned cases and is dependent on the sign of $\xi$ (i.e., for
leading waves and trailing waves the azimuthal magnetic perturbation
have a phase difference of $\pi$). In the case of $\xi=0$ for $y_1$
and most of $y_2$ branches, the ratio $S^g/(\mathrm{i}R)$ is
positive, i.e., the radial magnetic field perturbation lags behind
the gas surface mass density perturbation by $\pi/2$ in phase (no
azimuthal magnetic perturbation). As $\xi$ increases, the imaginary
part of this ratio decreases in general while its real part does not
vary significantly, which indicates that the phase of the radial
magnetic field perturbation is catching up with the gas surface mass
density perturbation (meanwhile the azimuthal magnetic perturbation
becomes ahead of $S^g$ from the original in-phase relation). In the
tight-winding approximation (i.e., $\xi\rightarrow\infty$), $R$ and
$S^g$ are in phase and the phase of azimuthal magnetic perturbation
tends to be $\sim\pi/2$ ahead of $S^g$. We also find that in the
$\beta=1/4$ case, for a small $\sigma=1$ and a large $q=1.5$, there
is an abrupt phase shift of $\pi$ in $y_2$ branch as $\xi$ increases
from 0. In the physical $y_3$ branch for SMDWs, the phase relation
between $S^g$ and $R$ or $S^g$ and $Z$ has an approximate 
difference of $\pi$ compared with that of the $y_1$ branch.

(\textrm{iii})\emph{Axisymmetric Marginal 
Stabilities and Magnetorotational Instabilities}

We explore the unaligned $m=0$ perturbations referred 
to as the marginal stability curves where two regimes 
corresponding to the Jeans collapse instability and 
the ring fragmentation instability appear. 
The criterion for the global ring fragmentation instability 
is an MHD generalization of the local Toomre Q-criterion.

In our fluid-magnetofluid model, we arrive at the same conclusions
as discussed by Shen \& Lou (2004b) and Shen et al. (2005)
that large $\sigma$ and large $\delta$ lower both the collapse and
ring fragmentation regions. In other words, the increase of $\sigma$
or $\delta$ reduces the danger of Jeans collapse instability but
makes the system more vulnerable to ring fragmentation instability.
The enhancement of magnetic field reduces the danger of ring
fragmentation, suppresses Jeans collapse instabilities for
$-1/4<\beta<1/4$, and aggravates Jeans collapse instabilities for
$1/4<\beta<1/2$.

An interesting result is that when we examine the $m=0$ case
for both aligned and unaligned perturbations, we find that
the aligned $m=0$ case is just the limit of the unaligned
$m=0$ case with the radial wavenumber $\xi\rightarrow0$
(i.e., the breathing mode) which does not merely represent
a rescaling of the background MHD rotational equilibrium.

We discuss briefly on the well-known magneto-rotational 
instabilities (MRIs) in a weakly magnetized disc system. In general,
for a disc of finite thickness, if its angular velocity decreases
outward with a weak magnetic field, the disc rotational profile 
becomes unstable caused by MRIs. Such MRIs are widely invoked to 
generate and sustain MHD turbulence in the disc and result in an 
outward transport of angular momentum. Both local and global MRIs 
appear in extensive numerical simulations. Although MRIs are initially 
explored with a vertical magnetic field, it can also manifest when 
the background magnetic field is azimuthal \citep{b0b,b7e1}. 
Although a purely azimuthal magnetic field is less effective in 
producing an MHD turbulence, numerical simulations do show that 
a weak vertical magnetic field can be produced from an azimuthal 
magnetic field and the entire system then behaves similar to a 
system with a purely vertical magnetic field \citep{b2b1,b0b}. 
Such MRIs are suppressed when the magnetic field becomes so strong 
such that magnetic and rotational energies approach an equipartition 
condition \citep{b7e1}. Since only razor-thin discs are considered 
in our current formulation, MRIs do not appear in our model. 
Nevertheless, in reality, the finite thickness of a gas disc can 
induce MRIs and result in an outward transport of angular momentum. 
Moreover, locally produced MRIs in a galactic disc may further 
complicate the appearance of large-scale spiral patterns 
discussed here.

(\textrm{iv})\emph{ Partial Discs with a Dark Matter Halo}

We apply previous analysis to a composite system of partial discs
with $F<1$. For the $m=1$ case, the potential ratio parameter $F$
plays an important role to affect the behaviour of $D_s^2$ solutions.
In the aligned $m=1$ cases, when $\beta$ is not too small (e.g.,
$\beta>0$), there is a `divergent point' $F_c$ determined by $\beta$
where the $y_1$ branch diverges and physical solutions appear only
for $F\geq F_c$.
When $\beta$ is very small (e.g., $\beta=-0.24$) together with other
parameters (i.e., $\sigma=1.2,\ q=0.4$), the only physical solutions
lie in the interval $0<F\leq F_{c2}$ [see panel (b) of Fig.\ \ref{fig:11}].
For stationary logarithmic spirals with $m=1$, the value $F_c$ depends
on both $\beta$ and $\xi$ and we would propose that spiral galaxies
with stationary patterns of one arm and a large proportion of dark
matter halo should have large radial wavenumber (i.e., tightly wound
one-arm spiral).


When $m\geq2$ in both aligned and unaligned cases, the dependence
of $D_s^2$ solutions on the potential parameter $F$ is generally weak.
Nevertheless, $F$ may change phase relations of perturbation variables
significantly. First, we discuss the aligned cases. When $\beta$ is
small (e.g., $\beta=-0.24$), the ratio $S^g/S^s$ for the $y_1$ branch
can be positive for small $\xi$ and small $\sigma$ although for a
full disc this ratio remains always negative. For the phase relation
between $S^g$ and $R$, one feature is that when $q$ is large, $R$ can
be zero [i.e., a singularity in the ratio $S^g/(\mathrm{i}R)$] for a
certain $\xi_c$. Meanwhile, there are also some variations in the
phase difference for partial discs which are clearly shown in panel
(c) of Fig.\ \ref{fig:19}. For $\beta=1/4$ case and in full discs,
the $y_2$ branch has negative $S^g/S^s$ ratio when $\sigma=1$.
In a partial composite disc system with $F=0.5$, for the entire $y_2$
branch the ratio $S^g/S^s$ is positive. Meanwhile, for both $y_1$ and
$y_3$ branches, the magnitude ratio $|S^g/S^s|$ is much larger than
that of full disc cases. For the phase relation between $S^g$ and $R$,
we no longer have $R=0$, but the phase variation is still evident
(see Fig.\ \ref{fig:20} for a comparison).

Finally, in the $m=0$ case with purely radial oscillations in
the composite MHD system, the presence of a dark matter halo
mass greatly reduces the danger of both Jeans collapse and
ring fragmentation instabilities. In other words, a composite
system of partial discs appear more stable than full discs.

\section*{Acknowledgments}
This research has been supported in part by the ASCI Centre for
Astrophysical Thermonuclear Flashes at the University of Chicago,
by the Special
Funds for Major State Basic Science Research Projects of China,
by the Tsinghua Centre for Astrophysics, by the Collaborative
Research Fund from the National Science Foundation of China (NSFC)
for Young Outstanding Overseas Chinese Scholars (NSFC 10028306) at
the National Astronomical Observatories, Chinese Academy of Sciences,
by the NSFC grants 10373009 and 10533020 at the Tsinghua University,
and by the SRFDP 20050003088
and the Yangtze Endowment from the Ministry of Education at the Tsinghua
University. Affiliated institutions of Y-QL share this contribution.


\appendix
\section[]{A Further Discussion for\\
the Aligned Axisymmetric Case}

As already noted in the main text, the $m=0$ case should be
calculated in the process of first setting $m=0$ and then
taking the limit $\omega\rightarrow0$. We therefore start from
equations $(\ref{eq:48})-(\ref{eq:52})$ for our calculation. Now
with $S,\ \Sigma_0\propto r^{-1-2\beta}$ in equations (\ref{eq:48})
and (\ref{eq:52}), we have $U\propto r$; it then follows from
equations (\ref{eq:50}) and (\ref{eq:52}) that $J\propto r^{1-\beta}$
and $\Psi\propto r^{-2\beta}$ as before.
Under such circumstances, the radial scalings of certain
perturbation variables are different from before and the
scale-free condition is invalid if $\omega\neq0$ holds. However, we
find that in the limit $\omega\rightarrow0$, we must require
$U\rightarrow0$. Now when the first term on the left-hand side of
equation (\ref{eq:49}) together with the first term on the left-hand
side of the second relation of equations (\ref{eq:52}) approach
zero, the scale-free condition becomes satisfied again. Meanwhile,
$U/\omega\neq0$ in such a limiting process, leading to a solution
for this case. We show the result of this limiting process below.
\begin{equation}
\begin{split}
&-\frac{2(1-\beta)\Omega_g^2r^2}{(1-2\beta)}S^g
=2\beta\Sigma_0^g\Psi^g-
\frac{C_A^2(4\beta-1)\beta}{(1-2\beta)}S^g\ ,\\
&-\frac{2(1-\beta)\Omega_s^2r^2}{(1-2\beta)}S^s
=2\beta\Sigma_0^s\Psi^s\ .
\end{split}\label{eq:A1}
\end{equation}
By a substitution of equations (\ref{eq:35}) and (\ref{eq:53}) into
equation (\ref{eq:A1}) and a further simplification, we obtain
\begin{equation}
\begin{split}
&\bigg[2\beta
a_g^2+\frac{2(1-\beta)\Omega_g^2r^2+C_A^2\beta(1-4\beta)}
{(1-2\beta)}\\ &\qquad
-2\pi Gr(2\beta\mathcal{P}_0)\Sigma_0^g\bigg]S^g
-2\pi Gr(2\beta\mathcal{P}_0)\Sigma_0^gS^s=0\ ,\\
&\bigg[2\beta a_s^2+\frac{2(1-\beta)\Omega_s^2r^2}
{(1-2\beta )}-2\pi Gr(2\beta\mathcal{P}_0)\Sigma_0^s\bigg]S^s
\\ &\qquad
-2\pi Gr(2\beta\mathcal{P}_0)\Sigma_0^sS^g=0\ .
\end{split}\label{eq:A2}
\end{equation}
The two relations of equation (\ref{eq:A2}) cannot be
simultaneously satisfied unless the coefficient determinant
of equation (\ref{eq:A2}) vanishes; for non-trivial solutions,
this requirement leads to the dispersion relation in the form of
\begin{equation}
\begin{split}
&[\mathcal{V}^g+\mathcal{F}-2\pi
Gr(2\beta\mathcal{P}_0)\Sigma_0^g][\mathcal{V}^s-2\pi
Gr(2\beta\mathcal{P}_0)\Sigma_0^s]
\\ &\qquad
=4\pi^2G^2r^2(2\beta\mathcal{P}_0)^2\Sigma_0^g\Sigma_0^s\ ,
\end{split}\label{eq:A3}
\end{equation}
where, for notational simplicity, we define
\begin{equation}
\begin{split}
&\mathcal{V}^i\equiv2\beta
a_i^2+{2(1-\beta)\Omega_i^2r^2}/{(1-2\beta )}\ ,\\
&\mathcal{F}\equiv {C_A^2\beta(1-4\beta^2)}/{(1-2\beta)}\ .
\end{split}\label{eq:A4}
\end{equation}
Substitutions of equations $(\ref{eq:20})-(\ref{eq:23})$
in equation (\ref{eq:A3}) yield
\begin{equation}
\begin{split}
&\bigg[\frac{2\beta}{1+2\beta}
-\frac{F(D_g^2+1+\mathcal{Q})\delta}{1+\delta}
+\frac{2(1-\beta)D_g^2-2\beta\mathcal{Q}}{1-2\beta}\bigg]\\
&\qquad\times\bigg[\frac{2\beta}{1+2\beta}-\frac{F(D_s^2+1)}
{1+\delta}+\frac{2(1-\beta)D_s^2}{1-2\beta}\bigg]
\\ &\qquad\qquad\qquad
=\frac{(D_s^2+1)(D_g^2+1+\mathcal{Q})F^2\delta}{(1+\delta)^2}\ ,
\end{split}\label{eq:A5}
\end{equation}
where the magnetic field effect is contained
in the following $\mathcal{Q}$ parameter
\begin{equation}
\mathcal{Q}\equiv {(4\beta-1)q^2}/{(4\beta+2)}\ .\label{eq:A6}
\end{equation}
By evaluating equation (\ref{eq:A5}) using radial
force balance relation (\ref{eq:22}), we obtain a quadratic
equation of $D_s^2$ which has two roots. To show the physical
meaning of this equation, we intend to compare this equation
with equation (\ref{eq:99}) where axisymmetric marginal
stabilities are discussed. We expect that by taking $\xi=0$
and $\beta=1/4$, equations (\ref{eq:A5}) and (\ref{eq:99})
should coincide and this is indeed the case by noting
$\mathcal{N}_m(1/4,\ 0)=\mathcal{P}_m(1/4)$,
$\mathcal{Q}(1/4)=0$, $\mathcal{A}_0^{S'}(0)=1/4$.
This result unambiguously shows that the aligned $m=0$ case does
actually correspond to the limiting case $\xi\rightarrow0$ of the
$m=0$ logarithmic spiral case. Since the potential-density
pairs carry different scales, only in the case $\beta=1/4$ do
they coincide.\footnote{In potential-density pair (\ref{eq:76})
for logarithmic spirals, the perturbed surface mass density does
not scale the same as the background surface mass density unless
$\beta=1/4$.}
Now it turns out that the aligned $m=0$ case is an extension of
the marginal stability curves in the limit $\xi\rightarrow0$.


This result may be understood as follows. In the aligned $m=0$ mode,
the equation of continuity is invalid unless in the limit of
$\omega=0$. Since $\Sigma_1(r)=\epsilon r^{-1-2\beta}
\mathrm{exp}(\mathrm{i}\omega t)$, the surface mass density of the
whole disc varies synchronously if $\omega\neq0$. However, we may
regard such mode as the limiting case of the spiral $m=0$ case with
radial wave number $\xi\rightarrow0$ and write
\begin{equation}
\Sigma_1(r)=\epsilon
r^{-1-2\beta}\mathrm{exp}(\xi\mathrm{ln}r+\mathrm{i}\omega t)\
.\label{eq:A7}
\end{equation}
Now when $\xi$ and $\omega$ is nonzero but extremely small, the
equation of continuity still holds. Here, the surface mass density
perturbation scale differs from that of the marginal stability curves
in Section 4.3 unless $\beta=1/4$. Thus in this perspective, the result
may be regarded as the marginal stability relation with perturbation
scale $\Sigma_1\propto r^{-1-2\beta+\mathrm{i}\xi}$ for small $\xi$
(i.e., widely open spirals) cases and is an extension of our
discussion on marginal stability problem in Section 4.3.

\section[]{Calculations\\
\qquad\qquad of Dispersion Relation\\
\qquad\qquad for the Aligned Cases}

We evaluate the MHD dispersion relation by computing the coefficient
determinant of equations (\ref{eq:55}), (\ref{eq:56}), (\ref{eq:57})
and (\ref{eq:59}) for a specified background equilibrium described
by equations $(\ref{eq:21})-(\ref{eq:23})$. The explicit coplanar
MHD perturbation equations in terms of characteristic parameters
$\beta,\ q,\ m,\ \delta,\ \eta,\ D_g,\ F$ are summarized below.
In the first step, we obtain
\begin{equation}
\begin{split}
&mD_gS^g-\frac{3A_g[D_g^2+1+{(4\beta-1)q^2}/{(4\beta+2)}]F\delta}
{2\pi G[2\beta\mathcal{P}_0(\beta)]r^{1+\beta}(1+\delta)}\mathrm{i}U^g\\
&+\frac{mA_g[D_g^2+1+{(4\beta-1)}q^2/{(4\beta+2)}]F\delta} {2\pi
G[2\beta\mathcal{P}_0(\beta)]r^{2+\beta}(1+\delta)}J^g=0\ ,
\end{split}\label{eq:b1}
\end{equation}
\begin{equation}
\begin{split}
&\bigg\{\frac{mA_gD_g}{r^{1+\beta}}+\frac{q^2A_g[(1-4\beta)^2-2m^2]}
{2m(1+2\beta)D_gr^{1+\beta}}\bigg\}\mathrm{i}U^g\\
&+\frac{2A_gD_g}{r^{2+\beta}}J^g -4\pi G\beta\mathcal{P}_m(\beta)S^s
-\frac{\pi G}{F\delta}\bigg\{4F\delta\beta\mathcal{P}_m(\beta)\\
&-\frac{[2\beta\mathcal{P}_0(\beta)]
[4\beta+(1-4\beta)q^2](1+\delta)}
{(1+2\beta)[D_g^2+1+{(4\beta-1)}q^2/{(4\beta+2)}]}\bigg\}S^g=0\ ,
\end{split}\label{eq:b2}
\end{equation}
\begin{equation}
\begin{split}
&\bigg[(1-\beta)\frac{A_gD_g}{r^{1+\beta}}+
\frac{q^2A_g(4\beta-1)}{(4\beta+2)D_gr^{1+\beta}}\bigg]\mathrm{i}U^g
\\ &+\frac{mA_gD_g}{r^{2+\beta}}J^g-2\pi Gm\mathcal{P}_m(\beta)S^s
-2\pi Gm\bigg\{\mathcal{P}_m(\beta)
\\ &-\frac{(1+2\beta)^{-1}[2\beta\mathcal{P}_0(\beta)](1+\delta)}
{[D_g^2+1+{(4\beta-1)}q^2/{(4\beta+2)}]F\delta}\bigg\}S^g=0
\end{split}\label{eq:b3}
\end{equation}
for the magnetized gas disc and
\begin{equation}
\begin{split}
&\bigg\{\frac{[m^2-2(1-\beta)]D_s^2}{[m^2+4\beta(1-\beta)]}
-\frac{1}{(1+2\beta )} \\ &\ \ \
+\frac{(D_s^2+1)\mathcal{P}_m(\beta)}
{[2\beta\mathcal{P}_0(\beta)](1+\delta)}\bigg\}S^g
+\frac{(D_s^2+1)\mathcal{P}_m(\beta)}
{[2\beta\mathcal{P}_0(\beta)](1+\delta)}S^s=0
\end{split}\label{eq:b4}
\end{equation}
for the stellar disc. In addition to the notations already
defined by equations (\ref{eq:65}) and (\ref{eq:A6}), we
here introduce
\begin{equation}
\begin{split}
&C(\beta)\equiv {2\beta}/{(1+2\beta)}\ ,\\
&H(D_g,\beta,q)\equiv {(D_g^2+1+\mathcal{Q})F}/{\mathcal{P}_0(\beta)}\ ,
\end{split}\label{eq:b5}
\end{equation}
and note from equation (\ref{eq:22}) that
\begin{equation*}
D_s^2=\eta H\mathcal{P}_0-1\ .
\end{equation*}
Using these newly defined notations, equations
$(\ref{eq:b1})-(\ref{eq:b4})$ take the forms of
\begin{equation}
mD_gS^g-\frac{3A_gH\delta \mathrm{i}U^g}{4\pi Gr^{1+\beta}(1+\delta)}
+\frac{mA_gH\delta J^g}{4\pi G\beta r^{2+\beta}(1+\delta)}=0\ ,\label{eq:b6}
\end{equation}
\begin{equation}
\begin{split}
&\bigg\{\frac{mA_gD_g}{r^{1+\beta}}
+\frac{q^2A_g[(1-4\beta)^2-2m^2]}
{2m(1+2\beta)D_gr^{1+\beta}}\bigg\}\mathrm{i}U^g
\\ &\ \
=2\pi G\beta\bigg\{2\mathcal{P}_m(\beta)
-\frac{[4\beta+(1-4\beta)q^2](1+\delta)}
{(1+2\beta)H\delta}\bigg\}S^g
\\ &\qquad\qquad\qquad\qquad
-\frac{2A_gD_g}{r^{2+\beta}}J^g +4\pi G\beta\mathcal{P}_m(\beta)S^s\ ,
\end{split}\label{eq:b7}
\end{equation}
\begin{equation}
\begin{split}
&\bigg [(1-\beta)\frac{A_gD_g}{r^{1+\beta}} +\frac{q^2A_g(4\beta-1)}
{(4\beta+2)D_gr^{1+\beta}}\bigg ]\mathrm{i}U^g
\\ &\qquad
=2\pi Gm\bigg[\mathcal{P}_m(\beta)-\frac{2\beta(1+\delta)}
{(1+2\beta)H\delta}\bigg]S^g
\\ &\qquad\qquad\qquad\qquad
-\frac{mA_gD_g}{r^{2+\beta}}J^g +2\pi Gm\mathcal{P}_m(\beta)S^s
\end{split}\label{eq:b8}
\end{equation}
for the magnetized gas disc and
\begin{equation}
\begin{split}
&\bigg\{\frac{\mathcal{B}_m[\eta
H\mathcal{P}_0(\beta)-1]}{\mathcal{A}_m}-1+\frac{\eta
H\mathcal{P}_m(\beta)} {C(1+\delta)}\bigg\}S^s
\\ &\qquad\qquad\qquad\qquad\qquad
+\frac{\eta H\mathcal{P}_m(\beta)}{C(1+\delta)}S^g=0
\end{split}\label{eq:b9}
\end{equation}
for the stellar disc. We rearrange the terms in these four equations
in the order of ($S^g, \mathrm{i}U^g, J^g, S^s$) to identify and
manipulate the coefficient determinant by drawing several factors
and by properly adding and subtracting lines and rows. The simplified
determinant appears as
\begin{equation}
 \left|
 \begin{array}{cccc}
 a_{11} & a_{12} & a_{13} & 0\\
 b_{11} & b_{12} & 1 & b_{14}\\
 c_{11} & c_{12} & 1 & c_{14}\\
 d_{11} & 0 & 0 & d_{14}
 \end{array}
 \right|
\end{equation}\label{eq:b10}

\noindent
where the eleven coefficients are given explicitly as
\begin{equation}
\begin{split}
&a_{11}\equiv D_g^2\ ,\qquad
a_{12}\equiv -\frac{3\delta D_g^2}{1+\delta}\ ,\qquad
a_{13}\equiv\frac{\delta}{2\beta(1+\delta)}\ ,\\
&b_{11}\equiv-H\beta\mathcal{P}_m
+{\beta(1+\delta)(C-\mathcal{Q})}/{\delta}\ ,\\
&b_{12}\equiv m^2D_g^2+\frac{(4\beta-1)^2-2m^2}{(4\beta-1)}\mathcal{Q}\ ,
\quad b_{14}\equiv -H\beta\mathcal{P}_m\ ,\\
&c_{11}\equiv -H\mathcal{P}_m+\frac{C(1+\delta)}{\delta}\ ,
\quad c_{12}\equiv 2(1-\beta)D_g^2+2\mathcal{Q}\ ,\\
&c_{14}\equiv -H\mathcal{P}_m\ ,\qquad
d_{11}\equiv {\eta H\mathcal{P}_m\mathcal{A}_m}/{[C(1+\delta)]}\ ,\\
&d_{14}\equiv \eta H\mathcal{P}_0\mathcal{B}_m
-\mathcal{A}_m-\mathcal{B}_m+\eta H\mathcal{P}_m\mathcal{A}_m/[C(1+\delta)]\ .
\end{split}\label{eq:b11}
\end{equation}
With these definitions, it is straightforward to evaluate the
determinant numerically and the result is shown in Section 3. We
note that in acquiring the coefficients of the polynomial equation
of $y$ in the form of equations (\ref{eq:67}) and (\ref{eq:68}), it
is somewhat onerous and tedious to combine fragment terms (e.g.,
$m^2,\ \beta^2,\ \beta$) to form extra coefficients $\mathcal{A}_m$
and $\mathcal{B}_m$ etc.

\section[]{Calculations\\
\qquad\qquad of Dispersion Relation\\
\qquad\qquad for Logarithmic Spirals }

Since the MHD perturbation equations for logarithmic spiral
perturbations parallel those for aligned perturbations, the
procedure of derivations is strikingly similar. We will not
repeat the steps but directly summarize the key results of
our calculations.

Similar to the aligned case, the main purpose is to express the
coefficient determinant of equations (\ref{eq:55}), (\ref{eq:56}),
(\ref{eq:57}) and (\ref{eq:59}) in reference to the background
equilibrium described by equations $(\ref{eq:21})-(\ref{eq:23})$ in
terms of the dimensionless parameters ($\beta,\ q,\ m,\ \delta,\
\eta,\ D_g,\ F$). Following the same steps of Appendix B and using
the same notations, we show the simplified determinant below.
\begin{equation}
 \left|
 \begin{array}{cccc}
 a_{11} & a_{12} & a_{13} & 0\\
 b_{11} & b_{12} & 2 & b_{14}\\
 c_{11} & c_{12} & 1 & c_{14}\\
 d_{11} & 0 & 0 & d_{14}
 \end{array}
 \right|
\end{equation}\label{eq:c1}

\noindent
where the eleven coefficients are explicitly defined by
\begin{equation}
\begin{split}
&a_{11}\equiv D_g^2\ ,\qquad
a_{12}\equiv -\frac{(1/2+\beta-\mathrm{i}\xi)\delta
D_g^2}{4\beta(1+\delta)}\ , \\
&b_{11}\equiv -(1-2\mathrm{i}\xi)H\mathcal{N}_m
+\frac{(1+\delta)[(1-2\mathrm{i}\xi)C-4\beta\mathcal{Q}]}{\delta}\ ,\\
&b_{12}\equiv m^2D_g^2-\frac{2m^2
+\mathrm{i}\xi(4\beta-1)+2\xi^2}{(4\beta-1)}\mathcal{Q}\ ,\\
&a_{13}\equiv\frac{\delta}{4\beta(1+\delta)}\ ,
\qquad b_{14}\equiv -(1-2\mathrm{i}\xi)H\mathcal{N}_m\ ,\\
&c_{11}\equiv -2H\mathcal{N}_m+{2C(1+\delta)}/{\delta}\ ,\\
&c_{12}\equiv (1-\beta)D_g^2+\mathcal{Q}\ ,\qquad\quad
c_{14}\equiv -2H\mathcal{N}_m\ ,\\
&d_{11}\equiv {\eta H\mathcal{N}_m\mathcal{A}_m}/{[C(1+\delta)]}\ ,\\
&d_{14}\equiv\eta
H\mathcal{P}_0\mathcal{B}_m-\mathcal{A}_m-\mathcal{B}_m
+\frac{\eta H\mathcal{N}_m\mathcal{A}_m}{C(1+\delta)}\ .
\end{split}\label{eq:c2}
\end{equation}
It is then straightforward to evaluate this determinant
numerically and the results are shown in Section 4 for
logarithmic spirals of MHD perturbations.

\bsp

\label{lastpage}

\end{document}